\documentclass[a4paper,11pt]{article}
\pdfoutput=1
\usepackage{jheppub}
\usepackage{amsmath,amssymb,amsfonts,graphicx,caption,subcaption,xcolor, slashed}
\usepackage[english]{babel}
\usepackage[colorlinks=true,linktocpage=true,linkcolor=blue,citecolor=blue]{hyperref}
\usepackage{marginnote}
\usepackage{enumitem}

\newcommand{\executeiffilenewer}[3]{%
\ifnum\pdfstrcmp{\pdffilemoddate{#1}}%
{\pdffilemoddate{#2}}>0%
{\immediate\write18{#3}}\fi%
}

\numberwithin{equation}{section}

\newcommand{\be}{\begin{equation}} \newcommand{\ee}{\end{equation}}
\newcommand{\bea}{\begin{eqnarray}} \newcommand{\eea}{\end{eqnarray}}
\newcommand{\beann}{\begin{eqnarray*}}  \newcommand{\eeann}{\end{eqnarray*}}
\newcommand{\bfig}{\begin{figure}} \newcommand{\efig}{\end{figure}}
\newcommand{\ba}{\begin{array}} \newcommand{\ea}{\end{array}}
\newcommand{\bcen}{\begin{center}} \newcommand{\ecen}{\end{center}}
\newcommand{\btab}{\begin{tabular}} \newcommand{\etab}{\end{tabular}}
\newcommand{\nn}{\nonumber}
\newcommand{\matt}{\left ( \begin{array}{ccc}}
    \newcommand{\ematt}{\end{array} \right )} \newcommand{\matf}{\left ( \begin{array}{cccc}}
    \newcommand{\ematf}{\end{array} \right )} \newcommand{\vect}{\left ( \begin{array}{c}}
    \newcommand{\evect}{\end{array} \right )}    \def\beqn{\begin{eqnarray}}
 \def\eeqn{\end{eqnarray}}  
     
\newtheorem{Proposition}{Proposition}[section]

\newtheorem{Theorem}{Theorem}[section]
\newtheorem{Lemma}{Lemma}[section]
\newtheorem{Corrolary}{Corrolary}[section]

\newcommand{\bp}{\begin{Proposition}}   \newcommand{\ep}{\end{Proposition}}
\newcommand{\bt}{\begin{Theorem}}   \newcommand{\et}{\end{Theorem}}
\newcommand{\bl}{\begin{Lemma}}     \newcommand{\el}{\end{Lemma}}
\newcommand{\bc}{\begin{Corrolary}} \newcommand{\ec}{\end{Corrolary}}

\newcommand{\sdc}{\sigma_{\rm DC}}

\title{Noisy Branes}
\author[a]{Mario Ara\'ujo,}
\author[a]{Daniel Are\'an,}
\affiliation[a]{Max-Planck-Institut f\"ur Physik (Werner-Heisenberg-Institut)\\ 
F\"ohringer Ring 6, D-80805 Munich, Germany}
\author[b]{and Javier M. Lizana}
\affiliation[b]{CAFPE and Departamento de F\'isica Te\'orica y del Cosmos\\
Universidad de Granada, E-18071 Granada, Spain}
\emailAdd{maraujo@mpp.mpg.de}
\emailAdd{darean@mpp.mpg.de}
\emailAdd{jlizan@ugr.es}

\preprint{MPP-2016-34}

\abstract{
We study the effects of disorder on strongly coupled compressible matter in 2+1 dimensions.
Our system consists of a D3/D5 intersection at finite temperature and in the presence of a disordered
chemical potential.
We first study the impact of disorder on the charge density and the quark condensate.
Next, we focus on the DC conductivity and derive analytic expressions for the corrections
induced by weak disorder.
It is found that disorder enhances the DC conductivity at low charge density,
while for large charge density the conductivity is reduced.
We present numerical simulations both for weak and strong disorder.
Finally, we show how disorder gives rise to a sublinear behavior for the conductivity as a function of the
charge density, a behavior qualitatively similar to predictions and observations for electric transport
in graphene.}

\makeatletter

\renewcommand*\env@matrix[1][c]{\hskip -\arraycolsep
  \let\@ifnextchar\new@ifnextchar
  \array{*\c@MaxMatrixCols #1}}
  
\def\@eqnnum{{\normalfont\normalcolor(\theequation)}}  

\makeatother

\begin{document}

\maketitle


\section{ Introduction}

The interplay of disorder and strong interactions is a challenging  problem in Condensed Matter,
with a wide range of potential applications from High-Tc superconductors~\cite{Pan_Nature,Renner_Nature} to 
graphene \cite{PhysRevLett.99.246803,PhysRevLett.98.186806}. At the theoretical level it poses important questions as 
the existence and nature of disordered quantum critical points~\cite{PhysRevB.27.413,0305-4470-39-22-R01}, and the
possibility of disorder-induced metal to insulator phase transitions for strongly interacting systems.

Gauge/Gravity duality is a promising venue to address strongly coupled problems in Condensed Matter, and the last years have
seen interesting progress towards a description of disordered strongly coupled systems.
These advances include holographic models of  disordered fixed points \cite{Hartnoll:2014cua,Garcia-Garcia:2015crx}, 
disordered superconductors \cite{Arean:2014oaa,Arean:2013mta}, and hyperscaling violating geometries,
which are promising candidates to duals of strange metals, deformed by disordered sources~\cite{Lucas:2014zea}.

A natural procedure to characterize the effects of disorder is to study the transport properties of the system,
and in particular the electrical conductivity. Compelling results for the transport properties of solutions dual to theories 
with disorder have been obtained, mainly through numerical solutions of Einstein plus matter 
theories \cite{Donos:2014yya,Hartnoll:2015rza}, but also
via analytic computations at weak disorder \cite{O'Keeffe:2015awa,Garcia-Garcia:2015crx}. Finally, hydrodynamic models of strongly coupled disordered systems
have led to promising results like the fitting of experimental data for graphene~\cite{Lucas:2015sya}, or the description
of phase disordered superconducting phase transitions~\cite{Davison:2016hno}.

All the holographic models dual to disordered theories we have described above, and the majority of those constructed
thus far, are of the so-called bottom-up type. They are effective models whose Lagrangian is not derived from a solution
of String Theory, or Supergravity, and thus lack a microscopic description. In this note we will instead implement disorder
in a top-down model that has been one of the workhorses of Gauge/Gravity duality applications  to QCD-like theories;
that of probe branes~\cite{Karch:2002sh}. We will consider a D5-brane probe embedded in the geometry generated by 
$N_c$ D3-branes: the D5 shares two spatial directions with the D3s, and 
introduces fundamental degrees of freedom, quarks, along a (2+1)-dimensional defect in the theory dual to $N_c$ D3-branes. 
More precisely, for $N_f$ D5-brane probes, the system is dual to ${\cal N}=4$ $SU(N_c)$ SYM with $N_f$ ${\cal N}=2$ matter 
hypermultiplets living on a (2+1)-dimensional defect~\cite{Karch:2000gx,DeWolfe:2001pq}. Since we are interested in
systems at finite temperature and charge density, we will work with the black hole background generated by black D3-branes, 
and add charge density by switching on the temporal component of the gauge field living on the worldvolume 
of the D5-brane~\cite{Kobayashi:2006sb,Evans:2008nf}.

The implementation of disorder via a probe brane model was considered for the first time in~\cite{Ryu:2011vq},
and subsequently
in~\cite{Ikeda:2016rqh}, where it was shown that the DC conductivity is bounded, and, as a consequence, an insulating
phase is
excluded from this scenario. 
However, it is in this work that for the first time disorder is implemented explicitly in a top-down holographic model
of probe branes.
We construct disordered embeddings of a probe D5-brane in a black D3-background by switching on a disordered chemical
potential
on the worldvolume of the probe. The analysis of those embeddings, and the study of their fluctuations
has produced the following results

\begin{itemize}
\item We construct both massive and massless inhomogeneous embeddings characterized by a disordered chemical potential,
and compute their charge density and, for the massive case, also the value of the quark condensate.

\item We study the effects of disorder on the charge density and the quark condensate using analytic and numerical methods.
Two regimes are found: at small charge density the quark condensate scales quadratically with the strength of disorder, while
the charge density is almost independent of that disorder strength. For large charge density the converse happens: the quark
condensate is largely unaffected by disorder while the charge density scales quadratically with the disorder strength.

\item We express the DC conductivity, $\sdc$, in terms of horizon data and obtain analytic expressions 
for $\sdc$ in a small
disorder regime. For small charge density $\sdc$ is enhanced by disorder. For large charge density,
$\sdc$ decreases with
the disorder strength. Numerical simulations, which agree with the analytic predictions, 
confirm this scenario.

\item We compute the dependence of $\sdc$ on the charge density via numerical simulations, and 
analytic approximations.
At weak disorder $\sdc$ scales linearly with the charge density (except at very low charge density),
with a slope that is reduced as the noise strength
is increased.
At strong disorder the dependence of $\sdc$ on the charge density becomes sublinear.
This last behavior shows similarities to that observed in graphene near the charge neutrality point.
Finally, we observe that the analytic, or semi-analytical, approximations agree very well with the numerical simulations.

\item We study the spectral properties of the system by considering a noise characterized by a 
Fourier power spectrum
of the form $1/k^{2\alpha}$.  The resulting power spectra for the charge density and the quark 
condensate are found to be
of the form $1/k^{2\alpha-2}$ and $1/k^{2\alpha+6}$ respectively. With respect to the input power 
spectrum, our holographic model smooths out the quark condensate, while it makes  the charge density 
more irregular.

\end{itemize}

The rest of this paper is organized as follows. 
Section~\ref{sec:disint} is devoted to the construction of disordered embeddings of a probe D5-brane: in 
Sec.~\ref{ssc:background} we introduce the background geometry, and in Sec.~\ref{ssc:embed} we write down
the action and asymptotics for the embedding of the probe D5-brane. The disordered chemical potential is described 
in Sec.~\ref{ssc:dismu}, and the numerical methods used to construct the embeddings are discussed in Sec.~\ref{ssc:numerics}. In Section~\ref{ssc:bkresults} we present the 
numerical results for the 
embeddings and discuss the effects of disorder on the charge density and the quark condensate. 
Sec.~\ref{sec:conductivity} is
dedicated to the study of the electrical conductivity. In Sec.~\ref{ssec:dccond} we compute the DC conductivity in terms 
of horizon data, and in Sec.~\ref{sssec:pertnoise} we derive analytic expressions for $\sdc$ in a small noise expansion. In Sec.~\ref{sssec:strongnoise} we introduce a semi-analytical approach to $\sdc$ valid at all orders in the strength of noise, and obtain predictions for the behavior of $\sdc$ at strong noise.
In Sec.~\ref{ssc:sgresults} we present our numerical results for $\sdc$ and compare them with the 
analytic predictions, paying special attention in Sec.~\ref{ssec:sdcvsrho0} to 
the behavior of $\sdc$ as a function of 
$\langle\rho\rangle$.
Finally, in Section~\ref{sec:spectral} we analyze the spectral properties of the system. We conclude 
in Sec.~\ref{sec:conclusions} with a summary of our results and a review of the ways forward this work opens.
We have included three appendices in this manuscript: App.~\ref{app:homcase} is dedicated to the homogeneous 
version of our model ({\emph i.e.} without disorder). In App.~\ref{app:numstability} we discuss  the reliability of our 
numerics against the lattice size, and present supplementary results for strong disorder.
In App.~\ref{app:corremb} we show numerical embeddings for the case of a disorder characterized by a 
Fourier power spectrum $\sim 1/k^{2\alpha}$.

\section{ Disordered D3/D5 intersection}
\label{sec:disint}
Our setup is built upon the supersymmetric intersection
of $N_c$ D3- and $N_f$ D5-branes along 2+1 spacetime dimensions, which is dual to (3+1)-dimensional
${\cal N} = 4$ SYM with $N_f$ fundamental hypermultiplets living on a (2+1)-dimensional 
defect~\cite{Karch:2000gx,DeWolfe:2001pq}. We work in the probe limit $N_f\ll N_c$, and hence consider a probe D5-brane
in the background generated by $N_c$ D3-branes. Moreover, we are interested in systems at finite temperature and with 
a nonzero density of the fundamental degrees of freedom realized by the strings stretched between the D3- and D5-branes.

\subsection{Background}
\label{ssc:background}
The metric of the geometry generated by $N_c$ black D3-branes can be written as
\bea
&ds^2=\frac{L^2}{z^2}\left(-\frac{f(z)^2}{h(z)}\,dt^2+h(z)\,d\vec{x}^2+dz^2
\right)+L^2\,d\Omega_5^2\,,\\
&{\rm with}\quad
f(z)=1-z^4\,,\qquad h(z)=1+z^4\,,
\label{eq:metric}
\eea
where we are following the conventions of \cite{Araujo:2015hna}, and we have performed a rescaling
that sends the horizon radius $z_0$ to the unity.\footnote{We have actually defined dimensionless
coordinates $(\tilde z,\tilde x_\mu)=1/z_0\,(z, x_\mu)$, and dropped the tilde for notational simplicity.}
Remember that the temperature of the black hole
is determined in terms of $z_0$ as $T=\sqrt{2}/(\pi\,z_0)$.
It is straightforward to check that this metric is asymptotic, as $z\to0$, to $AdS_5\times S^5$,
and one should recall that the AdS radius $L$, the number of D3-branes, $N_c$, the string tension 
$(2\pi\alpha')^{-1}$, and the coupling constant of the dual theory, $g_{\rm YM}$,
are related via $L^4/\alpha'^2=2g_{\rm YM}^2\,N_c\equiv2\lambda$,  where $\lambda$ is the 't Hooft coupling.
This background possesses a nonzero RR five form given by $d{\rm Vol} (S^5)+{\rm h.d.}$, which
will not play any role in this setup.

\subsection{Embedding}
\label{ssc:embed}
The probe D5-brane is extended along two Minkowski directions, say $(x,y)$, the radial coordinate
$z$, and wraps an $S^2$ inside the internal $S^5$ whose metric can be written as
\be
d\Omega_5^2=d\theta^2+\sin^2\theta\,d\Omega_2^2+\cos^2\theta\,d\tilde\Omega_2^2\,,
\label{eq:s5metric}
\ee
where $\Omega_2$ is the volume element of the $S^2$ wrapped by the probe brane,
and the D5 sits at a fixed point of the remaining $S^2$.
The embedding can then be described in terms of the coordinate $\theta$ determining
the radius of the $S^2$ wrapped by the probe. For simplicity we will work in terms of
$\chi=\cos\theta$.

We will study configurations with finite charge density of the fundamental fields introduced by the
D5-brane, and hence we must turn on the temporal component of the $U(1)$ worldvolume gauge field.
Moreover, we want to describe a system where the charge density depends on one of the spatial
directions, which we choose to be $x$.
Therefore, the embedding is described in terms of the fields
\be
\chi(z,x)\,,\quad A_t={L^2\over 2\pi\,\alpha'\,z_0}\,\phi(z,x)\,,
\label{eq:fcontent}
\ee
where as in \cite{Araujo:2015hna} we have written $A_t$ in terms of a conveniently
dimensionless field $\phi$.
The action is given by the DBI for the probe D5-brane, and takes the form
\begin{equation}
{{\cal S}\over N_f\,T_{D5}\,L^6}=-\int dt\,d^2x\,dz\,d\Omega_2\, f\,z^{-4}\,
\sqrt{h\, (1-\chi^2)\,(S_\chi+S_\phi + S_{\text{int}})}\,,
\label{eq:action}
\end{equation}
with
\begin{align}
& S_{\chi}=1-\chi^2+z^2 \chi'^2+\frac{z^2\,\dot\chi^2}{h}\,,
\nn \\
& S_\phi=-\frac{z^4(1-\chi^2)}{f^2}\left(h\,\phi'^2+\dot \phi^2\right)\,,
\nn \\
& S_{\text{int}}=-\frac{z^6(\dot{\chi}\phi'-\chi' \dot{\phi})^2}{f^2 }\,,
\end{align}
where the tilde and the dot denote a partial derivative with respect to $z$ and $x$ respectively. 
The equations of motion for $\chi(z,x)$ and $\phi(z,x)$ follow readily from this action, and they have
been written explicitly in the appendix of \cite{Araujo:2015hna}.
In the last part of this work we will consider massless embeddings corresponding to $\chi(z,x)=0$.
In that case the equation of motion for $\chi(z,x)$ is trivially satisfied, while that for $\phi(z,x)$
takes the form
\begin{align}
&z^3 \left[\dot\phi^2 \left(\phi ' \left(3 z\,h'-4 h\right)+2 h\,z\, \phi ''\right)
-4 h\, z\, \dot\phi\, \phi '\, \overset{.}{\phi '}
+2 h \left(\phi '\right)^2 \left(z\,\ddot\phi+\phi'\left(z\,h'-2 h\right)\right)\right]\nn\\
&
\quad-f^2 \left(2 \ddot\phi+3 h'\,\phi '+2 h\,\phi''\right)+2 f\, h\, f'\, \phi'=0\,.
\label{eq:masslessphi}
\end{align}

One can read the observables of the dual theory from the UV asymptotics of the embedding fields.
These result from solving the equations of motion in the $z\to0$ limit, and read
\begin{subequations}
\label{eq:embUV}
\begin{align} 
& \phi(z,x)=\mu(x)-\rho(x) z+{\cal O}(z^{2})\,, \label{eq:phiUV}\\
&\chi(z,x)=m(x) \,z+c(x)z^2+{\cal O}(z^{3})\,, \label{eq:chiUV}
\end{align}
\end{subequations}
where $\mu$, $\rho$, $m$, and $c$ determine the chemical potential, charge density, quark mass,
and quark condensate respectively. Proceeding as in \cite{Araujo:2015hna} we plug in the
dimensionful constants, and define the quark mass $M_q=\sqrt{\lambda}\,\bar M$,
where $\bar M = m/z_0$. We then arrive to
\be
m={2\sqrt{2}\over\pi\sqrt{\lambda}}\,{M_q\over T}\,,\qquad
\mu = {2\over\sqrt{\lambda}}\,{\bar\mu\over T}\,,\qquad
\rho = {2\sqrt{2}\over\pi\sqrt{\lambda}}\, {\bar \rho \over T^2}\,,
\label{eq:dimratios}
\ee
where $\bar\mu$ and $\bar\rho$ are the dimensionful chemical potential and charge density 
respectively.
Moreover, as discussed with detail in \cite{Mateos:2007vn}, $c$ is proportional to the condensate
of the supersymmetric version of the quark bilinear ${\cal O}_m=\bar\psi\,\psi+\dots$, (the
dots stand for terms including the superpartners of $\psi$)
\be
c=-{1\over N_f\,N_c}{\langle{\cal O}_m\rangle\over T^2}\,.
\label{eq:condensate}
\ee
We will work at fixed $\mu$ and $m$. The phase diagram of the homogeneous D3/D5
setup was studied in \cite{Evans:2008nf}, and reviewed in \cite{Araujo:2015hna} in terms of $\mu$ and $m$.
As first explained in \cite{Kobayashi:2006sb}, at finite charge density only black hole embeddings exist.
These are embeddings for which the probe brane intersects the horizon.

Finally, we shall study the IR asymptotics of our system. 
Imposing regularity at the black hole horizon, it is straightforward to check that
in its vicinity the solutions for the embedding fields take the form
\begin{subequations}
\label{eq:bhexpansion}
\begin{align}
& \phi(z,x)=a^{(2)}(x)\,(1-z)^2+{\cal O}((1-z)^{3})\,, \label{eq:phibh}\\
&\chi(z,x)=C^{(0)}(x)+C^{(2)}(x)\,(1-z)^2+{\cal O}((1-z)^{3})\,,\label{eq:chibh}
\end{align}
\end{subequations}
where $C^{(2)}(x)$ is a function of $C^{(0)}(x)$ and $a^{(2)}(x)$ as shown in \cite{Araujo:2015hna}.

\subsection{Disordered $\mu$}
\label{ssc:dismu}

To mimic a random on-site potential as that used originally by Anderson \cite{Anderson:1958vr},
we introduce a noisy chemical potential of the form
\begin{equation}
\mu(x)=\mu_0+{\mu_0\over25}\,w\,\sum_{k=k_0}^{k_*}\,\cos(k\,x+\delta_k)\,,
\label{eq:noisefunc}
\end{equation}
with $\delta_k$ being a random phase for each wave number $k$, and $w$ a parameter that determines
the strength of the noise. This chemical potential does not depend on $y$, hence our setup is
homogeneous along this remaining spatial direction. We discretize the space along $x$ and impose
periodic boundary conditions in that direction, which results in $k$ taking the values
\be
k_n={2\pi\over L_x}\,(n+1)\quad {\rm with}\quad 0\leq n< N=\frac{k_*}{k_0}\,,
\ee
where $L_x$ is the length of the (cylindrical) system along $x$. Notice that this noise is a truncated
version of Gaussian white noise.\footnote{As explained in \cite{Hartnoll:2014cua,Garcia-Garcia:2015crx},
in terms of the Harris criterion~\cite{Harris:1974zz} that generalizes the standard power-counting criterion to random couplings, our chemical potential disordered along one dimension introduces relevant disorder. One would have
to go beyond the probe limit to investigate the expected important effects of a relevant disorder in the IR of the theory.} The highest wave number, $k_*$, plays the role of the inverse of the correlation length for the chemical potential, while the lower wave number, $k_0$, is proportional
to the inverse of the system size.
Moreover, for a lattice with $N_x$ sites along the $x$ direction,
\begin{equation}
k_*\leq k_{\rm ns}
\equiv\frac{\pi}{L_x}\,(N_x-1),
\end{equation}
where $k_{\rm ns}$ is the Nyquist frequency for that lattice.\footnote{We refer to the frequency saturating the
Nyquist sampling rate, which is half the frequency of the sampling resulting from our lattice. In particular, to recover all the 
Fourier components of a periodic wave, one needs a sampling rate that is at least twice that of the highest
mode. Our system of length $L_x$ is sampled by a periodic lattice with $N_x-1$ points, hence 
the sampling wave number is ${2\pi\over L_x}(N_x-1)$.}
The properties of this choice of disorder
have been discussed in more detail in \cite{Arean:2014oaa}. We will specify our choice
of parameters when discussing the numerical integration.

\subsection{Numerics}
\label{ssc:numerics}

To construct the embeddings we will solve numerically the equations of motion resulting from (\ref{eq:action}).
These are two coupled PDEs depending on $z$ and $x$. To solve them numerically we discretize the space
along both directions, and impose periodic boundary conditions along $x$. As for the radial direction,
in the UV we have the asymptotic solution (\ref{eq:embUV}), while in the IR ($z\to1$) the
requirement of regularity at the black hole horizon imposes that $\phi$ and $\chi'$ vanish there
(see Eq. \eqref{eq:bhexpansion}).
Therefore, we impose the following boundary conditions
\begin{subequations}
\label{eq:bcs}
\begin{align}
&\phi(0,x)=\mu(x)\,,\qquad \chi'(0,x)=m\,,\\
&\phi(1,x)=0\,, \qquad\quad\;\; \chi'(1,0)=0\,.
\end{align}
\end{subequations}
We use pseudospectral methods implemented in Mathematica, discretizing the plane $(z,x)$ on a rectangular
lattice of size $N_z\times N_x$, with $N_z$ and $N_x$ being, respectively, the number of points in
the $z$ and $x$ directions. We use a Chebyshev grid along $z$, and a planar one in the $x$ direction;
and employ a Newton-Raphson iterative algorithm to solve the resulting nonlinear algebraic equations.
For all the numerical simulations apart from those in Sec.~\ref{sec:spectral} we set 
$L_x=20\pi$, so that $k_0/T\ll 1$,  
and take $k_*=1$ which corresponds to truncating the sum in \eqref{eq:noisefunc} at 10 modes.
Finally, we average over different realizations of the noisy chemical potential. When
it is visible we show the error of the average
$\sigma_N / \sqrt{N}$, where $\sigma_N$ is the standard deviation, and $N$ the number of realizations.
Only in Fig.~\ref{fig:sigmavsmuUP} we plot the standard deviation instead of the error of the average.

\subsection{Results}
\label{ssc:bkresults}

Let us first consider massless embeddings; for these the embedding field $\chi$ vanishes, and
we need only solve the equation of motion for $\phi$.
\begin{figure}[bht]
\begin{center}
\includegraphics[width=0.49\textwidth]{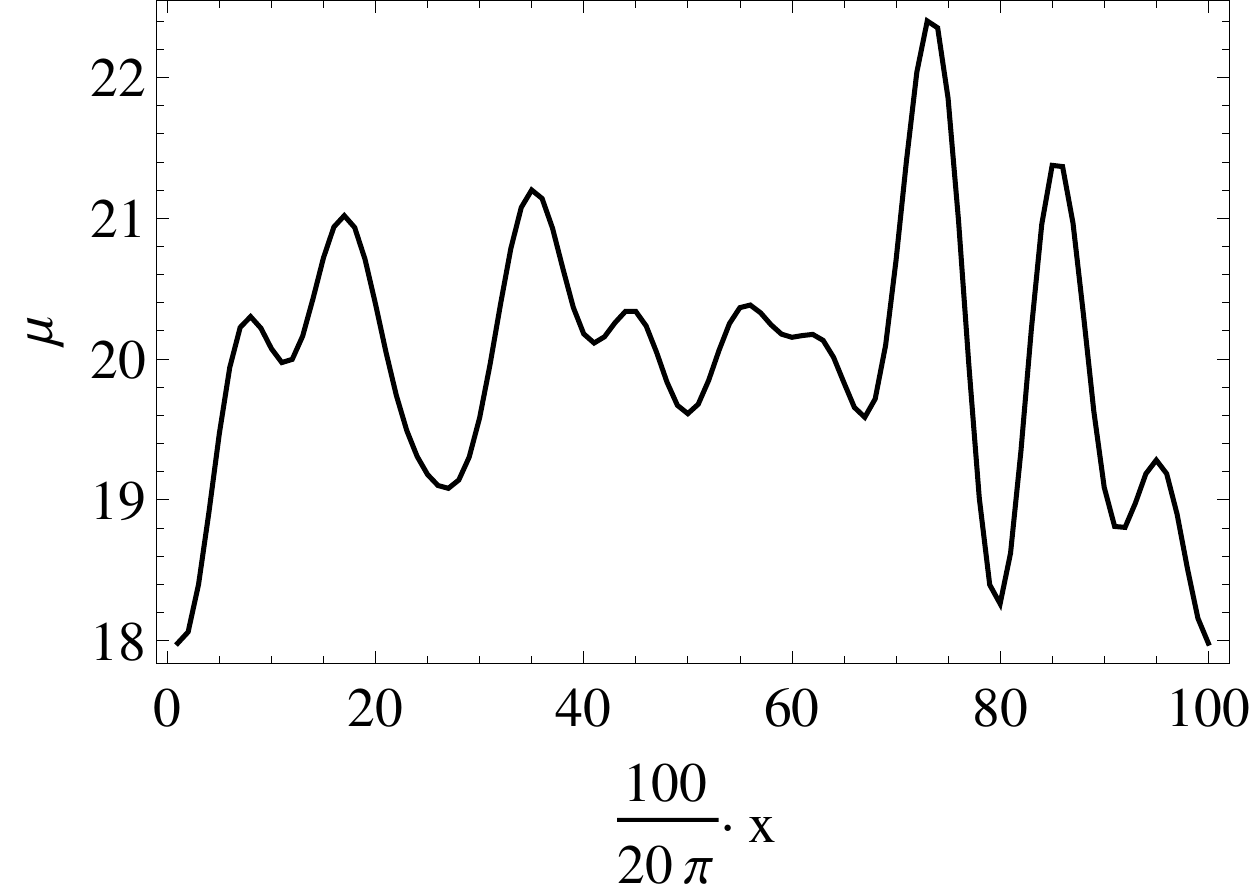}
\includegraphics[width=0.49\textwidth]{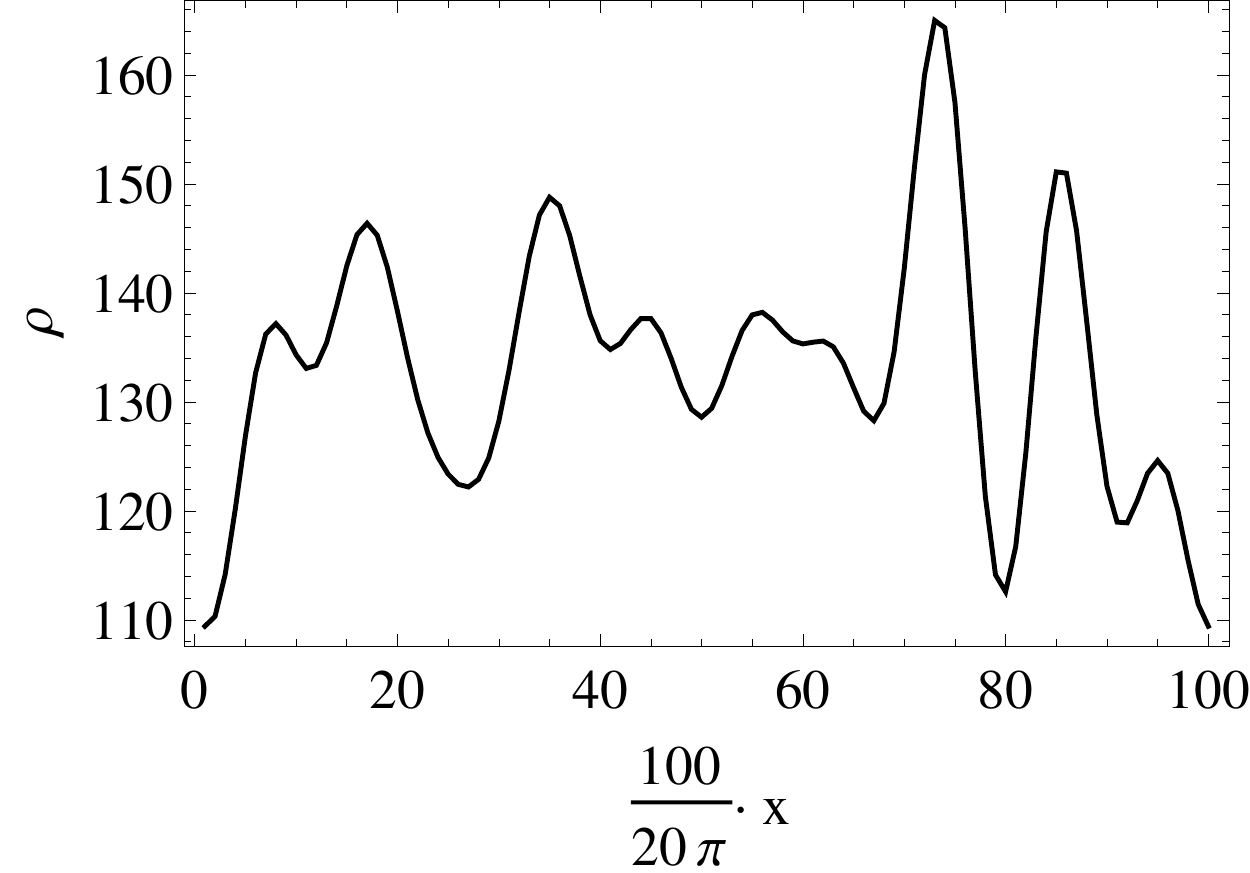}
\caption{\label{fig:murhom0}
Massless embedding. On the left panel we plot an example of one realization of $\mu(x)$ 
with $\mu_0=20$ and $w=0.5$. On the right we present the corresponding charge density for a massless
embedding with that chemical potential. We have employed a lattice of size $100\times100$ and set
$L_x=20\pi$, $k_*=1$.}
\end{center}
\end{figure}
In Fig. \ref{fig:murhom0} we plot the result of a single simulation of our noisy system.
We present the random chemical potential we plug in as our boundary condition
together with the resulting charge density which we read from the asymptotic behavior of the worldvolume
field $\phi$. 

\subsubsection*{Charge Density}

It is interesting to study how the charge density depends on the strength of disorder.
An important observable of our setup is given by the spatial average of the charge density,
$\langle\rho\rangle$ (we will denote by $\langle\cdot\rangle$ the average over the spatial direction $x$).
Therefore, 
for a noisy chemical potential as \eqref{eq:noisefunc}, 
we will analyze how $\langle\rho\rangle$ depends on the strength of the noise, parametrized by $w$.
The expected behavior can be anticipated by considering how the charge
density depends on the chemical potential in the homogeneous case.
In App.~\ref{app:homcase} we have reviewed the homogeneous D3/D5 intersection, and in particular we have
shown that the function $\mu(\rho)$ can be computed analytically, and is given by Eq.~\eqref{eq:murhoev}.
That function interpolates between two distinct behaviors as shown in \eqref{eq:rhomulim}; while
$\rho\sim\mu^2$ for large $\rho$, $\rho\sim\mu$ in the small $\rho$ limit. Assuming this behavior holds
for an $x$ dependent noisy $\mu$ as \eqref{eq:noisefunc}, it is easy to predict how $\langle\rho\rangle$
should behave as a function of $w$. First, in the low $\rho$ regime, where $\rho\sim\mu$, 
taking into account that the noisy $x$ dependent part of \eqref{eq:noisefunc} averages to zero, 
one concludes that $\langle\rho\rangle$ must be independent of $w$. Instead, at large $\rho$, where we
assume $\rho\sim\mu^2$, one expects $\langle\rho\rangle\sim w^2$ (we will show this explicitly in 
Eq.~\eqref{eq:avrhovsrho0} below).
\begin{figure}[hbt]
\begin{center}
\includegraphics[width=0.46\textwidth]{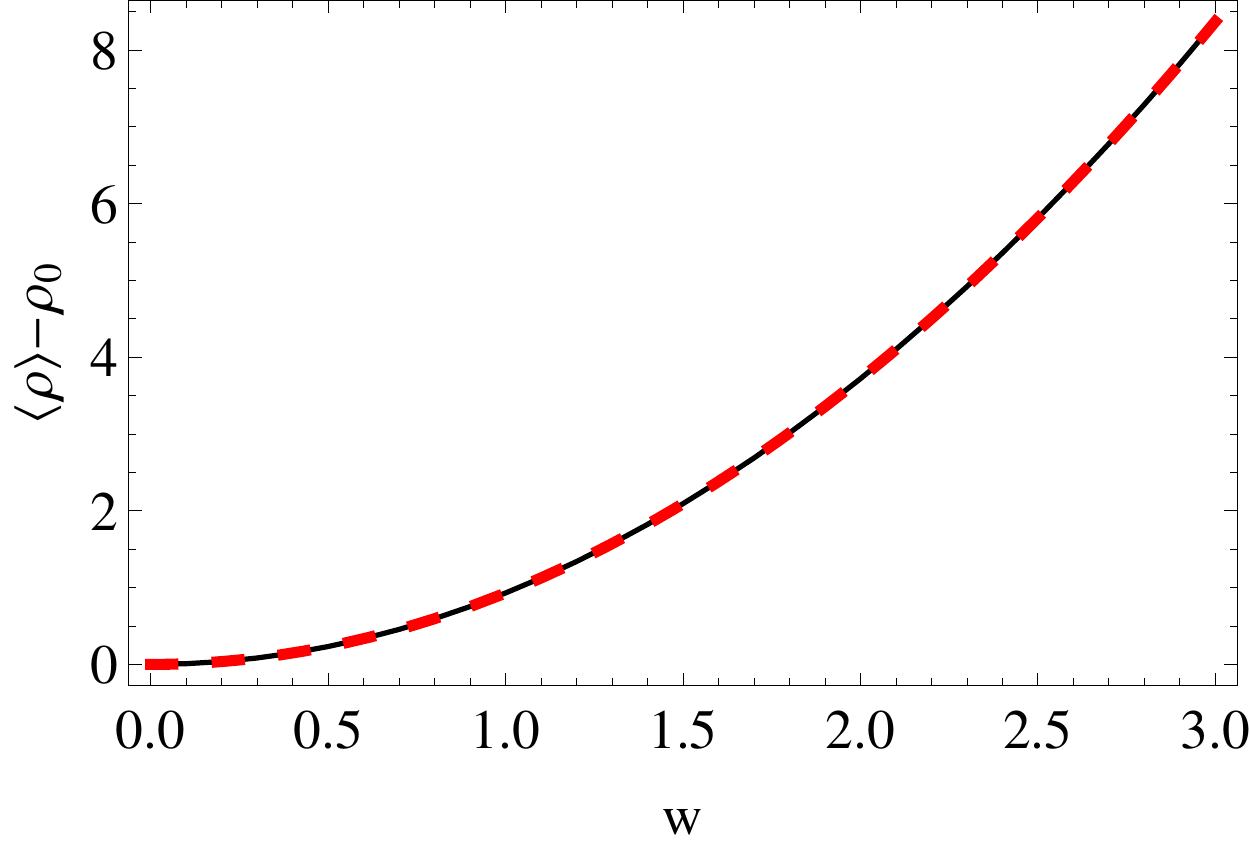}
\includegraphics[width=0.485\textwidth]{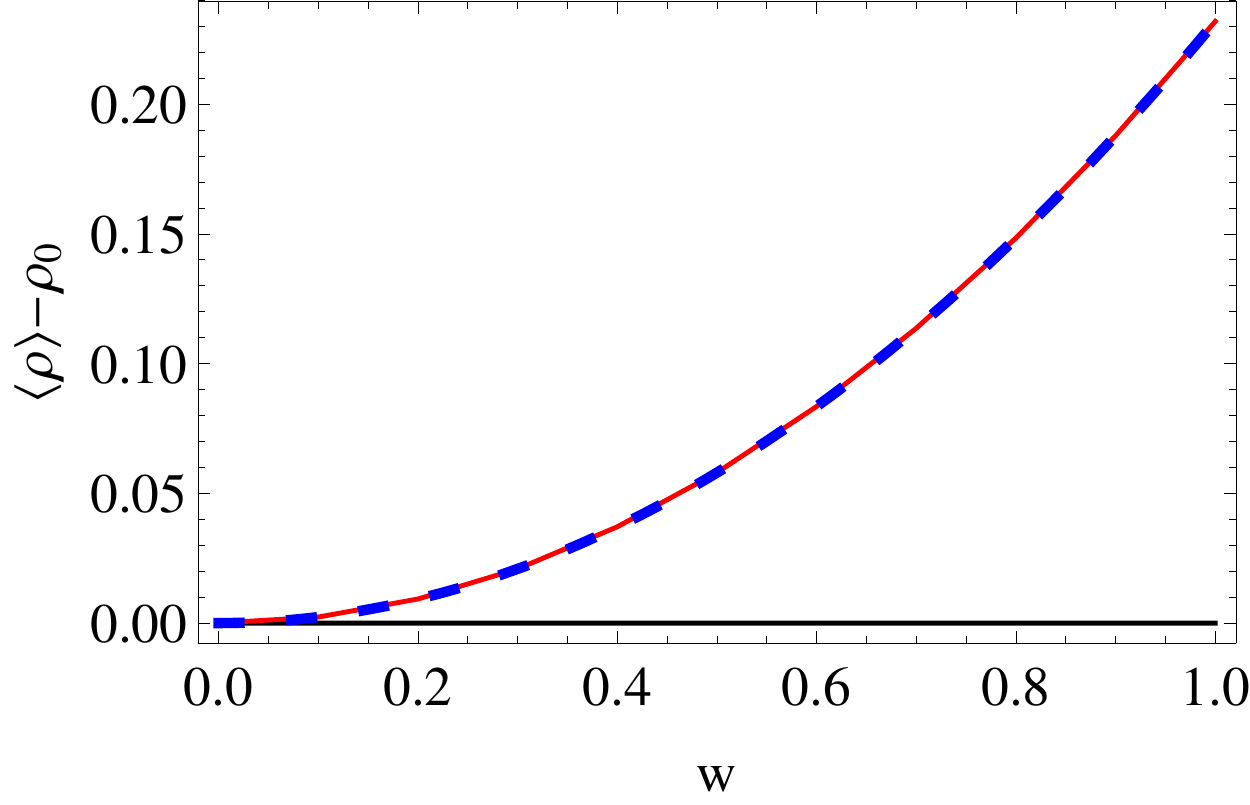}
\caption{
\label{fig:rhovsmum0}
$\langle \rho \rangle$ versus $w$. On the left we plot $\langle \rho \rangle-\rho_0$ for a system
with $\mu_0=20$, $\rho_0=133.398$ is the value of $\rho$ for the clean case.
The black line shows the result of the numerical
integration, and the red dashed line the fit 
$\log(\langle \rho \rangle-\rho_0) = -0.0726 +2.000 \log(w)$.
On the right panel we plot the subtracted charge density for $\mu_0=10$ (red line, $\rho_0=38.897$),
and $\mu_0=0.1$ (black line, $\rho_0=0.141$). The blue dashed line results from the fit 
$\log(\langle \rho \rangle-\rho_0) = -1.462 +2.000 \log(w)$ to the $\mu_0=10$ data.
For both graphs we have averaged over 10 realizations on a lattice of size $100\times100$
and set $L_x=20\pi$, $k_*=1$.
}
\end{center}
\end{figure}

In Fig.~\ref{fig:rhovsmum0} we present the results of our numerical simulations for the evolution of
the averaged charge density as a function of the disorder strength.
First, on the left panel we plot $\langle\rho\rangle$ versus $w$ for a system with $\mu_0=20$, which corresponds
to $\rho_0\approx133.398$, and places the setup in the large charge density regime.
As expected $\langle\rho\rangle$ scales quadratically with $w$ as the fit in the graph demonstrates.
Next, on the right plot we show two cases corresponding to lower charge density, namely $\mu_0=10$ (red line),
and $\mu_0=0.1$ (black line), for which $\rho_0$ takes the values 38.897, and 0.141 respectively.
While for $\mu_0=10$, the quadratic scaling of $\langle\rho\rangle$ is still observed, for $\mu_0=0.1$ we
see that the averaged charge density is independent of the noise strength.

\begin{figure}[htb]
\begin{center}
\includegraphics[width=0.31\textwidth]{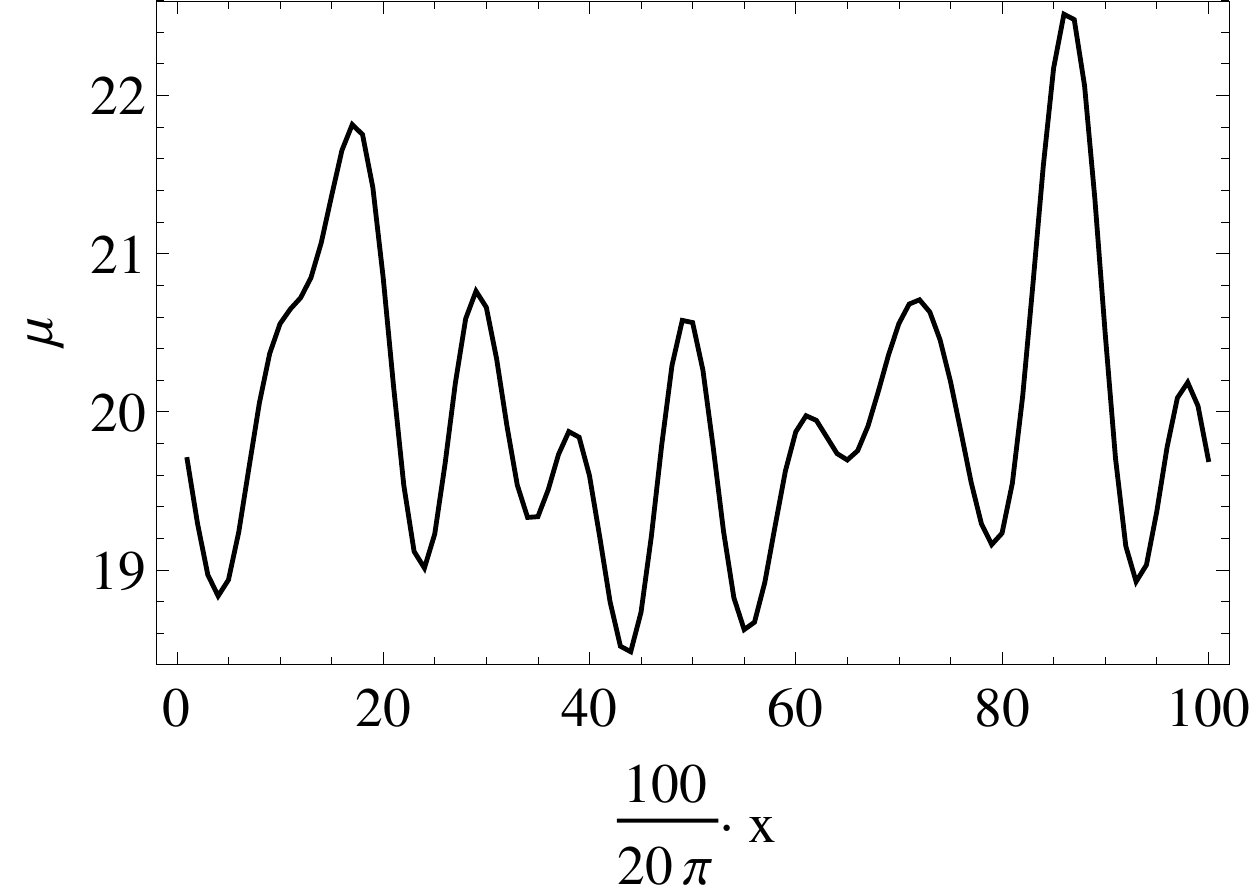}
\includegraphics[width=0.315\textwidth]{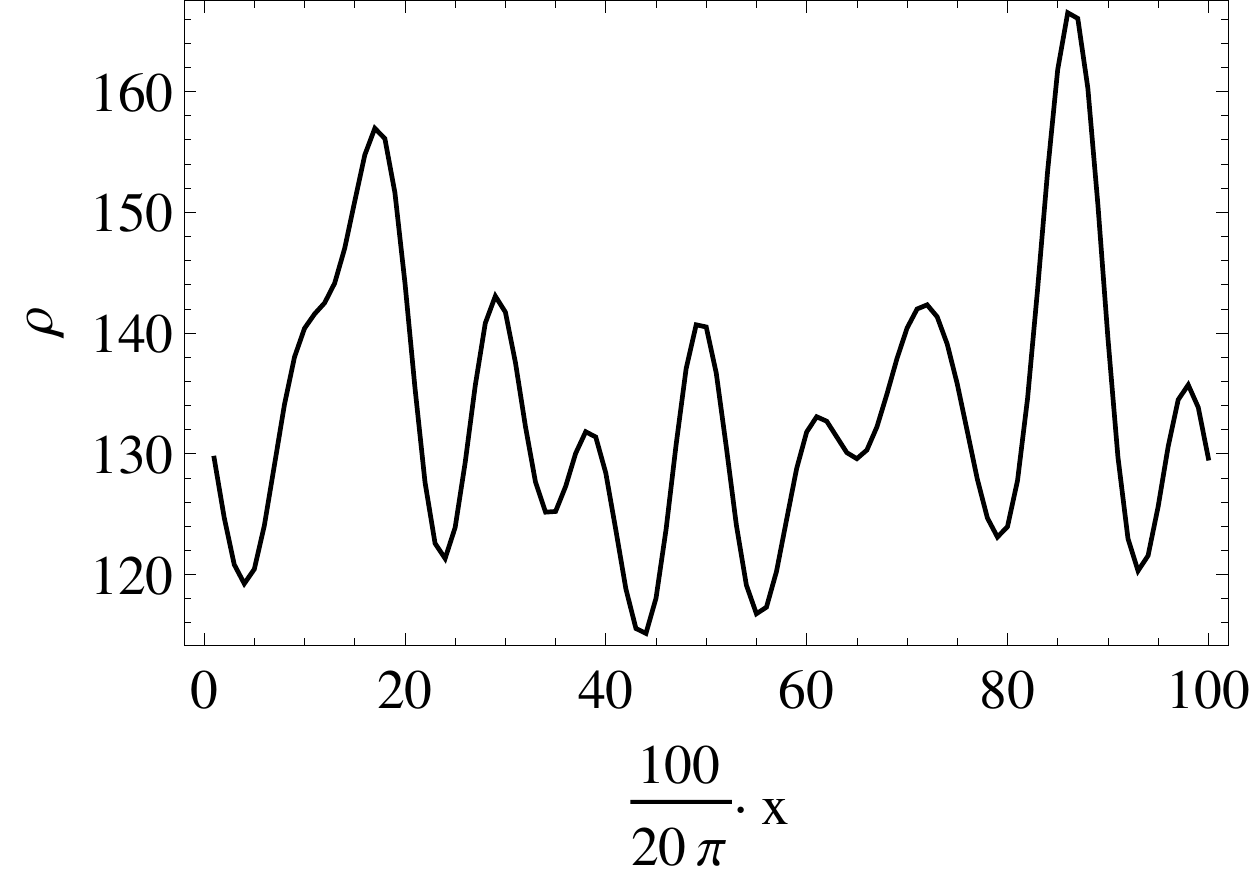}
\includegraphics[width=0.34\textwidth]{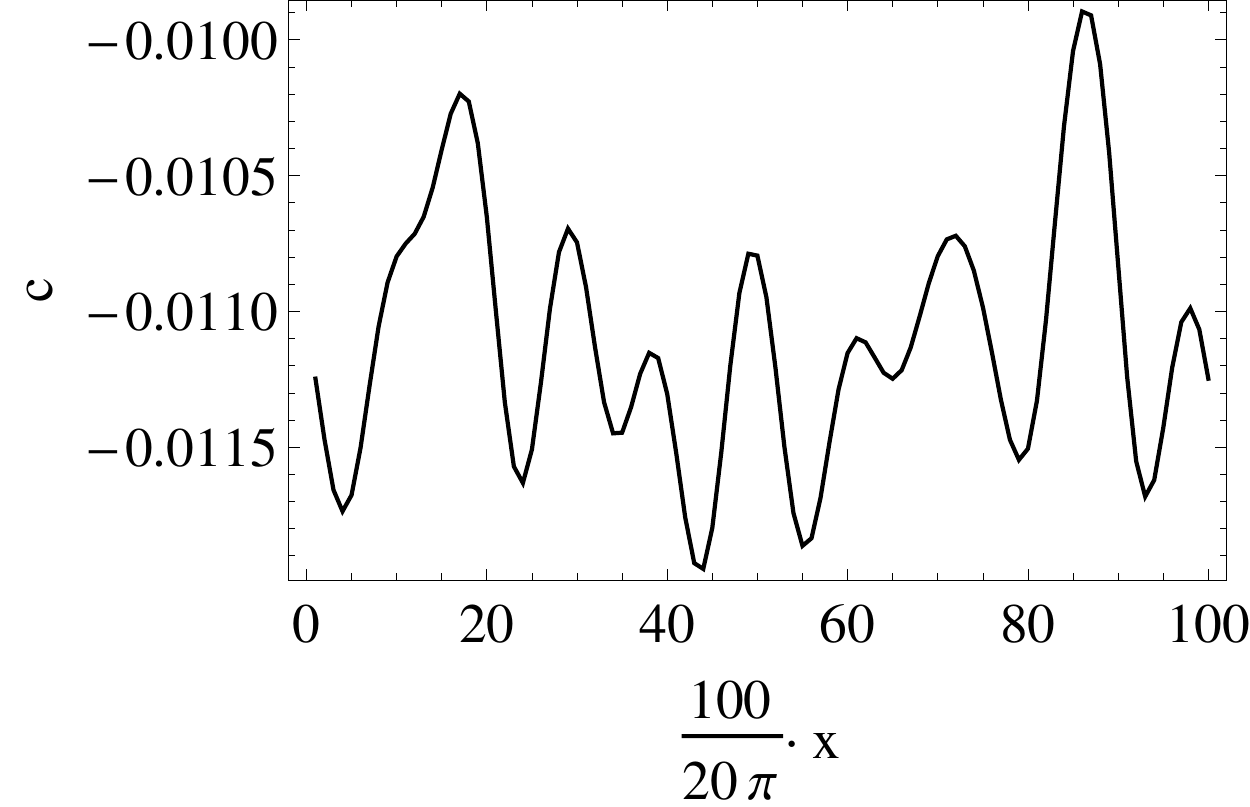}
\caption{\label{fig:massemb}
Massive embedding. The three panels display, from left to right, the chemical potential 
$\mu(x)$, the charge density $\rho(x)$, and the quark condensate $c(x)$ for a simulation  
with $\mu_0=20$, $m=0.5$, $w=0.5$, $L_x=20\pi$, and $k_*=1$ (corresponding to 10 modes)
on a lattice of size $100\times 100$.} 
\end{center}
\end{figure}

In the remaining of this section we consider the case of nonzero mass. 
We then have to solve the two coupled PDEs for $\phi$ and $\chi$.
When choosing the value of the mass $m$ one has to take into account the phase diagram
for black hole embeddings at finite charge density and nonzero mass, which was reviewed
in \cite{Araujo:2015hna}. In particular, notice that for $m\gtrsim 1.5$, black hole embeddings
exist only for $\mu\gtrsim m-1.41$. We will restrict our analysis to cases where the space 
dependent chemical potential never reaches that forbidden region.
In Fig.~\ref{fig:massemb} we plot the result of a simulation for a massive embedding, showing
the noisy chemical potential that we impose as boundary condition, and the resulting charge density
and quark condensate we read from the solution of the PDEs.

\subsubsection*{Quark Condensate}

\begin{figure}[htb]
\begin{center}
\includegraphics[width=0.47\textwidth]{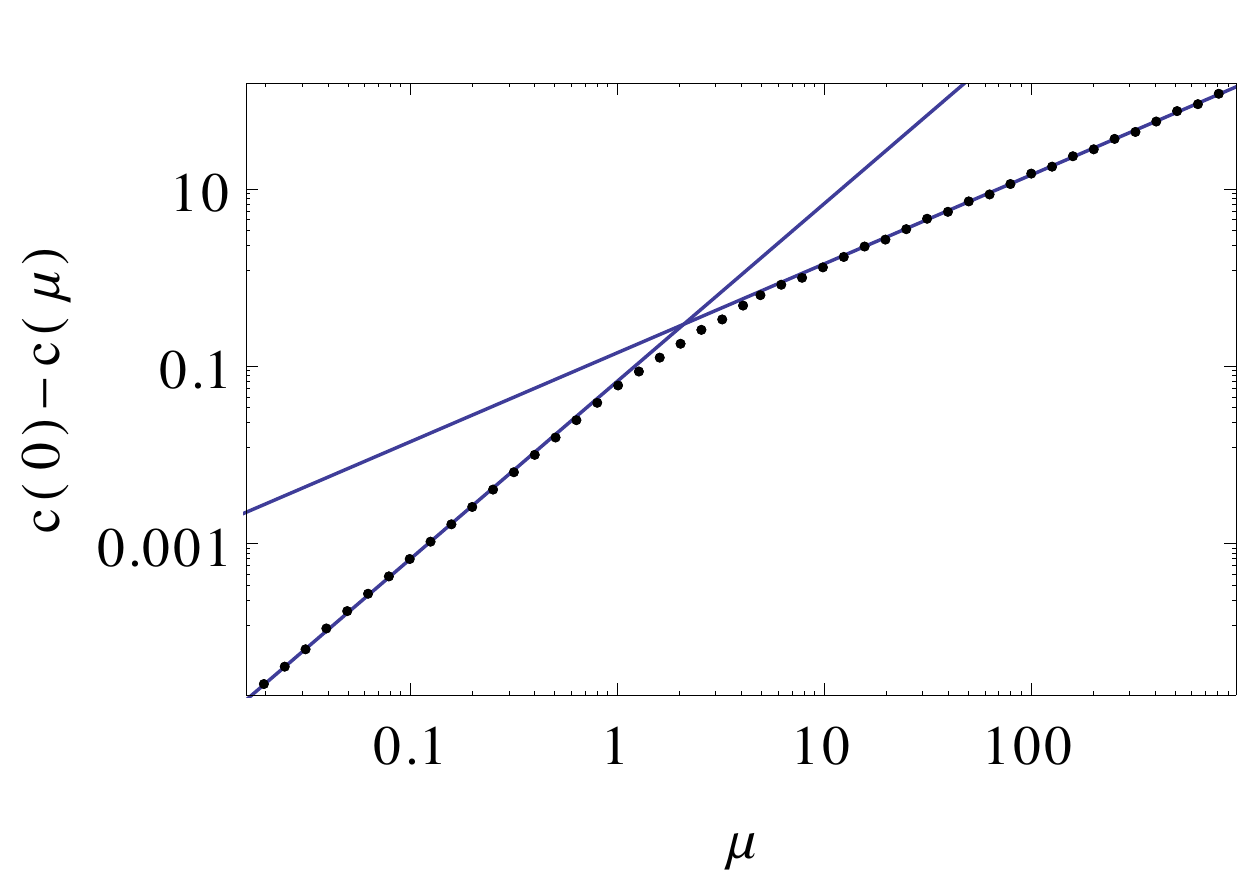}
\includegraphics[width=0.48\textwidth]{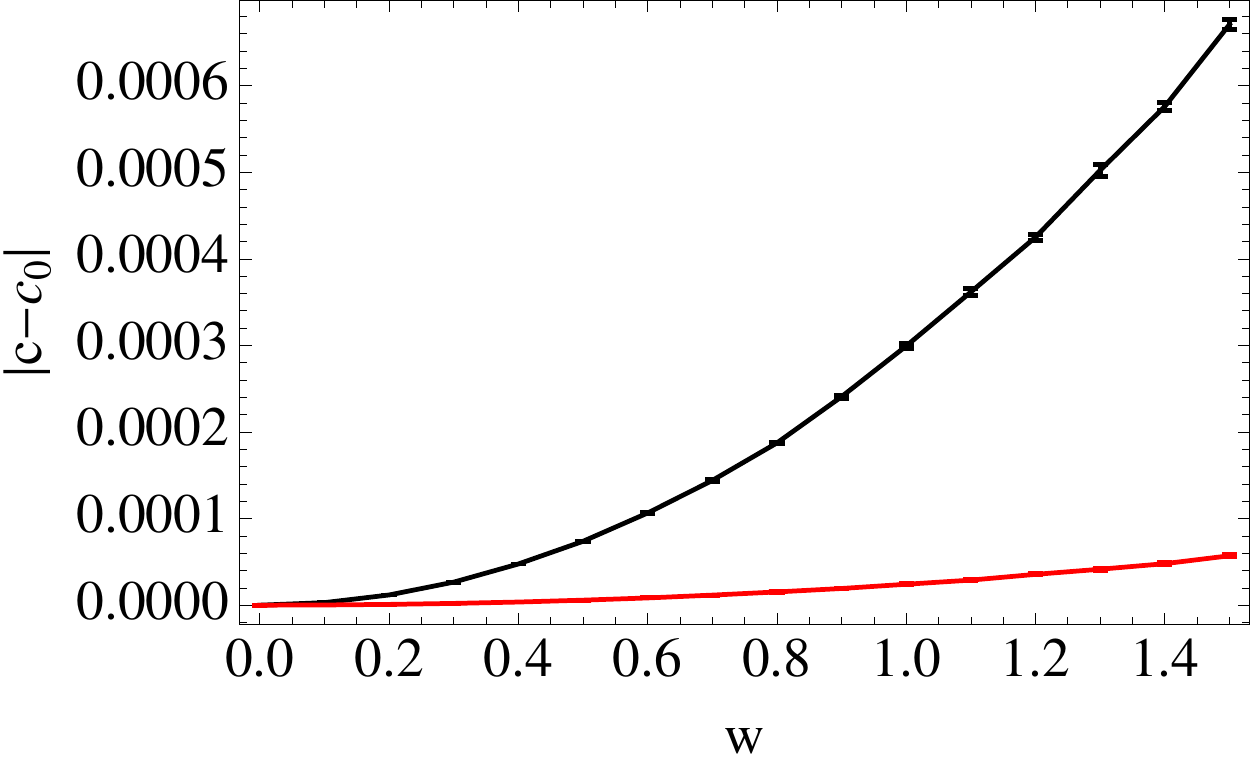}
\caption{\label{fig:massembrhoc}
Massive embedding. On the left we plot the behavior of the homogeneous condensate $c$ as a 
function of the chemical potential $\mu$ for a fixed mass $m=0.5$. The fits are 
$\log_{10}(c(0)-c(\mu)) =  2.004\log_{10}\mu-1.170$,
and $\log_{10}(c(0)-c(\mu))=1.003\log_{10}\mu-0.847$. 
The right panel shows the subtracted condensate $c-c_0$ versus the noise strength $w$ for $\mu_0=1.5$ (black) 
and $\mu_0 = 20$ (red), with $m=0.5$ in both cases. For the clean system ($w=0$) the condensate takes 
the value $c_0=-0.431$ ($c_0=-3.114$) for the case $\mu_0=1.5$ ($\mu_0=20$).
We have averaged over 10 realizations on a grid of size $100\times 100$, and set $L_x=20\pi$, $k_*=1$.
}
\end{center}
\end{figure}

The construction of disordered massive embeddings offers us the possibility of looking into the
behavior of the quark condensate $c$ in presence of disorder. 
As above, much can be inferred from the behavior of the homogeneous brane intersection.
For massive embeddings there is no analytical solution that allows us to express $c$ in terms of 
$\mu$, and instead, as we show in App.~\ref{app:homcase},
one needs to solve numerically a single ordinary differential equation for $\chi(z)$. 
However, one can obtain analytic (or semi-analytical) expressions for the dependence of $c$ on $\mu$ in the
limits of large and small chemical potential. These are Eqs.~\eqref{eq:cvsmusmmu} for $\mu\ll1$, and 
\eqref{eq:cvsmulgmu} for $\mu\to\infty$. Those equations reflect two different scaling regimes:
$c$ scales as $\mu^2$ for small $\mu$, while it becomes linear in $\mu$ as $\mu\to\infty$.
The numerical integration confirms these two  regimes 
as is illustrated on the left panel of Fig.~\ref{fig:massembrhoc}.
Notice that for an homogeneous embedding with mass $m=0.5$,
Eqs.~\eqref{eq:cvsmusmmu} and \eqref{eq:cvsmulgmu}
result in the following behavior of the quark condensate as a function of $\mu$
\begin{subequations}
\label{eq:cvsmum05}
\begin{align}
&\log_{10}(c(0)-c(\mu)) =  2\,\log_{10}\mu-1.170\,;\qquad (\mu\to 0)\,,
\label{eq:cvsmum05smrho}\\
&\log_{10}(c(0)-c(\mu))=\log_{10}\mu-0.837\,;\qquad  (\mu\to \infty)\,,
\label{eq:cvsmum05lgrho}
\end{align}
\end{subequations}
which agrees with the fit to the numerical data 
in Fig.~\ref{fig:massembrhoc} (see caption).
Therefore, an opposite scenario to that of the charge density above is anticipated for the dependence
of $\langle c \rangle$ on the noise strength $w$. While for low $\mu_0$ we expect $\langle c \rangle$
to scale quadratically with $w$, for large $\mu_0$, the averaged quark condensate should become largely
independent of the noise strength.
These expectations are confirmed by our numerical simulations presented on the right hand side of 
Fig.~\ref{fig:massembrhoc}. There we plot the subtracted quark condensate (denoting $c_0$ the value of $c$
at zero disorder) for systems with $\mu_0 = 1.5$ (black line) and $\mu_0 = 20$ (red line). 

One could repeat the analysis for the dependence of the charge density on the strength of disorder
for the case of massive embeddings. However, Eqs.~\eqref{eq:muvsrhosmrhomass} and \eqref{eq:muvsrholgrhomass}
show that the scaling of $\rho$ with $\mu$ in the homogeneous case is the same as for massless embeddings. 
Therefore, the behavior of $\rho$ as a function of $w$ is qualitatively the same as for the massless 
case examined above.

\section{Noisy Conductivity}
\label{sec:conductivity}

In this section we will study the electrical transport properties of our setup.
We will focus on the study of the electrical conductivity in the direction parallel to the
noise, namely $\sigma_{xx}\equiv\sigma$, and in particular its DC (zero frequency) value.
To compute that conductivity we need to consider fluctuations of the worldvolume gauge field
$A_x$ realizing an electric current along $x$. In general those fluctuations will source
other fields in the model, and one has to solve a system of coupled linear PDEs.

The ansatz for the fluctuating field is
\be
A_\mu=a_\mu(z,x)\,e^{i\omega\,t}\,,
\label{pertdef}
\ee
and we will require that $f_{tx}(0,x)=i\omega\,e^{i\omega\,t}$, so that on the boundary
the fluctuation is sourcing an oscillating electric field with constant modulus.
The AdS/CFT dictionary tells us to compute the conductivity\footnote{As in \cite{Araujo:2015hna} we are 
rescaling the conductivity by the  dimensionless constant appearing in front of the action 
\eqref{eq:action}, \emph{i.e.} $\sigma\to\sigma/(N_f T_{D5}\,L^6)$.} as
\be
\sigma=\frac{ J_x}{E_x}=\lim_{z \to 0} \frac{f_{xz}}{f_{tx}} \,.
\ee
Hence, we need to solve the equations of motion resulting from the expansion of the DBI action up to
quadratic order in the fluctuations. These equations couple $a_x(z,x)$ to the fluctuation of the
temporal component of the gauge field, $a_t(z,x)$, and in the massive case also to the fluctuation of the
embedding field $\chi$, which we will denote $c(z,x)$. 
Moreover, we choose to work in the gauge $a_z(z,x)=0$.
The quadratic action for these fluctuating fields
has been written in the appendix of \cite{Araujo:2015hna}.
To compute the AC conductivity one then needs to solve the resulting system
of linear PDEs. Since the background is periodic along $x$, one has to impose
periodic boundary conditions in that direction.
As for the radial direction, in the UV one requires
\begin{align}
&f_{tx}(0,x)=i\omega\,e^{i\omega\,t}\,,\qquad c'(0,x) =0\,,\nonumber\\
&i \omega\, \partial_z a_t(0,x)- \partial_x \partial_z a_x(0,x)=0\,.
\label{eq:cuv}
\end{align}
The first condition corresponds to switching on a constant electric on the boundary,
while the second ensures that no fluctuation of the mass of the quarks is sourced.
The third equation is a constraint resulting from the equation of motion for $a_z$,
and upon substituting the form of the UV asymptotics amounts to the equation
for charge conservation on the boundary \cite{Araujo:2015hna}. In particular, that equation
implies that at zero frequency the electric current in the $x$ direction, $J_x= -\partial_z a_x(0,x)$,
is independent of $x$. Consequently, the DC conductivity in that direction is a constant.

In the IR one must impose infalling boundary conditions at the horizon. The IR behavior of the fluctuations
was also studied in \cite{Araujo:2015hna}, and it is straightforward to check that the resulting
conditions to impose read
\begin{align}
&\tilde a_t(1,x) = 0\,,\quad \tilde a_x'(1,x)=-\frac{i\omega}{4 \sqrt{2}}\,\tilde a_x(1,x)\,,\nonumber\\
&\tilde c'(1,x)=-\frac{i\omega}{4 \sqrt{2}}\,\tilde c(1,x)\,,
\label{eq:irbcs}
\end{align}
where we have redefined the fields as $\tilde \Phi = (1-z)^{-i\omega\over2\sqrt{2}}\,\Phi$.

In this work we will not present results for the AC conductivity, and will instead
focus our attention on the $\omega=0$ DC conductivity. Crucially,
as was shown in~\cite{Araujo:2015hna}, the DC conductivity for this system can be computed without
having to solve for the fluctuations we have just described.

\subsection{DC Conductivity}
\label{ssec:dccond}

The DBI action governing the fluctuations allows us to follow the procedure of~\cite{Iqbal:2008by,Ryu:2011vq}, 
and express the DC conductivity, $\sigma_{\rm DC}$,
in terms of the background functions, $\phi$ and $\chi$, evaluated at the horizon.
In \cite{Araujo:2015hna} that computation was particularized to a D3/D5 intersection
like the one in our setup,\footnote{The intersection analyzed in \cite{Araujo:2015hna}
had a homogeneous chemical potential, and an $x$ dependent mass. However, the analysis
in section 3.2 of that paper applies to a generic inhomogeneous D3/D5 intersection.} obtaining for $\sigma_{\rm DC}$
\be
\sigma_{\rm DC}={L_x\over \int_{0}^{L_x}\frac{dx}{{\cal F}(1,x)}}\,,
\label{eq:sigdcdef2}
\ee
where ${\cal F}$ is the following function of the embedding fields $\phi$ and $\chi$, and the metric functions
$f$ and $h$
\begin{equation}
{\cal F}(z,x)= f \left(1-\chi^2\right)^{3/2} \sqrt{\frac{h}{\Gamma}}\,,
\label{eq:Fdef}
\end{equation}
with
\begin{align} 
\Gamma =& -z^4 h \bigg\{\phi'^2 \left[h(1-\chi^2)+z^2\, \dot\chi^2\right]
-2 z^2 \phi'\,\dot\phi\, \chi' \, \dot\chi\ +\dot\phi^2 (1-\chi^2+z^2\, \chi'^2)\bigg\} 
\nonumber \\ &
- f^2 \left[ h\left(\chi^2-1-z^2\, \chi'^2\right)-z^2\,\dot\chi^2\right].
\end{align}
Notice that for the case of massless embeddings where $\chi = 0$, ${\cal F}(z,x)$
simplifies to
\be
\tilde{\cal F}(z,x)={1\over \sqrt{1-{z^4\over f^2}(h\,\phi'^2+\dot\phi^2)}}\,.
\ee
Moreover, by substituting the IR asymptotics of $\phi$ given by \eqref{eq:phibh} one arrives
at the following expression for the DC conductivity in the massless case
\be
\sigma_{\rm DC}={L_x\over\int_{0}^{L_x} dx\,\sqrt{1-(a^{(2)}(x))^2/2}}\,.
\label{eq:sigdcm0}
\ee

Therefore, by making use of \eqref{eq:sigdcdef2} we can compute the DC conductivity of our disordered
brane intersection without having to solve for the fluctuations of the gauge fields.
The conductivity is indeed determined in terms of
the behavior of the 
embedding fields $\chi$ and $\phi$ at the horizon. Furthermore, as we detail below,
in some interesting limits
we will be able to obtain analytic expressions for the DC conductivity as a function
of the charge density and the noise strength. 
Finally, notice that already the form of Eq.~\eqref{eq:sigdcm0}
makes clear that $\sdc>1$, and therefore even in the presence of noise the massless
setup is always metallic.

\subsection{$\sdc$ at Weak Disorder}
\label{sssec:pertnoise}

We shall now compute the effect a perturbatively small noise has on the DC conductivity in the large and small
charge limits. We will restrict the analysis to massless embeddings, and 
consider a scenario where the disorder is introduced as a small $x$ dependent
perturbation of the otherwise homogeneous chemical potential for a massless D3/D5 intersection.
This will allow us to build our analysis upon key results of the homogeneous case which we review
in Appendix \ref{app:homcase}. 

\subsubsection*{Small Charge}
We will assume that, as shown in Eq. \eqref{eq:rhomusrho} for the homogeneous case,
the charge density grows linearly with the chemical potential 
\be
\rho(x)= d\,\mu_0 (1 + \tilde w\, n(x))\,,
\label{eq:rholownoise}
\ee
where $d$ is a positive proportionality constant, $\tilde w$ is a small parameter,  and $n(x)$
is a noisy function with vanishing spatial average. In terms
of our previous definition of $\mu(x)$ in \eqref{eq:noisefunc} one can identify
\be
\tilde w = {w\over 25}\,,\qquad n(x)=\sum_{k=k_0}^{k_*}\,\cos(k\,x+\delta_k)\,,
\label{eq:wtdef}
\ee
while from Eq.~\eqref{eq:murhosmrho} $d=\sqrt{2}$. However we shall keep $\tilde w$ as an unknown small
parameter, and $n(x)$ an unknown function whose average over $x$ vanishes.
We will also assume that the relation \eqref{eq:a2rho} between the charge density and the second radial
derivative
of $\phi$ at the horizon that we derived for the homogeneous system, holds in the presence of a perturbatively
small noise, namely
\be
a^{(2)}(x) = {\sqrt{2}\rho(x)\over\sqrt{4+\rho(x)^2}}\,,
\label{eq:a2snoise}
\ee
and therefore, by means of Eq.~\eqref{eq:sigdcm0} one can write the resistivity $\varrho\equiv1/\sdc$ as
\be
\varrho={1\over L_x} \int_{0}^{L_x} dx\,{2\over\sqrt{4+\rho(x)^2}}\,.
\label{eq:resistivity}
\ee
Notice that this expression is valid in a regime where we are considering the setup as a succession of homogeneous
systems, one at each value of $x$. We are therefore neglecting
the effect of the gradients of the embedding field $\phi$ in the $x$ direction.
Next, substituting Eq.~\eqref{eq:rholownoise} into Eq.~\eqref{eq:resistivity}, and expanding the integrand 
up to order $\tilde w^2$ we arrive to
\be
\varrho={2\over\sqrt{ 4+d^2\mu_0^2}}\left[1+\tilde w^2\,{B(\mu_0)\over L}\,\int_0^L dx\,n(x)^2
+O(\tilde w^3)\right]\,,
\label{eq:pertsigmalc}
\ee
where
we have taken into account that the integral along $x$ of $n(x)$ 
vanishes,\footnote{For $n(x)$ as in \eqref{eq:wtdef}, upon averaging over realizations the integral along $x$ of any 
odd power of $n(x)$ vanishes too.} and have defined
\be
B(\mu_0)={d^2\mu_0^2\left(-2+d^2\,\mu_0^2\right)\over(4+d^2\,\mu_0^2)^2}\,.
\label{eq:bdef}
\ee
In the limit of small charge density, $\mu_0\ll1$, \eqref{eq:pertsigmalc} becomes
\be
\varrho= 1-{d^2\mu_0^2\over8}\left(1+\tilde w^2\,{1\over L}\int_0^L dx\,n(x)^2\right)+O(\mu_0^4,\,\tilde w^3)\,.
\label{eq:sdcsnslc}
\ee
Finally, 
substituting $\tilde w$ from \eqref{eq:wtdef} one obtains for  $\sdc$
\begin{align}
\sdc&= 1+{d^2\mu_0^2\over8}\left(1+{w^2\over 25^2}\,{1\over L}\left<\int_0^L dx\,n(x)^2\right>_{\rm noise}\,\right)+O(\mu_0^4,\,\tilde w^3)\nonumber\\
&\approx1+{\mu_0^2\over4}\left(1+{w^2\over 25^2}\,{\#({\rm modes})\over2}\,\right)\,,
\label{eq:sdcvswlrho}
\end{align}
where we have introduced the notation $\langle\cdot\rangle_{\rm noise}$ to denote the average 
over realizations of disorder;
and in the second line we have taken into account that for $n(x)$ of the form \eqref{eq:wtdef}, 
the integral ${1\over L}\int_0^L dx\,n(x)^2$ 
is nothing else than half the number of modes in the sum ($\#({\rm modes})/2$), and that as 
shown in \eqref{eq:rhomusrho} $d=\sqrt{2}$.
Hence, in the small charge density limit one expects an enhancement of the DC conductivity that grows 
quadratically with the strength of noise. We will show in Sec.~\ref{ssc:sgresults} that our numerics 
confirm this prediction.

\subsubsection*{Large Charge}
We will assume that, as in the homogeneous case \eqref{eq:rhomulrho}, the charge density grows
quadratically with  the chemical potential 
\be
\rho(x)= c\,\mu_0^2\left(1 + \tilde w\, n(x)\right)^2\,,
\label{eq:rhohgnoise}
\ee
where $c$ is a positive proportionality constant.
Proceeding as before, it is straightforward to arrive to an expression for the resistivity analogous to Eq. 
\eqref{eq:pertsigmalc}
\be
\varrho={2\over \sqrt{4+c^2\mu_0^4}}\left[1+\tilde w^2\,{C(\mu_0)\over L}\,\int_0^L dx\,n(x)^2+O(\tilde w^3)
\right]\,,
\label{eq:pertsigmahc}
\ee
where
\be
C(\mu_0)=3c^2\mu_0^4\,{c^2\mu_0^4-4\over(4+c^2\mu_0^4)^2}\,.
\label{eq:cdef}
\ee
In the limit $\mu_0\gg1$ we can write
\be
\varrho={2\over c\,\mu_0^2}+{6\over L\,c\,\mu_0^2}\,\tilde w^2\,\int_0^L dx\,n(x)^2+O(\mu_0^{-6},\,\tilde w^3)\,,
\ee
and thus the conductivity reads
\begin{align}
\sdc&={c\,\mu_0^2\over2}\left(1-w^2\,{3\over25^2}\,{1\over L} \left<\int_0^L dx\,n(x)^2\right>_{\rm noise}\,\right)
+O(\mu_0^{-2},\,w^3)\\
&\approx {0.291\mu_0^2\over2}\left(1-w^2\,{3\over25^2}{\#({\rm modes})\over2}\right)\,,
\label{eq:sdcsnshc}
\end{align}
where we have read $\tilde w$ from \eqref{eq:wtdef}, and as indicated in \eqref{eq:rhomulrho} $c\approx0.291$. 
We observe that in the regime of large charge density the introduction of a noisy perturbation of the
chemical potential results in a decrease of the DC conductivity.

In Sec.~\ref{ssc:sgresults} we will check this weak disorder analysis against our numerical computations, showing
a very good agreement in both the large and small charge density regimes.

\subsection{$\sdc$ at Strong Disorder}
\label{sssec:strongnoise}
In this section we will generalize the analytic approach of Sec.~\ref{sssec:pertnoise} to the case of strong noise.
We will start with a generic analysis which is valid for any value of the noise strength $w$ at the price of introducing
numerical integrals. This procedure will nevertheless allow us to determine the behavior of $\sdc$
in two interesting limits of strong noise. First, we will be able to determine the dependence of $\sdc$ on 
$\langle\rho\rangle$ in the large $\mu_0$ limit,  which in view of Eq.~\eqref{eq:noisefunc}  
corresponds to large disorder too.
Second, we will consider the case of a disordered chemical potential with vanishing $\mu_0$, thus describing noisy
oscillations around the charge neutrality point,  and predict the behavior of $\sdc$ as a function of noise strength
in the limit of strong noise.

Approximating the setup by a succession of homogenous systems with a different charge density at each point $x$,
we have written in Eq.~\eqref{eq:resistivity} 
the resistivity $\varrho$ as a function of the inhomogeneous charge density
$\rho(x)$. Moreover, in the homogeneous case $\mu(\rho)$ is given by the analytic function written in 
Eq.~\eqref{eq:murhoev}.
We can therefore write the following expression determining the conductivity in terms of the inhomogeneous chemical
potential $\mu(x)$
\be
\sdc=\left<{1\over{1\over L} \int_{0}^{L_x} dx\,{2\over\sqrt{4+\rho(x)^2}}}\right>_{\rm noise}
\,,\qquad {\rm with} \quad \rho(x)= {\cal G}^{-1}(\mu(x))\,,
\label{eq:sdcall}
\ee
where ${\cal G}$ is the function relating the chemical potential and the charge density in the homogeneous case,
namely $\mu = {\cal G}(\rho)$, which was computed in Eq.~\eqref{eq:murhoev} and takes the form
\be
{\cal G}(u)={u\over\sqrt{2}}\,_{1}F_2\left({1\over4},{1\over2},{5\over4};-{u^2\over4}\right)\,,
\label{eq:curgdef}
\ee
in terms of the hypergeometric function. 

In principle, by inserting $\mu(x)$ as given in Eq.~\eqref{eq:noisefunc} into Eq.~\eqref{eq:sdcall} one can compute
$\sdc$ at all orders in $w$. However, as \eqref{eq:sdcall} makes clear, one would first need to invert the relation
between $\mu$, and $\rho$, and already this first step can be done only numerically. Therefore, we can compute
$\sdc$ via numerical evaluation of the integral in that equation. 
It is worth stressing that within this approach we can compute the conductivity without having to solve the PDE
for the embedding field $\phi(z,x)$.


\subsubsection*{ Weak Noise Limit}

Before examining the scenarios of strong noise
 we shall now proceed as in Sec.~\ref{sssec:pertnoise} and particularize the analysis above to the case of 
weak noise, but this time keeping corrections up to order $w^4$.
First, we assume the large charge limit, and from the expression \eqref{eq:rhohgnoise} 
for the charge density in presence of noise, averaging over $x$, we obtain
\be
\langle\rho\rangle=c\,\mu_0^2\left(1+{w^2\over25^2}\,{1\over L}\int dx\,n(x)^2\right)\,,
\label{eq:avrhovsrho0}
\ee
where $w=25\tilde w$ as in \eqref{eq:wtdef}.
Next, we substitute $\rho(x)$ from Eq.~\eqref{eq:rhohgnoise} into Eq.~\eqref{eq:resistivity} for the resistivity, 
expand the integrand up to 
$O(w^4)$, and take the large $\mu_0$ limit, arriving to the following expression for the conductivity
\begin{align}
\sdc&=\left<\varrho^{-1}\right>_{\rm noise}\\
&=c\,\mu_0^2\left[1-3\left(w^2/25^2\right)\,\Delta+\left(9\Delta^2/25^4-5\Delta_4/25^4\right)w^4\right]+O(w^6,\mu_0^{-6})\,,
\end{align}
where $\Delta$ and $\Delta_4$ are
\begin{align}
&\Delta=\left<{1\over L}\int dx\,n(x)^2\right>_{\rm noise}={{\rm \#modes}\over2}\,,\label{eq:deltadef}\\
&\Delta_4=\left<{1\over L}\int dx\,n(x)^4\right>_{\rm noise}=\frac{3}{4} \left({\rm \#modes}\right)^2-\frac{3}{8} {\rm \#modes}\,.
\end{align}
Finally, using \eqref{eq:avrhovsrho0} to express $\mu_0$ in terms of $\langle \rho\rangle$ we  have
\be
\sdc\approx{1-3\left(w^2/25^2\right)\,\Delta+\left(9\Delta^2/25^4-5\Delta_4/25^4\right)w^4\over
1+(w^2/25^2)\,\Delta}\, {\langle\rho\rangle\over2}\,,
\label{eq:sdcvsavrho}
\ee
which predicts a conductivity linear in the charge density. The slope is always lower than the value of
the clean system ($\sdc\approx\langle\rho\rangle/2$) and decreases with increasing $w$.
Moreover, we will see in Fig.~\ref{fig:sdcvsrho} below, that this expression agrees very 
well with the numerical data also for a moderate noise with $w=3$ (where $\mu$ has oscillations 
$\approx70\%\,\mu_0$).

\subsubsection*{ Strong Noise}
When considering the case of generic noise strength $w$, for which the perturbative treatment above is not valid,
it is important to distinguish two scenarios: that where $w$ is small enough for $\rho(x)$ to stay positive along
the whole system, which we call `moderate noise'; and that where $w$ is large enough for $\rho(x)$ to become negative in some regions, which we denote `strong noise'.
 
We begin this analysis by rewriting the noisy chemical potential~\eqref{eq:noisefunc} in the generic form
\begin{equation}
\mu(x) =\mu_0[1+\tilde w\, n(x)]\,,
\label{eq:gennoisef}
\end{equation} 
and considering the moderate noise case where $\rho(x)$ stays positive along the system. In this situation,
substituting the large $\rho$ approximation~\eqref{eq:rhohgnoise} in Eq.~\eqref{eq:sdcall}, $\sdc$ becomes
\begin{equation}
\sdc \sim c\, \mu_0^2 \left<{1\over{1\over L_x} \int_{0}^{L_x} dx\,{2\over{\eta (x)}}}\right>_{\rm noise}
=\frac{\langle 
\rho \rangle}{1+\tilde w^2 \Delta} \left<{1\over{1\over L_x} \int_{0}^{L_x} dx\,{2\over{\eta (x)}}}\right>_{\rm noise},
\quad(\langle\rho\rangle\gg1)\,,
\label{eq:sdcallmodw}
\end{equation} 
where in the last equality 
we have used Eq.~\eqref{eq:avrhovsrho0}, $\Delta$ is given by Eq.~\eqref{eq:deltadef} above,
and $\eta(x)$ is defined as
\begin{equation}
\eta(x)=(1+\tilde w\, n(x))\,.
\label{eq:etadef}
\end{equation}
Therefore, we see that for a moderate noise, the DC conductivity grows linearly with $\langle\rho\rangle$.
Notice that the slope given by the expression \eqref{eq:sdcallmodw}
constitutes an all order (in $w$) correction to the
result in Eq.~\eqref{eq:sdcvsavrho} where we have kept contributions up to $O(w^4)$.

We shall now study the strong noise case where $w$ is large enough for regions of negative charge to appear
in the system.
In this scenario the regions around the zeros of $\rho(x)$, where the charge density is low even in the large
$\mu_0$ limit, will dominate the integral in Eq.~\eqref{eq:sdcall},
and determine the behavior of $\sdc$ in the large $\mu_0$ limit.
Let us denote $\{x_1,...x_i...,x_{N_{0}}\}$ the set of points where $\mu(x)$ has a simple 
zero,\footnote{The set of random phases resulting in noise profiles with double or higher order zeros has zero measure, and thus we can neglect the contribution of these realizations.}
and let $f(\mu_0)$ be a monotonically increasing function with the following properties
\begin{equation}
{\rm (a)}\;\lim_{\mu_0\to\infty} f(\mu_0)=\infty\,,\qquad
{\rm (b)}\;\lim_{\mu_0\to\infty} f(\mu_0)/\mu_0=0\,.
\label{eq:fmuproperties}
\end{equation}
A function that does the job is $f(\mu_0)=\log(\mu_0)$. 
Next, we split the integration domain of Eq.~\eqref{eq:sdcall} into the regions
$I^{>}_{\mu_0}$ and $I^{<}_{\mu_0}$, defined as those zones 
with $|\mu(x)|\geq f(\mu_0)$, and $|\mu(x)|<f(\mu_0)$ respectively. 
We split the resulting contributions to the resistivity $\varrho=\sigma^{-1}$ accordingly as $\varrho=\varrho_{>}+
\varrho_{<}$, so that $\sdc = \langle (\varrho_{>}+ \varrho_{<})^{-1}\rangle_{\rm noise}$. Let us focus first on the regions where $|\mu(x)|>f(\mu_0)$; 
in view of \eqref{eq:fmuproperties} (a), in the large $\mu_0$ limit one can implement the large charge limit
\eqref{eq:rhomulrho} in the integral in Eq.~\eqref{eq:sdcall} arriving to 
\begin{equation}
\varrho_{>} \sim \frac{2}{0.291\mu_0^2\,L_x} \int _{I^{>}_{\mu_0}} dx\, (\eta (x))^{-1}\,,
\label{eq:varrhor}
\end{equation} 
which shows that the contribution of the domain $I^{>}_{\mu_0}$ to the resistivity goes as $1/\mu_0^2$.
As for the domain $I^{<}_{\mu_0}$ (where $|\mu(x)|<f(\mu_0)$), it can be decomposed as the union of intervals 
$I^{<}_{\mu_0,i}$ localized around the zeros of $\mu(x)$, $x_i$. Expanding $\eta (x)$ around those points
as $\eta (x)=\eta'(x_i) (x-x_i)+ O((x-x_i)^2)$,
and taking into account that $\eta(x)=\mu(x)/\mu_0$, 
one can see that the diameter of each $I^{<}_{\mu_0,i}$
is order $O(f(\mu_0)/\mu_0)$, namely
\begin{equation}
|x-x_i|<\frac{f(\mu_0)}{|\eta'(x_i)|\,\mu_0} 
+ O\left(\left(\frac{f(\mu_0)}{\mu_0}\right)^2\right)~~~\mathrm{if}~x\in I^{<}_{\mu_0,i}\,\,.
\label{eq:Iminint}
\end{equation} 
Due to \eqref{eq:fmuproperties}(b) the length of these intervals goes to zero in the large $\mu_0$ limit. 
Hence for high enough $\mu_0$ all the $I^{<}_{\mu_0,i}$
are disjoint, and thus the contribution to the resistivity $\varrho_{<}$
can be written as the sum
\begin{equation}
\varrho_{<}=\frac{2}{L_x}\sum_i \int _{I^{<}_{\mu_0,i}} \frac{dx}{\sqrt{\mathcal{G}^{-1}[\mu(x)]^2+4}}.
\end{equation} 
In order to evaluate these integrals we change variables 
to $s=\mathcal{G}^{-1}[\mu(x)]$, with $\mathcal{G}(u)$ as defined in \eqref{eq:curgdef}.
This change of variables is well defined and invertible in the large  $\mu_0$ limit.\footnote{This can be 
seen using that $\mathcal{G}$ is bijective, and the diameter of $I^{<}_{\mu,i}$ will only cover an arbitrary small region around the simple zero $x_i$ of $\eta(x)$.} 
Denoting by
$\check x_i(s)$ the inverse of $s(x)$ inside $I^{<}_{\mu,i}$, we obtain
\begin{equation}
\varrho_{<}=\frac{2}{L_x} \sum_i  \int_{-\mathcal{G}^{-1}[f(\mu_0)]}^{\mathcal{G}^{-1}[f(\mu_0)]} \frac{ds}{\sqrt{s^2+4}} \,\frac{\mathcal{G}'(s)}{\mu_0|\eta'[\check x_i(s)]|}\,,
\label{eq:varrholeft}
\end{equation}
where we have taken into account that $\mathcal{G}(u)$ is odd. Keeping the leading contribution
in the large $\mu_0$ limit this integral becomes
\begin{equation}
\varrho_{<} \sim \frac{1}{\mu_0\,L_x} \int_{-\infty}^{\infty} \frac{2\,\mathcal{G}'(s)\, ds}{\sqrt{s^2+4}}  \left[ \sum_i  \frac{1}{|\eta'(x_i)|}  \right]  \approx  \frac{7.083 }{\mu_0\,L_x}  \sum_i  \frac{1}{|\eta'(x_i)|}\,.
\label{eq:varrholeftlgmu0}
\end{equation} 
In view of \eqref{eq:varrhor} and \eqref{eq:varrholeftlgmu0} we observe that 
the leading contribution to $\varrho=\varrho_{>}+\varrho_{<}$
comes from $\varrho_{<}$, and is of order $1/\mu_0$.
Taking the inverse and averaging over noises
we arrive to the following expression for $\sdc$ at large $\mu_0$,
\begin{equation}
\sigma_{DC}\sim\frac{\mu_0\,L_x}{7.083}\left< \frac{1}{\sum_i \frac{1}{|\eta'(x_i)|}} \right>_{\mathrm{noise}}\,,\label{eq:SigmaLargeMu}
\end{equation}
and to write it as a function of $\langle \rho \rangle$ we make use of
\begin{equation}
\langle \rho \rangle = \frac{1}{L_x}\int_0^{L_x} dx \,\mathcal{G}^{-1}[\mu(x)] \sim \frac{0.291\,\mu_0^2}{L_x}\int_0^{L_x} dx \,\mathrm{sign}[\eta(x)]\, \eta(x)^2,
\label{eq:avrhostnoise}
\end{equation} 
where as in Eq.~\eqref{eq:avrhovsrho0}, we have substituted the large $\mu$ approximation of
$\rho(x)= \mathcal{G}[\mu(x)]$ given by Eq.~\eqref{eq:rhomulrho}. 
Notice that the factor $\mathrm{sign}[\eta(x)]$ accounts for the regions where $\mu(x)$ (and $\rho(x)$)
becomes negative, and that the contribution from the regions where $|\mu(x)| \sim 0$ is subleading (and
remember that as explained above the length of these regions vanishes in the large $\mu_0$ limit).

Finally, combining Eqs.~\eqref{eq:SigmaLargeMu} and \eqref{eq:avrhostnoise} we arrive
 to the following expression for $\sdc(\langle\rho\rangle)$ in the large $\mu_0$ limit,
\begin{equation}
\sigma_{DC}\sim \frac{L_x^{3/2}}{3.820} \, \sqrt{\langle \rho \rangle}
\left< \frac{1}{\sum_i \frac{1}{|\eta'(x_i)|}} \right>_{\mathrm{noise}}
\left<\int_0^{L_x} dx \,\mathrm{sign}[\eta(x)] \eta(x)^2 \right>_{\mathrm{noise}}^{-1/2}.
\label{eq:SigmaLargeRho}
\end{equation} 
Notice that this result implies that in contrast to what happens in the weak and moderate noise cases
of Eqs.~\eqref{eq:sdcvsavrho} and \eqref{eq:sdcallmodw} 
where the conductivity is linear in $\langle\rho\rangle$, in the strong noise scenario 
$\sdc$ becomes a sublinear function of $\langle\rho\rangle$ in the large charge limit.
We will successfully check this prediction against our numerical, and semi-analytical, simulations
in the next section.

We end this section by analyzing the limit of small charge density.
This limit is simpler than its large charge counterpart since when $\mu_0\ll1$
the low charge approximation of $\mathcal{G}^{-1}(\mu (x))$ given by Eq.~\eqref{eq:rhomusrho} 
can be used independently of the value of $w$.
Inserting such approximation into Eq.~\eqref{eq:sdcall}, and taking the small $\mu_0$ limit, we find
\begin{equation}
\sigma_{DC}  \approx 1+\frac{\mu_0^2}{4} \left(1+\tilde w ^2 \Delta\right).
\end{equation} 
Moreover, as discussed in Sec.~\ref{ssc:bkresults},
in the small $\mu_0$ limit
the averaged charge density does not depend on the noise strength $\tilde w$, 
and Eq.~\eqref{eq:rhomusrho} implies that $\langle\rho\rangle\approx\sqrt{2}\mu_0$,
which allows us to write
\begin{equation}
\sigma_{DC}\approx1+\frac{\langle \rho \rangle^2}{8} \left(1+\tilde w ^2 \Delta\right)\,.
\label{eq:SigmaLowRho}
\end{equation} 
Therefore, in the small charge limit we expect a quadratic growth of the DC conductivity which will be checked
against our numerical analysis in the next section.

\subsection{Results}
\label{ssc:sgresults}

In this subsection we present the results of our numerical simulations 
for the conductivity. 
We will restrict ourselves to massless embeddings, and focus on the study of the DC conductivity.
We will start by studying the regimes of small and large charge density $\rho$ for weak 
to moderate noise,
and compare our numerics with the analytic expressions derived
in Sec.~\ref{sssec:pertnoise}.
Subsequently, we will present more general results for the DC conductivity,
considering a wide range of values for the chemical potential $\mu_0$,
and the noise strength $w$.
One should recall that, as shown in Eq.~\eqref{eq:dimratios}, both $\rho$ and $\mu$ are 
proportional to the dimensionless ratios $\bar\rho/T^2$ and $\bar\mu/T$ respectively. Therefore, when
we plot quantities like the conductivity as a function of $\mu$ (or $\rho$) one can think of $\mu$
as the inverse temperature of the system when working in the grand canonical ensemble for which
the physical chemical potential $\bar\mu$ is kept fixed
(equivalently, $\rho$ would be proportional to the temperature squared in the canonical ensemble).
In the last two subsections we will consider two scenarios of particular interest: the evolution of $\sdc$
as a function of the spatial average of the charge density $\langle\rho\rangle$, and the case of a
noisy chemical potential with vanishing spatial average ($\mu_0=0$).
In these two last cases we will pay special attention to the situations where regions of positive and negative
charge density appear in the system, and compare the numerical results with the 
analytic predictions obtained
in Sec.~\ref{sssec:strongnoise} for the strong noise case.

Two competing effects of the disorder  on the conductivity are expected \cite{Ryu:2011vq}. On one hand, 
we have seen in Sec.~\ref{ssc:bkresults} that disorder increases the charge density, or strictly speaking,
its spatial average (except at very low charge density).
In a homogeneous system the DC conductivity grows as the charge density increases
(see Eq.~\eqref{eq:sdchom}), 
and therefore, if one were to ignore any effect of the spatial inhomogeneities, 
an enhancement of $\sdc$ would be expected.
On the other hand, disorder gives rise to random spatial gradients of the charge density, and on general
grounds these impede conductivity.
However, we have already seen in the previous section that even if one ignores the effects of those
spatial gradients, and just computes the corrections due to a disordered perturbation to the chemical potential,
two opposite effects are found. While the noise indeed enhances $\sdc$ at low $\rho$, it has the opposite 
effect at large $\rho$, and $\sdc$ decreases.\footnote{This is visible in how the $O(\tilde w^2)$
correction in Eqs.~\eqref{eq:pertsigmalc} and \eqref{eq:pertsigmahc} changes sign as $\rho_0$ becomes
large or small. Notice that for this to be true it is crucial that the $O(w)$ correction vanishes
due to the perturbation being a noisy function with vanishing spatial average.}
One would expect that at strong disorder the effect of the gradients of the charge
density enhances this decrease of $\sdc$ for large charge density. We look into this scenario in
Sec.~\ref{ssec:sdcvsrho0}.

\begin{figure}[htb]
\begin{center}
\includegraphics[width=0.49\textwidth]{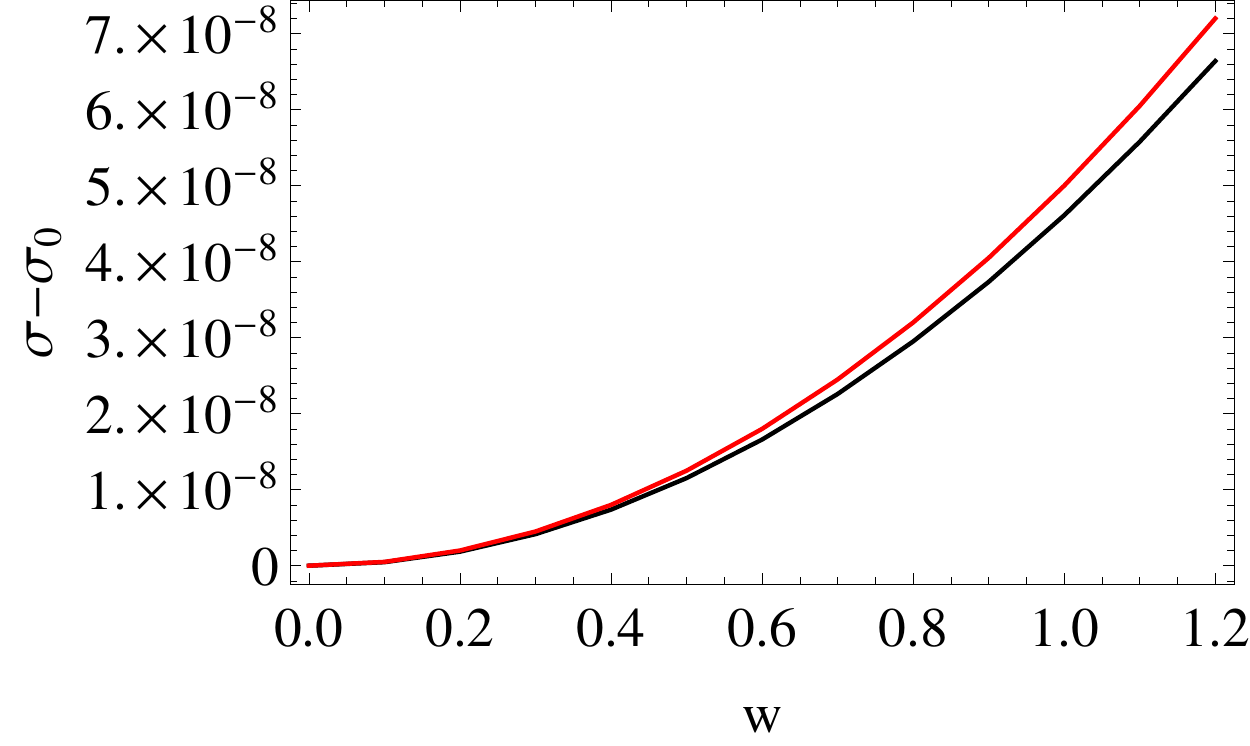}
\includegraphics[width=0.49\textwidth]{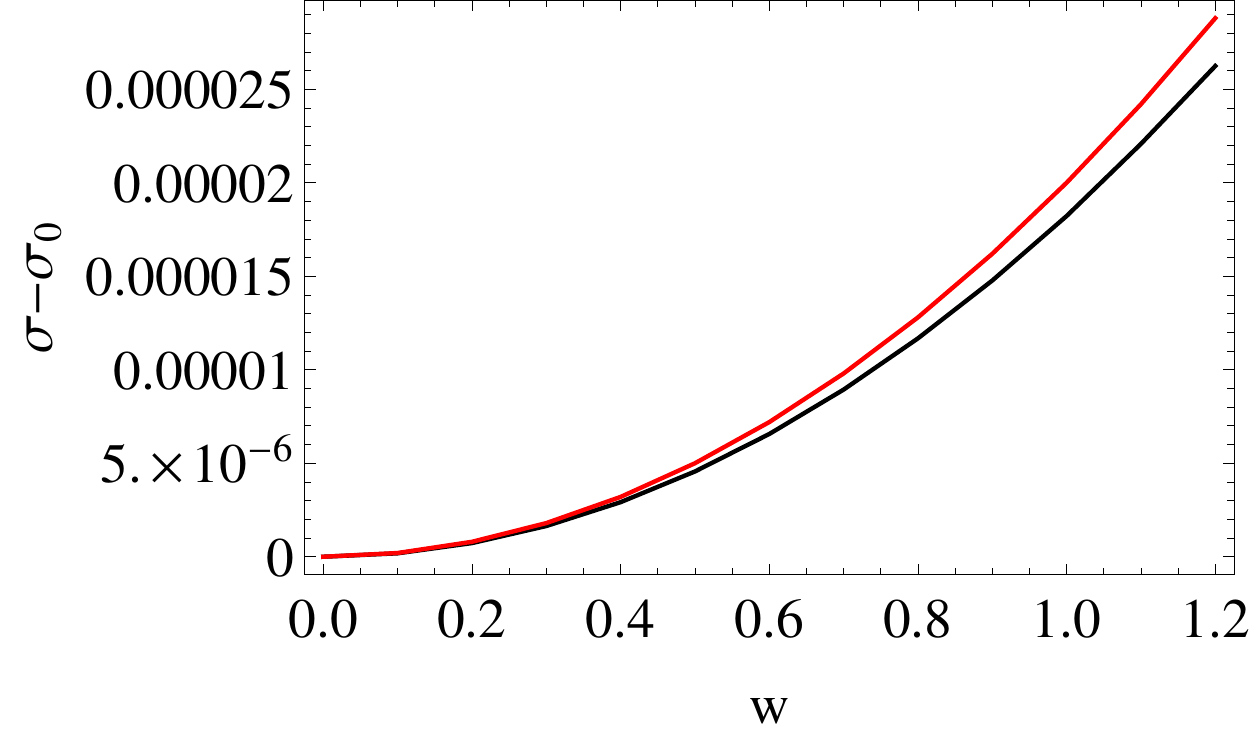}
\caption{\label{fig:sigcorrsmrho}
Enhancement of $\sdc$ at low $\rho$. We plot the subtracted $\sdc$ as a function of the noise strength $w$.
The black line corresponds to the numerical simulation, while the red line to the analytic expression
for small noise \eqref{eq:sdcvswlrho}. The left panel shows the results for $\mu_0=0.005$, and the right 
one for $\mu_0=0.1$.  The numerical results follow from
averaging over 10 simulations on a grid of size $100\times 100$, with $L_x=20\pi$ and $k_*=1$.} 
\end{center}
\end{figure}

We begin by considering the effect of noise on the conductivity in the limits of large and small
charge density.
In Fig.~\ref{fig:sigcorrsmrho} we compare our analytic prediction \eqref{eq:sdcvswlrho}
with numerical results at very low charge density. We plot the subtracted DC conductivity $\sigma -\sigma_0$,
where $\sigma_0$ is the value of $\sdc$ in the homogeneous case, versus the strength of noise parametrized
by $w$ (see Eq.~\eqref{eq:noisefunc}). We present results for systems with $\mu_0=0.005$, 
and $\mu_0=0.01$, for which the charge density (at zero noise) is respectively $\rho_0 = 0.00707$, and 
$\rho_0=0.141$.
We are then well within the region where the low $\rho$ analysis of Sec.~\ref{sssec:pertnoise} applies, and 
indeed one can see a good agreement of our numerics (black lines) with the analytic prediction of 
Eq.~\eqref{eq:sdcvswlrho} (red lines), specially at low $w$. 

\begin{figure}[htb]
\begin{center}
\includegraphics[width=0.49\textwidth]{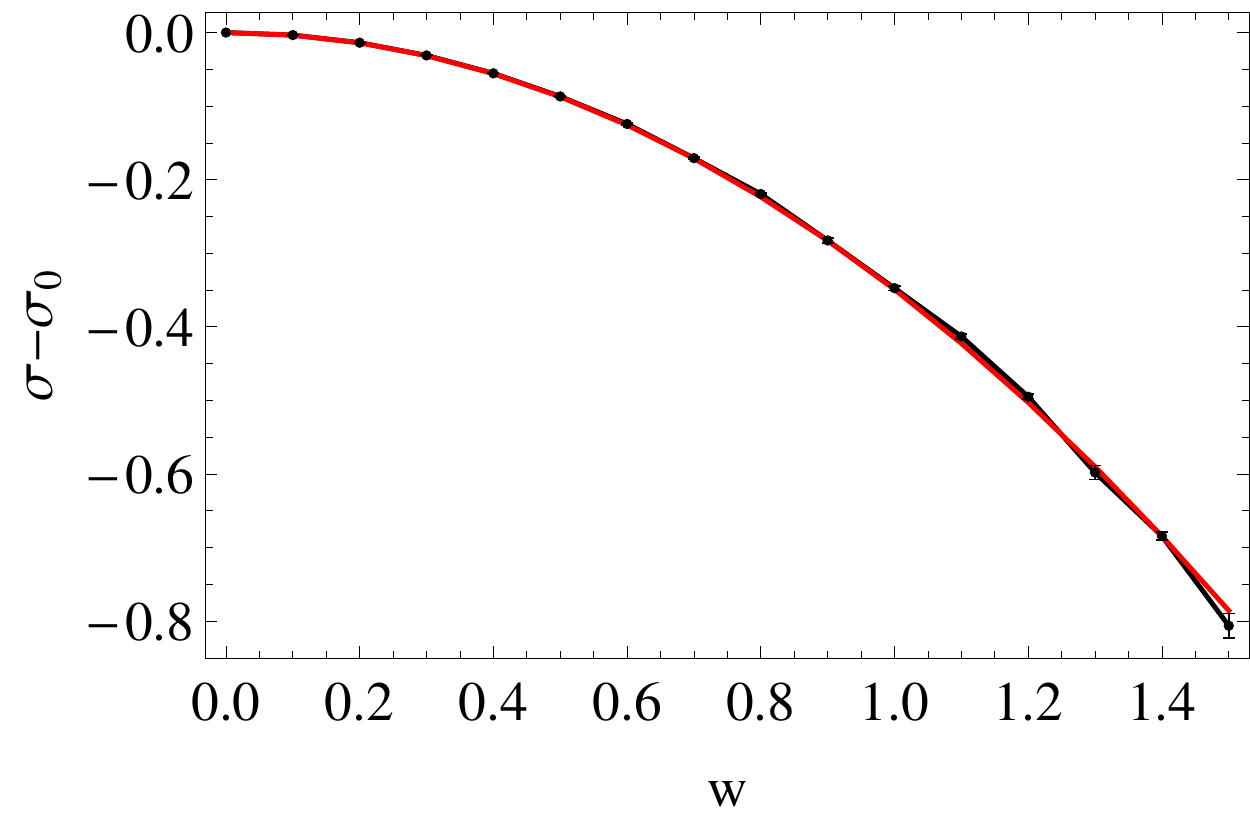}
\includegraphics[width=0.49\textwidth]{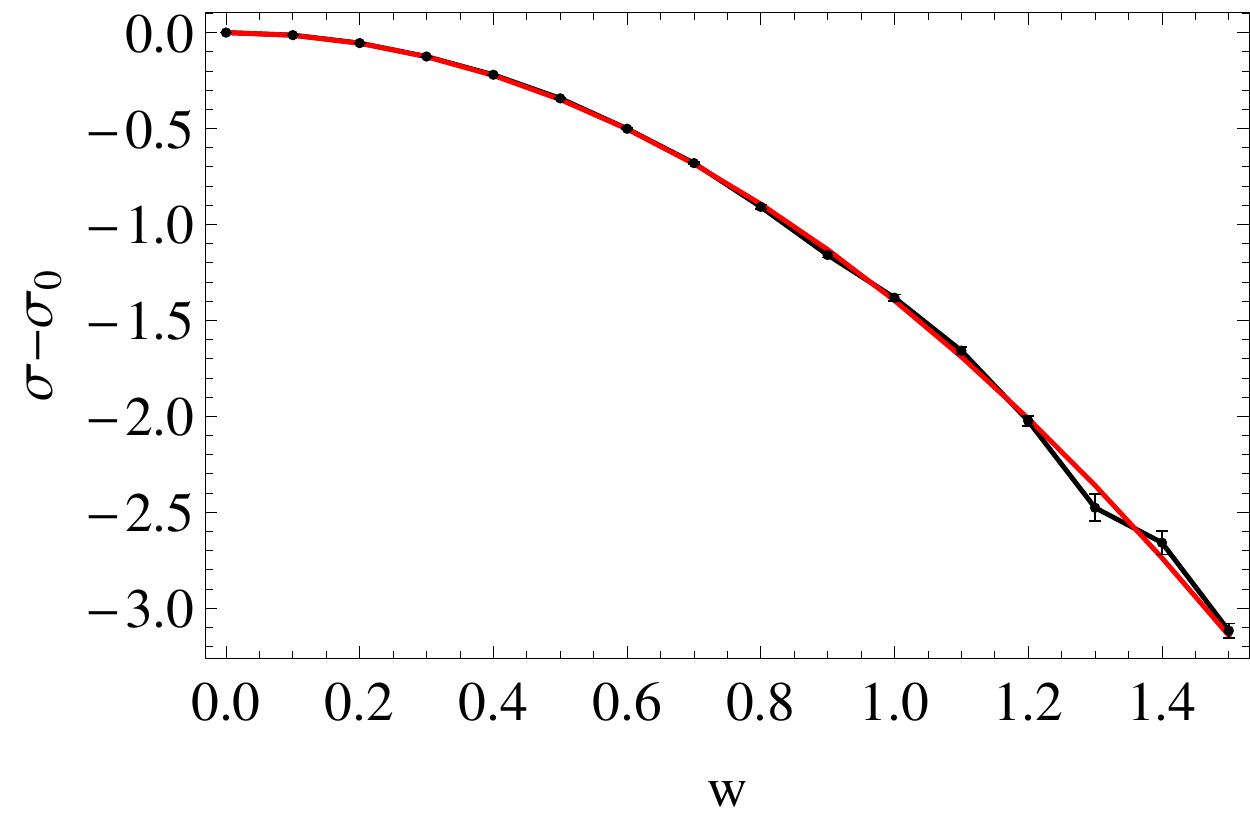}
\caption{\label{fig:sigcorrlgrho}
Decrease of $\sdc$ at large $\rho$ as a function of noise.
We plot the subtracted $\sdc$ as a function of the noise strength $w$. The black line shows the numerical
results, and the red line the prediction \eqref{eq:sdcsnshc} for small noise.
On the left panel we have set $\mu_0=10$, and on the right $\mu_0=20$. The numerical results follow from
averaging over 10 realizations on a grid of size $100\times 100$, with $L_x=20\pi$ and $k_*=1$.} 
\end{center}
\end{figure}

In Fig.~\ref{fig:sigcorrlgrho} we turn our attention to the case of large $\rho$, and again plot the
subtracted conductivity as a function of the noise strength $w$. We present both the numerical results
(black lines) and the analytic prediction of Eq.~\eqref{eq:sdcsnshc} (red lines). We consider systems with
$\mu_0=10$, and $\mu_0=20$, which at $w=0$ correspond respectively to $\rho_0=37.897$, and $\rho_0=133.398$.
The plots show a good agreement between analytic and numeric results, and confirm the prediction that
disorder decreases the conductivity in the limit of large charge density.

\begin{figure}
\centering
\def\svgwidth{\columnwidth}
\includegraphics[width=0.7\textwidth]{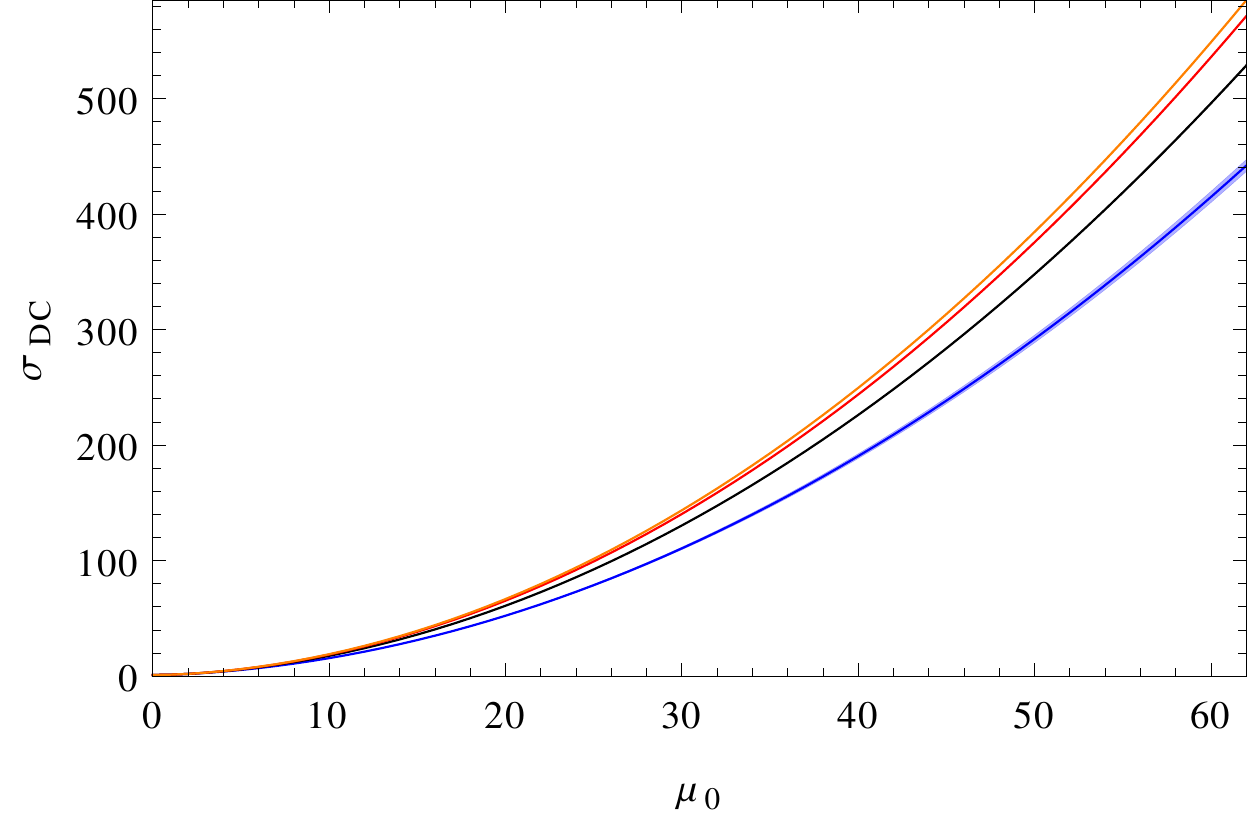}
\caption{$\sdc$ versus $\mu_0$. The different lines correspond, from top to bottom, to noisy chemical potentials
with $w = 0, 1, 2, 3$. 
The numerical results follow after averaging over 36 simulations on a grid of size $100\times 100$, 
with $L_x=20\pi$ and $k_*=1$. We plot with color bands the error of the average, which is only visible in the $w=3$ case.} 
\label{fig:sigmavsmu}
\end{figure}

\begin{figure}
\centering
\def\svgwidth{0.47\columnwidth}
\includegraphics[width=0.47\textwidth]{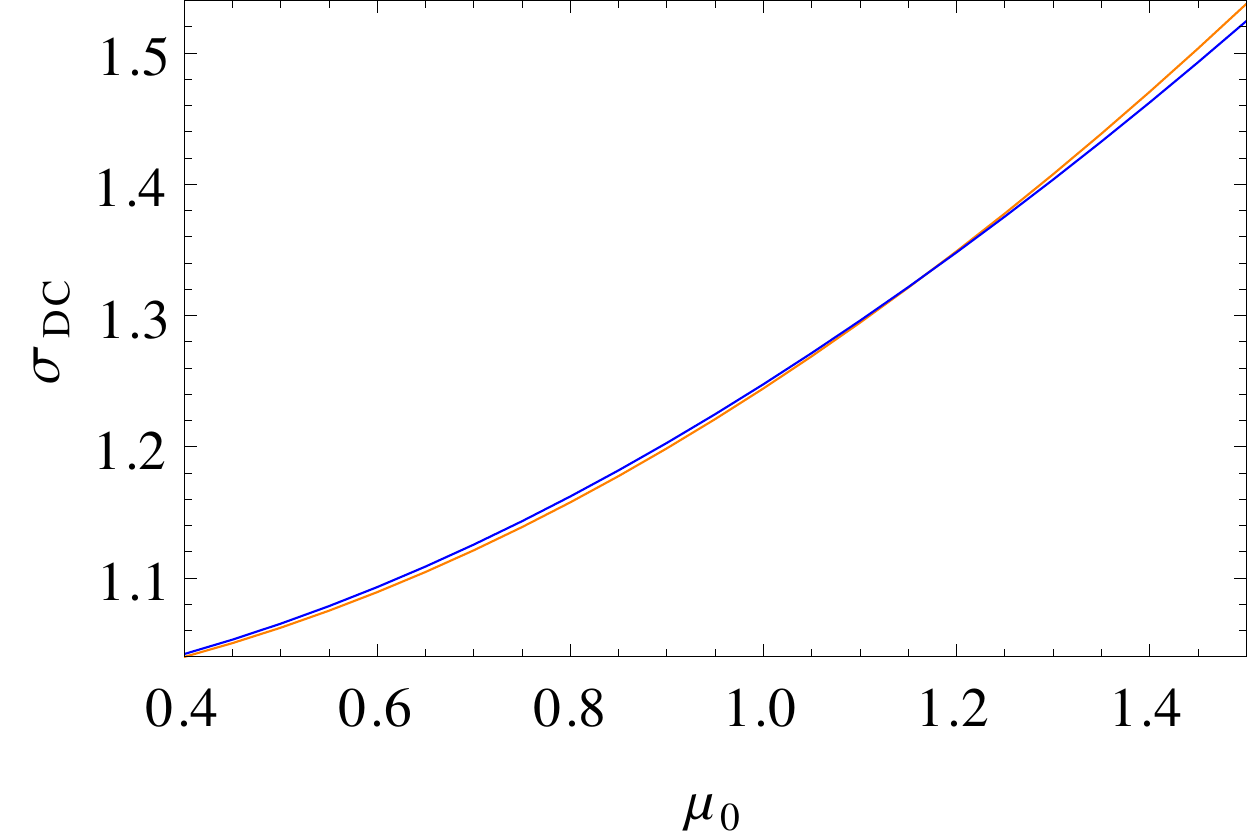}
\includegraphics[width=0.51\textwidth]{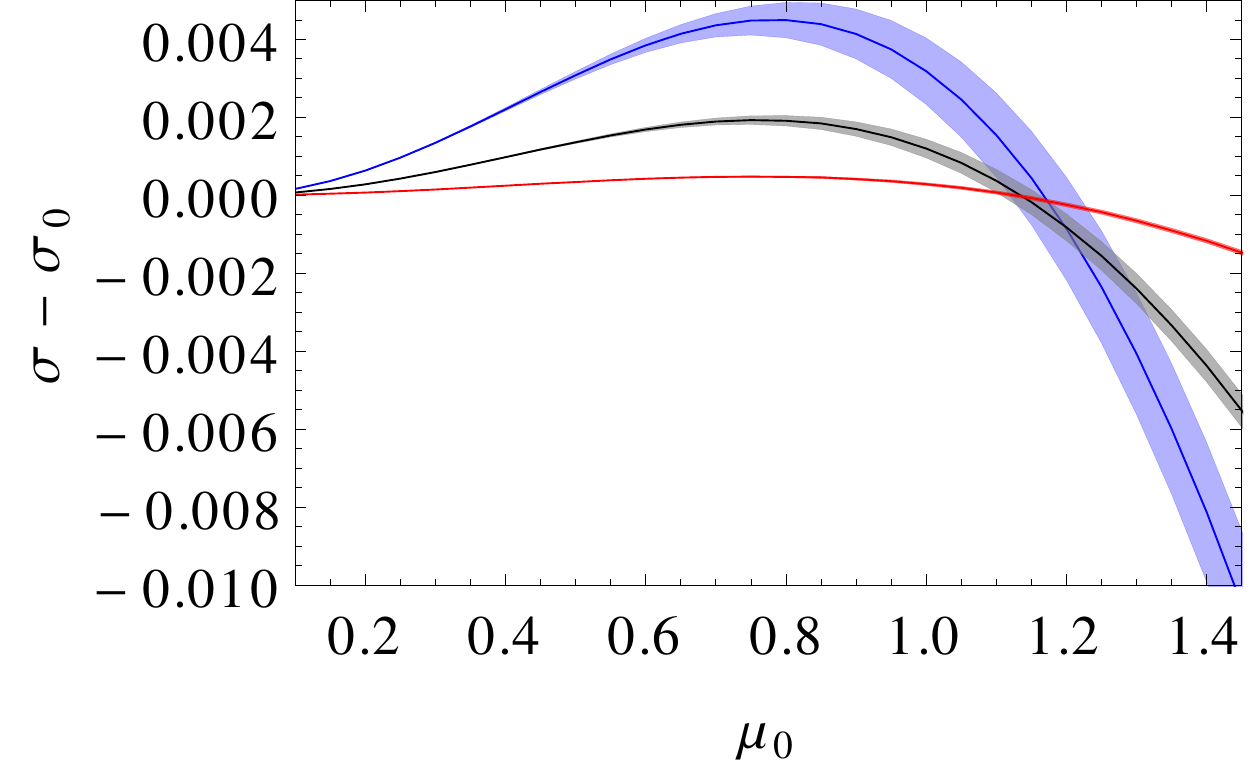}
\caption{Evolution of $\sigma_{DC}$ versus $\mu_0$; comparison of the clean and
noisy systems. On the left the orange line shows the clean case, and the blue line 
the noisy system with $w=3$. 
On the right we plot the difference between the clean and the noisy cases, $\sigma-\sigma_0$, versus $\mu_0$. 
The three lines correspond to $w=1,2,3$ (red, black, and blue respectively).
The value of the chemical potential at which $\sigma-\sigma_0$ 
changes sign lies in the interval $(1.15,\,1.20)$ for $w=3$ and $(1.10,\,1.15)$ for $w=1,\,2$.
We have averaged over $36$ simulations on a grid of size $100\times 100$, with $L_x=20\pi$ and $k_*=1$. 
The color bands correspond to the standard deviation.\label{fig:sigmavsmuUP}}
\end{figure}

We now proceed to study the disordered conductivity for the whole range of the chemical potential, and in 
Figs.~\ref{fig:sigmavsmu} and \ref{fig:sigmavsmuUP} we plot $\sdc$ versus $\mu_0$ for noisy chemical 
potentials of the form \eqref{eq:noisefunc}.
First, in Fig.~\ref{fig:sigmavsmu} we present the results of our simulations along a wide range of $\mu_0$
for four different values of the strength of noise: from $w=0$ (orange line) corresponding to the clean
system, up to $w=3$ (blue line) which, when setting $k_*=1$ in the sum \eqref{eq:noisefunc},
corresponds to a noisy chemical potential whose maximum oscillations have amplitudes $\approx70\%\,\mu_0$.
Except for the low $\mu_0$ region, which we will resolve in Fig.~\ref{fig:sigmavsmuUP}, the plot shows that 
the noise has the effect
of reducing the conductivity, and this effect increases with $\mu_0$. Notice however, that even for the
strongest noise studied, the conductivity is still increasing towards higher $\mu_0$, and is
always larger than the conformal value $\sdc=1$. Actually, one can see from Eq.~\eqref{eq:sigdcm0}
that $\sdc>1$ for any noise, and thus the system behaves always as a metal.

We have shown above, when studying the weak disorder case, that an enhancement of
$\sdc$ is observed in the small charge density limit.
To examine more closely the effect of disorder at low values of $\mu_0$, and therefore resolve the leftmost
region of the plot in Fig.~\ref{fig:sigmavsmu},
in Fig.~\ref{fig:sigmavsmuUP} left we plot $\sdc$ for $w=3$, and compare it with the homogeneous result. 
We observe that for values of $\mu_0\lesssim 1$ the conductivity is slightly enhanced 
in presence of noise. To better illustrate this fact, and pinpoint the value of $\mu_0$ above which 
the noise stops enhancing the conductivity and starts decreasing it,  
on the right panel of Fig.~\ref{fig:sigmavsmuUP} we plot the subtracted conductivity 
(\emph{i.e.} the difference between the noisy and the homogeneous results). 
We observe that the maximum enhancement occurs at $\mu_0 \approx 0.8$, and that the critical
value at which the enhancement ceases is in the interval $(1.1,\,1.2)$ (see caption of Fig.~\ref{fig:sigmavsmuUP}).
It is not difficult to estimate that the change of sign in the right panel of Fig.~\ref{fig:sigmavsmuUP} should  occur 
at $\mu_0\approx 1$.
Indeed from the expression \eqref{eq:pertsigmalc} for the resistivity in the small charge limit
one can see that the noisy correction proportional to $\tilde w^2$ changes sign at 
$\mu_0\approx 2/d^2=1$.\footnote{If 
one plays the same game in the large charge limit, from \eqref{eq:cdef} it follows that the $O(w^2)$ noisy correction to the
resistivity changes sign at $\mu_0\approx2.62$. Notice however, that in view of \eqref{eq:rhomulrho}, 
the approximation $\rho_0\approx c\,\mu_0^2$ is rather inaccurate at low values of $\mu_0$.}


\subsubsection{DC conductivity as a function of the charge density}
\label{ssec:sdcvsrho0}
In this section we will study how the DC conductivity behaves as a function of the charge density in presence
of disorder. This analysis will allow us to compare the qualitative behavior of our system to that
of semimetal graphene close to its charge neutrality point. 
Experimental results~\cite{PhysRevLett.99.246803} show that the DC conductivity of clean enough graphene samples
displays the following features: a nonzero minimal conductivity at vanishing charge density, a linear behavior
$\sdc\propto \rho$
up to some critical charge $\rho_*$ that is sample dependent (namely, determined by the amount of impurities), 
and a sublinear behavior for larger values of the charge density.

In Sec.~\ref{sssec:strongnoise} we have computed a semi-analytical prediction for the evolution of
 $\sdc$ as a function of $\langle\rho\rangle$.
We shall now present results following both from the purely numerical solution of the system, and the semi-analytical approach given by Eq.~\eqref{eq:sdcall}, compare them, and show their agreement.
First, we will show numerical results for $\sdc$ at moderate noise, that is for
a value of the strength of noise $w$ such that 
the chemical potential becomes very small at its minima, but neither 
$\mu(x)$ nor the charge density $\rho(x)$ ever become negative.
Next, we will present the numerical results for $\sdc$ vs $\langle\rho\rangle$ for strong noise.  This is a scenario
in which
the oscillations of the chemical potential are so large that regions of negative charge appear in the system,
bringing the setup to a configuration resembling that of graphene near the Dirac point, when puddles of charge of different sign are expected to be present in the system~\cite{Lucas:2015sya,PhysRevLett.107.156601}.

\begin{figure}[htb]
\begin{center}
\includegraphics[width=0.7\textwidth]{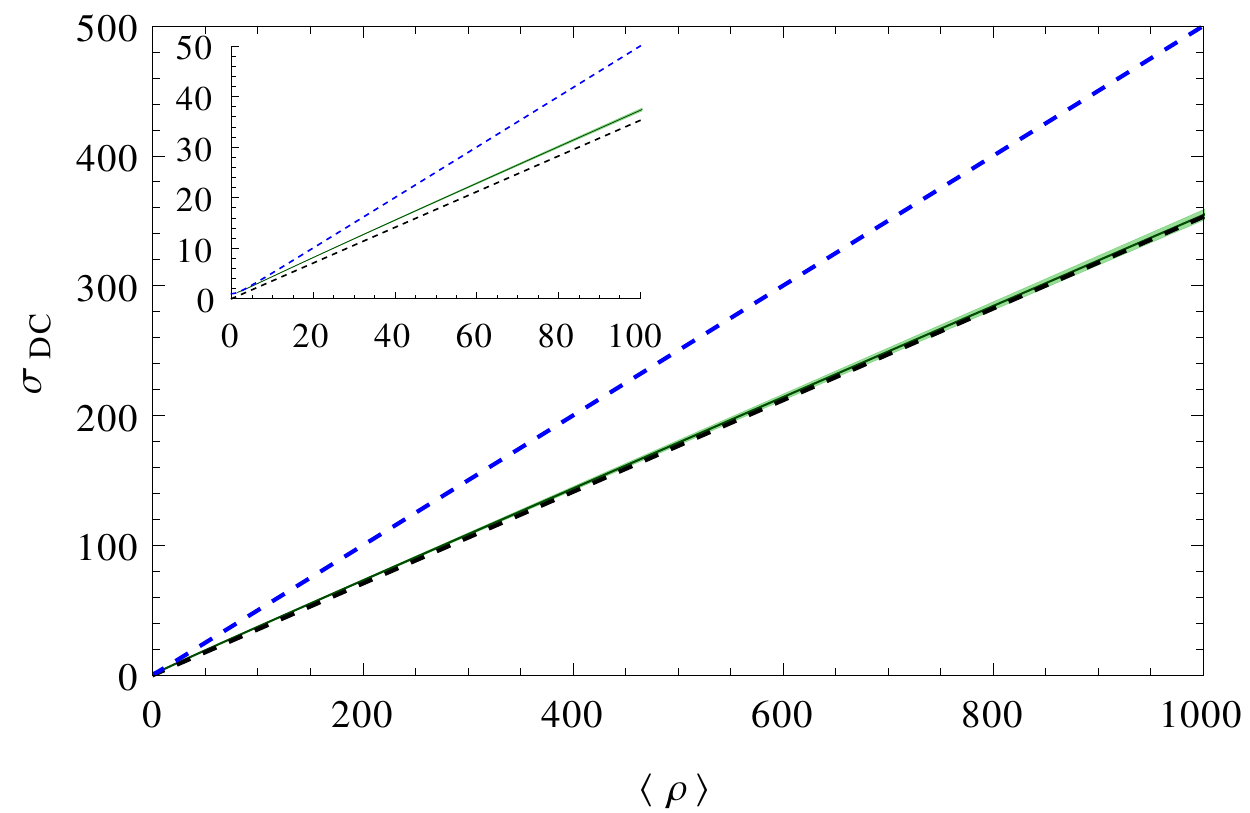}
\caption{\label{fig:sdcvsrho}DC conductivity as function of the charge density. The blue dashed upper line corresponds 
to the clean case. 
The solid green line is the numerical simulation for the noisy case with $\omega=3$, and the green band 
shows the error of the average.
The black dashed line corresponds to the analytic prediction \eqref{eq:sdcvsavrho}. 
We have used a grid of size $100\times 100$, set $L_x=20\pi$, $k_*=1$, and averaged over 36 realizations.} 
\end{center}
\end{figure}

In Fig.~\ref{fig:sdcvsrho} we consider the case of moderate noise and
plot the numerical results for 
$\sdc$ as a function of $\langle\rho\rangle$ for a fixed strength of noise $w=3$ (green line). 
For this value of $w$ the maximum oscillations of the chemical potential are of order 70\% $\mu_0$,
and both $\mu(x)$ and $\rho(x)$ stay positive for all $x$.
One can observe that, as for a clean system, $\sdc$ is linear in $\langle\rho\rangle$ (except for very low
values of $\langle\rho\rangle$). However, the slope is much lower than that of the clean system (blue dashed line),
and agrees very well with that predicted in Eq.~\eqref{eq:sdcvsavrho} above (black dashed line).

\begin{figure}[htb]
\begin{center}
\includegraphics[width=0.7\textwidth]{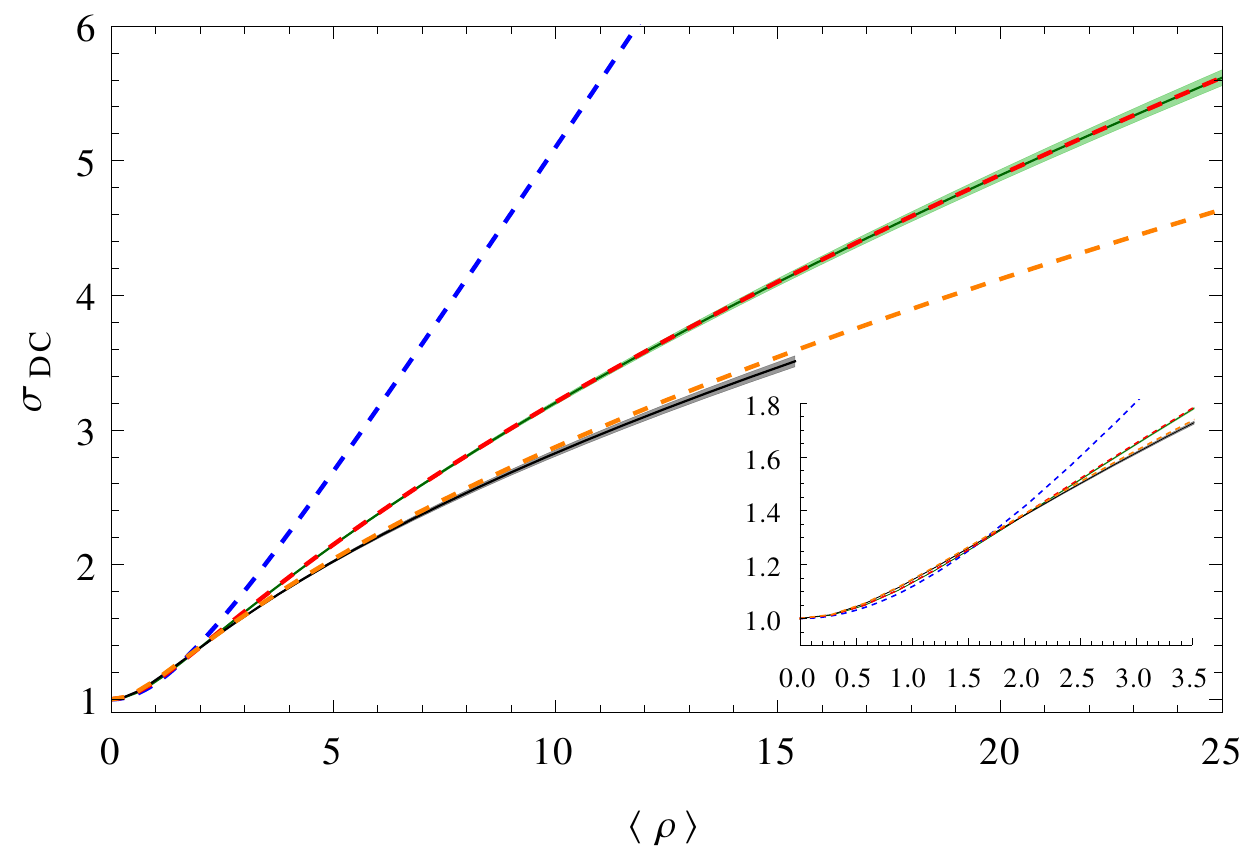}
\caption{\label{fig:sdcvsrhoStw}DC conductivity as function of the charge density. 
The blue dashed upper line corresponds to the clean case.
The solid green, and solid black lines are the numerical data for $w=6$ and $w=8$ respectively.
The dashed red line for $w=6$, and the dashed orange line for $w=8$ show the semi-analytical prediction 
of Eq.~\eqref{eq:sdcall} after averaging over 1000 realizations. As before, shaded bands depict the error of the average.
We have used a grid of size $120\times 120$, set $k_*=1$, $L_x=20\pi$, and averaged over 43 and 57 realizations for $w=6$ and $w=8$ respectively.} 
\end{center}
\end{figure}

\noindent{\bf Strong noise}: We focus now on the scenario where the strength of noise is 
large enough for $\mu(x)$, and $\rho(x)$, to change sign repeatedly along the system.
In Fig.~\ref{fig:sdcvsrhoStw} we present numerical results for $w=6$, and 8 (green and black solid lines respectively).
The numerical computation of $\sdc$ for these large values of $w$ becomes quite delicate,\footnote{At the largest
oscillations of $\phi(z,x)$, the radicand of the square root in Eq.~\eqref{eq:sigdcm0} becomes very close to zero.
One needs to resolve $a^{(2)}(x)$ with high accuracy in these situations, which demands thinner lattices
as the oscillations of the field $\phi$ get larger.}
and as a consequence,
the range of $\langle\rho\rangle$ where we can trust our results is significantly restricted with respect
to the case of $w=3$ in Fig.~\ref{fig:sdcvsrho}. In particular, notice that for $w=8$ we have numerical results only 
up to $\langle\rho\rangle\approx 16$. We have studied the stability of our results against the lattice size
in Fig.~\ref{fig:numstability} in the appendix.

For large values of $w$ the analytic weak noise prediction \eqref{eq:sdcvsavrho} is not reliable, and we instead 
compare our numerics to the prediction \eqref{eq:sdcall} where all orders in $w$ are kept.
First, it is remarkable how well our semi-analytical prediction works at $w=6$ (red dashed line) and $w=8$ 
(orange dashed line). 
Notice that the approximation  that led us to that expression, which effectively neglects the effect of gradients 
along $x$, should become less reliable as one increases $w$,  $\mu_0$, or both,
and those gradients become larger.\footnote{It follows from the form of $\mu(x)$ in
\eqref{eq:noisefunc} that the gradients along $x$ of the chemical potential are proportional to $w$ and $\mu_0$.}
One indeed expects that for large enough $w$ and $\mu_0$, the effect of the gradients become 
important and our semi-analytical approach fail. A closer inspection of Fig.~\ref{fig:sdcvsrhoStw} 
reveals that for the case of $w=8$, at the largest values of $\langle\rho\rangle$ 
we have reached the numerical results fall 
slightly below the prediction of Eq.~\eqref{eq:sdcall}. This effect seems to be enhanced for the case $w=10$ 
we present in Fig.~\ref{fig:sigvsrhow10} of the appendix, although a more ambitious numerical study that
would allow larger values of $w$ and $\langle\rho\rangle$ would be needed to settle the question.
A decrease of $\sdc$ due to the $x$ dependent gradients 
is to be expected on general grounds, since gradients of the charge density  result in a reduction of 
the diffusion constant~\cite{Ryu:2011vq}.
Additionally, increasing $k_*$, thus reducing the correlation length of our noise \eqref{eq:noisefunc},
one also expects the effects of the gradients to become more important.
However, we postpone a more thorough numerical study, including an analysis of the effect of $k_*$,
to a future work where a more realistic two-dimensional noise, which would get us closer to the
situation in graphene, will be implemented.

\begin{figure}[htb]
\begin{center}
\includegraphics[width=0.49\textwidth]{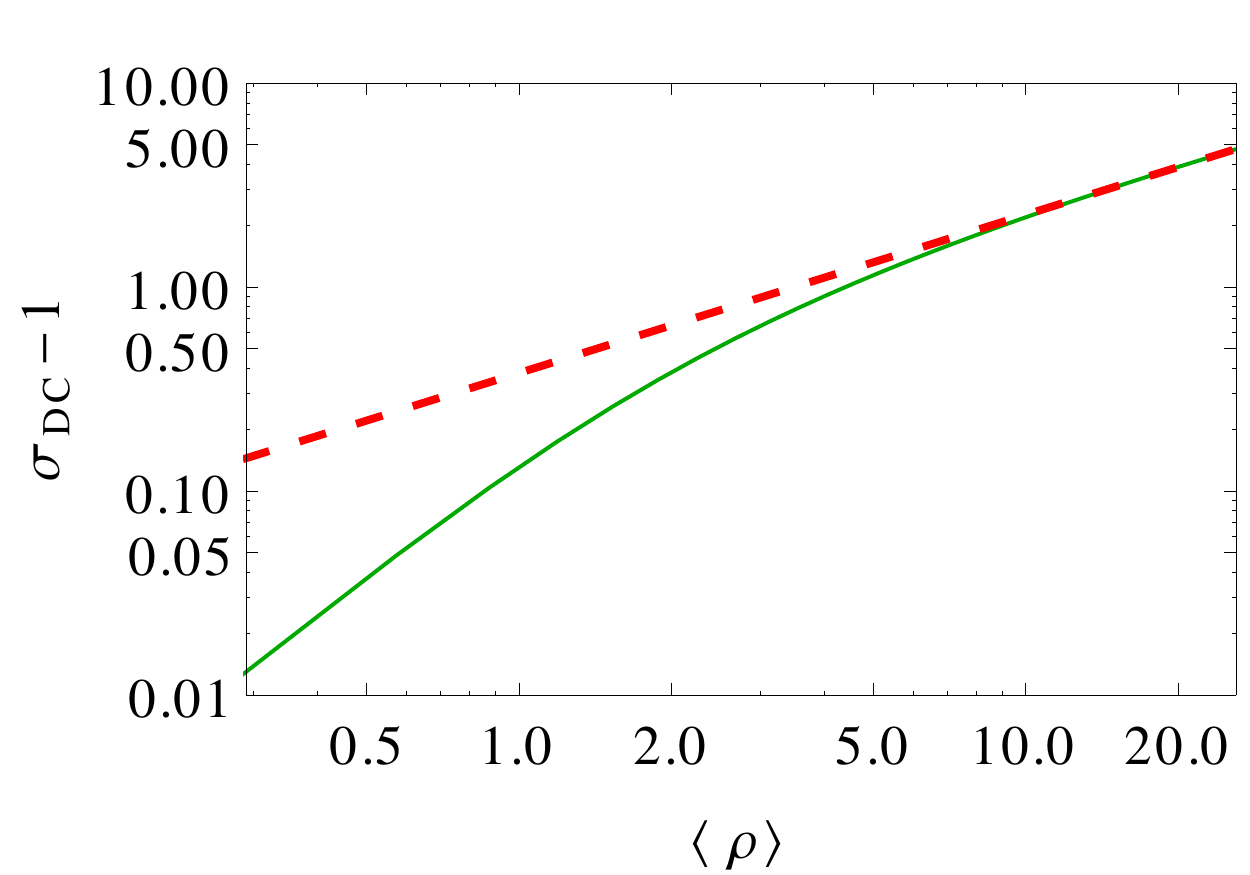}
\includegraphics[width=0.49\textwidth]{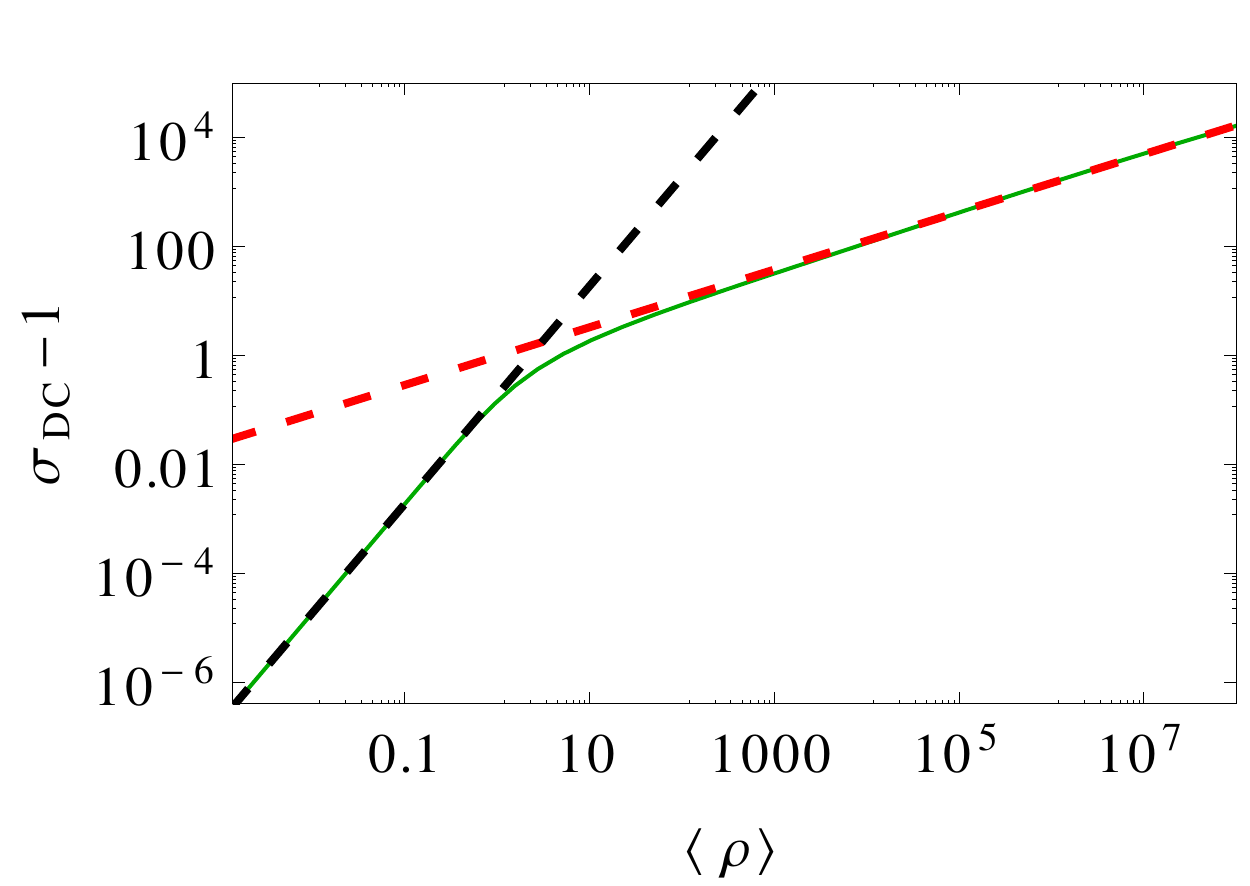}
\caption{\label{fig:subtrend}log-log plot of $\sdc$ versus $\langle \rho\rangle$.
On the left we present our numerical data for $\sdc$ at $w=6$
featured in Fig.~\ref{fig:sdcvsrhoStw} (green solid line). 
The red dashed line depicts a linear fit to the $\langle \rho \rangle >15$ region, resulting
in $\log_{10}(\sigma_{DC}-1)=0.78\, \log_{10} \langle \rho \rangle - 0.42$. On the right we plot $\sdc$ computed from Eq.~\eqref{eq:sdcall} with $w=8$ (green solid line), having set $k_*=1$ and $L_x=20\pi$. The black dashed line shows the linear fit
$\log_{10}(\sigma_{DC}-1)=2.00\, \log_{10} \langle \rho \rangle-0.72$ in the region $\langle\rho\rangle <0.1$,  while the red dashed line corresponds to the fit $\log_{10} (\sigma_{DC}-1)=0.53\, \log_{10} \langle \rho \rangle-0.01$ in the region
$10^5\lesssim\langle\rho\rangle\lesssim 10^8$. We have averaged over 33 realization of the noise.} 
\end{center}
\end{figure}

Finally, one readily notices that in Fig.~\ref{fig:sdcvsrhoStw}, after a short region at very low 
$\langle\rho\rangle$ where they largely agree with the clean case  (dashed blue line), the noisy conductivities 
become sublinear. To characterize this sublinear behavior, 
in Fig.~\ref{fig:subtrend} we examine closely the asymptotic behavior of $\sdc$ in the limits of small and large
$\langle\rho\rangle$.
By employing the semi-analytical expression \eqref{eq:sdcall}, we compute $\sdc$ for $w=8$ for a wide range
of values of $\langle\rho\rangle$ (up to $\sim10^8$).
The resulting data is shown as the green solid line on the right panel of Fig.~\ref{fig:subtrend}.
Also in that plot we present logarithmic fits in the low (black dashed line) and high (red dashed line) charge
regions. These fits reveal a quadratic growth $\sdc\propto \langle\rho\rangle^2$ in the small $\langle\rho\rangle$
region, and a sublinear trend $\sdc\propto\langle\rho\rangle^{0.53}$ as $\langle\rho\rangle$ becomes large.
Notice that the result at small $\langle\rho\rangle$ agrees perfectly with the prediction 
\eqref{eq:SigmaLowRho} resulting from
the strong noise analysis. Moreover, the behavior at large charge density is very close to that predicted in
Eq.~\eqref{eq:SigmaLargeRho}, pointing towards an asymptotic behavior $\sdc\propto\sqrt{\langle\rho\rangle}$.
Finally, in Fig.~\ref{fig:subtrend} left we present a logarithmic fit of the numerical results for $w=6$.
The fit is realized in the region of largest charge density available ($15\lesssim\langle \rho\rangle\lesssim25$),
where $\langle\rho\rangle$ is not large enough for the system to be properly in the large charge regime.
Nevertheless, the data displays a clearly nonlinear behavior, with the fit resulting in a growth
trend $\sdc\propto\langle\rho\rangle^{0.78}$.
In view of these results it is worth mentioning that theoretical models of 
electric transport in graphene  predict this sublinear behavior as a result of the presence of charged impurities in
the system~\cite{PhysRevLett.98.186806,PhysRevLett.107.156601},
with emphasis on this disorder creating puddles of charge of opposite sign.\footnote{In particular, in the model 
\cite{PhysRevLett.107.156601} it is important the fact that the disorder is correlated. 
Note that our disorder is, on one hand perfectly correlated along one spatial direction, and on the other it 
also possesses a nonzero correlation length in the $x$ direction, proportional to $1/ k_*$, 
since we have truncated the sum in \eqref{eq:noisefunc} to a finite number of modes.}


\subsubsection{Noisy DC at vanishing $\langle\rho\rangle$}
\label{ssec:sdcvsw}
We will finish this section by considering the case of a noisy chemical potential with vanishing spatial average 
of the form
\begin{equation}
\mu(x)=\hat w\,\sum_{k=k_0}^{k_*}\,\cos(k\,x+\delta_k)\,,
\label{eq:noisefuncmu0}
\end{equation}
which corresponds to Eq.~\eqref{eq:noisefunc} in the limit $\mu_0\to0$, $w\to\infty$ with $\hat w\propto \mu_0\,w$
fixed. We will then be studying a system where the regions of positive and negative charge average to zero.

To find asymptotic expressions at low and high $\hat w$, we can again use Eq.~\eqref{eq:sdcall}. For low values of $\hat w$, expanding in $\hat w$  and using Eq.~\eqref{eq:murhosmrho}, which gives $\rho(\mu)$
in the low charge limit, it is straightforward to check that the conductivity grows quadratically with $\hat w$:
\begin{equation}
\sigma_{DC} \approx 1+\frac{\hat w^2}{4}\Delta.\label{eq:SigmaLowW}
\end{equation} 
On the other hand, for high values of $\hat w$, the situation is analogous to the strong noise case at large $\langle \rho \rangle$. 
Zones of positive and negative charge are present in the system for any $\hat w$, and thus we can
apply the analysis of Sec.~\ref{sssec:strongnoise} based on the existence of zeros of $\mu(x)$.
Modifying suitably the definition of $\eta(x)$ as
\begin{equation}
\eta(x)=\sum_{k=k_0}^{k_*}\,\cos(k\,x+\delta_k)\,, 
\end{equation} 
and assigning  to $\hat w$ the roll played by $\mu_0$ in that analysis,
the asymptotic expression for the conductivity at large $\hat w$ is analogous to Eq.~\eqref{eq:SigmaLargeMu}:
\begin{equation}
\sigma_{DC}\sim\frac{\hat w L_x}{7.083}\left< \frac{1}{\sum_i \frac{1}{|\eta'(x_i)|}} \right>_{\mathrm{noise}},\label{eq:SigmaLargeW}
\end{equation}
and thus, linear in $\hat w$.

\begin{figure}
\centering
\def\svgwidth{0.47\columnwidth}
\includegraphics[width=0.48\textwidth]{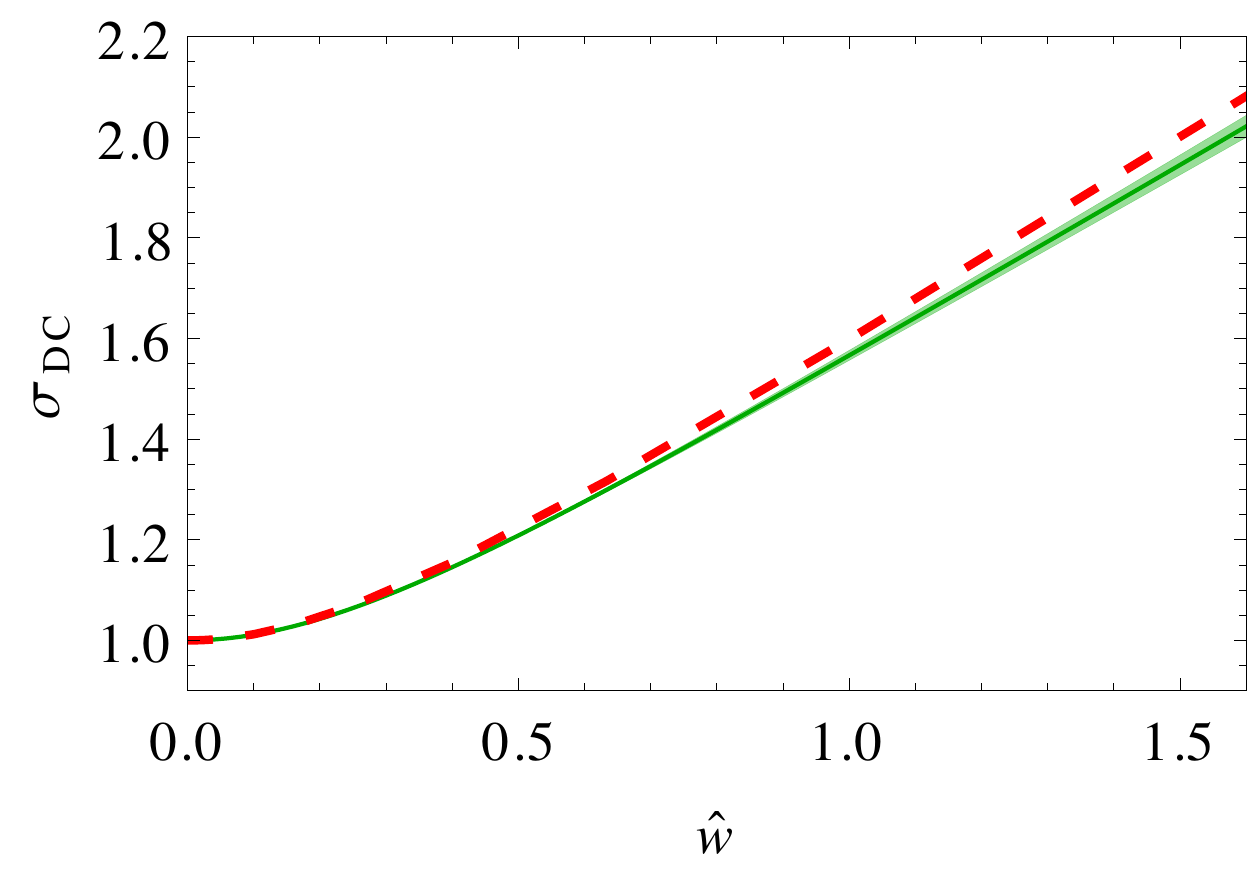}
\includegraphics[width=0.51\textwidth]{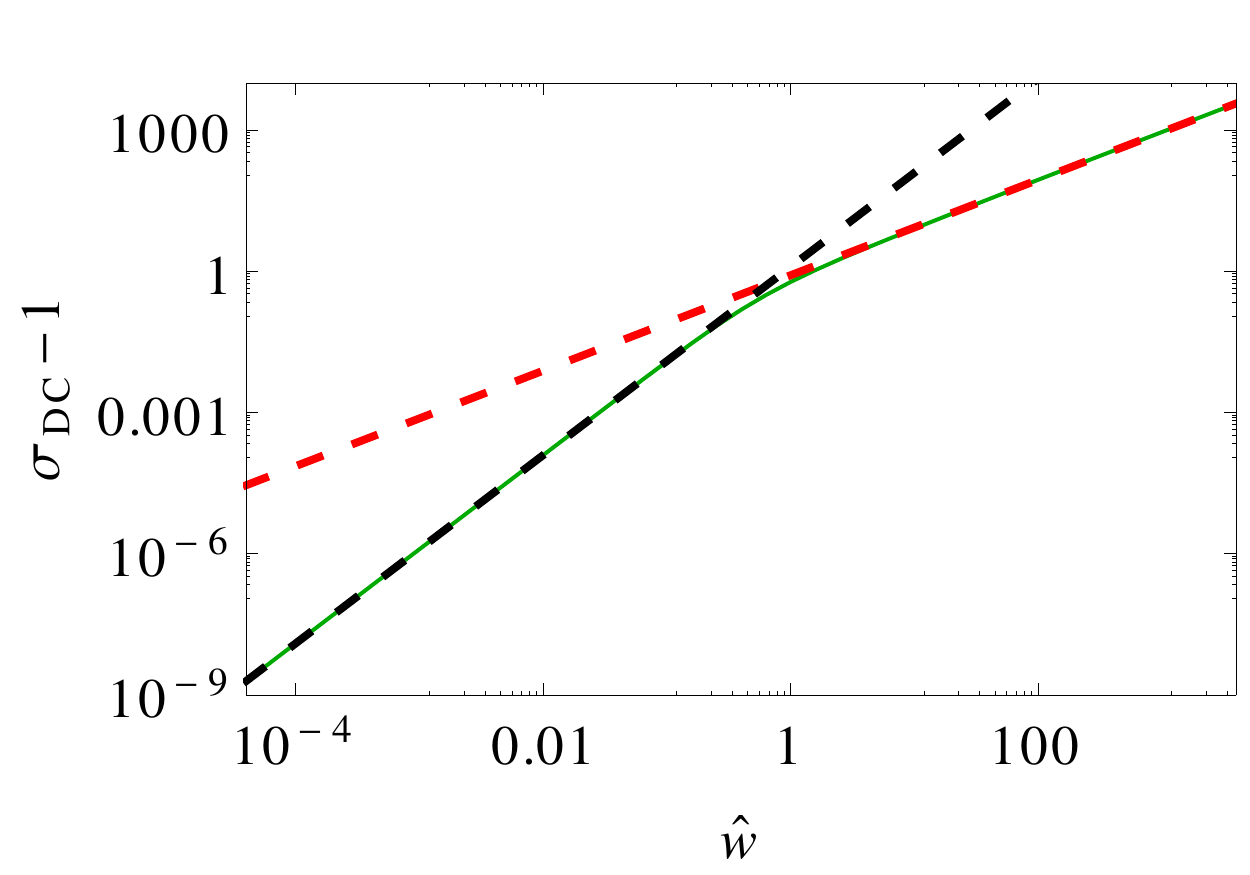}
\caption{Evolution of $\sigma_{DC}$ versus $\hat w$. On the left we plot our numerical results (solid green line) after averaging
over 22 realizations on a lattice $100\times100$ and a noise with $k_*=1$ and $L_x=20\pi$. We also plot the error with shaded bands. The dashed red line shows the semi-analytical prediction 
of Eq.~\eqref{eq:sdcall} after averaging over the same noise realizations.
On the right we present a logarithmic plot of the same semi-analytical prediction in a broader range (solid green
line).  The black dashed line shows the linear fit
$\log_{10}(\sigma_{DC}-1)=2.00\, \log_{10} \hat w +0.10$ in the region $\hat w <3\cdot 10^{-3}$, while the red dashed line corresponds to the fit $\log_{10} (\sigma_{DC}-1)=1.01\, \log_{10} \hat w-0.08$ in the region
$50 < \hat w<10^4$.
\label{fig:sigmavswmu0}}
\end{figure}

In Fig.~\ref{fig:sigmavswmu0} we present both numerical and semi-analytical results for $\sdc(\hat w)$. 
First, on the left panel we plot our numerical data (solid green line), and compare them with the result of
the semi-analytical expression \eqref{eq:sdcall} (dashed red line). As it happens in Fig.~\ref{fig:sdcvsrhoStw} for 
the strongest noise there ($w=8$),
our numerical results at the highest $\hat w$ available are slightly below the semi-analytical prediction,
again showing a likely onset of the effects of the gradients along $x$ (which are neglected in the semi-analytical
approximation).
Next, in Fig.~\ref{fig:sigmavswmu0} right we study the asymptotic behavior of the semi-analytical result
(solid green line) both in the small, and large $\hat w$ limits. We perform logarithmic fits in the small (black
dashed line), and large (red dashed line) $\hat w$ region. The fits (see caption)
confirm the expected quadratic behavior at small $\hat w$, and linear trend in the large $\hat w$ limit. In particular, for a system as that of Fig.~\ref{fig:sigmavswmu0}, Eq.~\eqref{eq:SigmaLowW} gives $\log_{10}(\sigma_{DC}-1)=2\log_{10}\hat w+0.10$,
while Eq.~\eqref{eq:SigmaLargeW} using the same noise realizations of Fig.~\ref{fig:sigmavswmu0} results
in  $\log_{10}(\sigma_{DC}-1)=\log_{10}\hat w-0.03$. Notice the perfect agreement at small $\hat w$, while
at large $\hat w$, as it happened in the last section for the large charge limit, we expect the semi-analytical
data to eventually agree with the prediction in the asymptotic $\hat w\to\infty$ limit (we have observed
how the fit of the data gets closer to the prediction \eqref{eq:SigmaLargeW} as larger $\hat w$ data points
are included in the fit).

\section{Spectral properties} 
\label{sec:spectral}

In this section we  consider a realization of disorder characterized by a power spectrum. Namely,
a chemical potential of the form
 \bea
\mu(x)&=&\mu_0+{\mu_0\over25}\,w\,\sum_{k=k_0}^{k_*}\sqrt{S(k)}\,\cos(k\,x+\delta_k)\,,
\label{eq:noisecorr}
\eea
where
\be
S(k)={1\over k^{2\alpha}}
\label{eq:powsp}
\ee
is the power spectrum of our noise. Notice that the disordered chemical potential considered
in previous sections corresponds to a flat spectrum with $\alpha =0$. As discussed in \cite{Arean:2014oaa}, 
for a spectrum with
$\alpha>0$ the typical length scale of the noise is of the order of the system size, 
and hence we will denote this case as correlated noise. 

We  will construct massive embeddings with a noisy chemical potential of the form \eqref{eq:noisecorr}
and analyze the power spectra of the observables of the system, namely the charge density $\rho(x)$, and the 
quark condensate $c(x)$. For a given input chemical potential with power spectrum $k^{-2\alpha}$ as above,
we will compute the power spectra of the charge density $S_\rho(k)$, and the quark condensate $S_c(k)$, and 
study how they behave as functions of $\alpha$. Let us then define
\be
S_\rho(k)={1\over k^{2\Gamma(\alpha)}}\,,\qquad
S_c(k)={1\over k^{2\Delta(\alpha)}}
\label{eq:outspec}
\ee
as the spectra of the charge density and quark condensate respectively.

In \cite{Arean:2013mta,Arean:2014oaa}, in the context of models of holographic superconductors,
it was found a certain universal behavior for the spectra of the VEVs of the model as functions of the
input power spectrum defining the chemical potential. In particular, for an input power
spectrum $~k^{2\alpha}$
as above, the output power spectra (those of the VEVs) were of the form $\sim k^{2\alpha+d}$, with $d$ an
integer coefficient different for each operator. This was interpreted as a kind of renormalization of
small wave lengths in which higher harmonics of the VEVs of the corresponding operators are suppressed or 
enhanced, depending on the sign of the coefficient $d$ in the output power spectra.

In order to determine the power spectra of the charge density and quark condensate ($S_\rho$ and $S_c$),
we construct massive embeddings characterized by a chemical potential of the form \eqref{eq:noisecorr}
for different values of $\alpha$. We keep fixed $m=0.5$, $\mu_0=1$, and $w=1$.\footnote{We have checked
that the results for the spectra are independent of the values of  $\mu_0$, $m$, and $w$.}.
From the asymptotic behavior of the embedding fields $\chi$ and $\phi$ we
can read $\rho$ and $c$, and determine their power spectra, namely $\Gamma$ and $\Delta$.
\begin{figure}[htb]
\begin{center}
\includegraphics[width=0.47\textwidth]{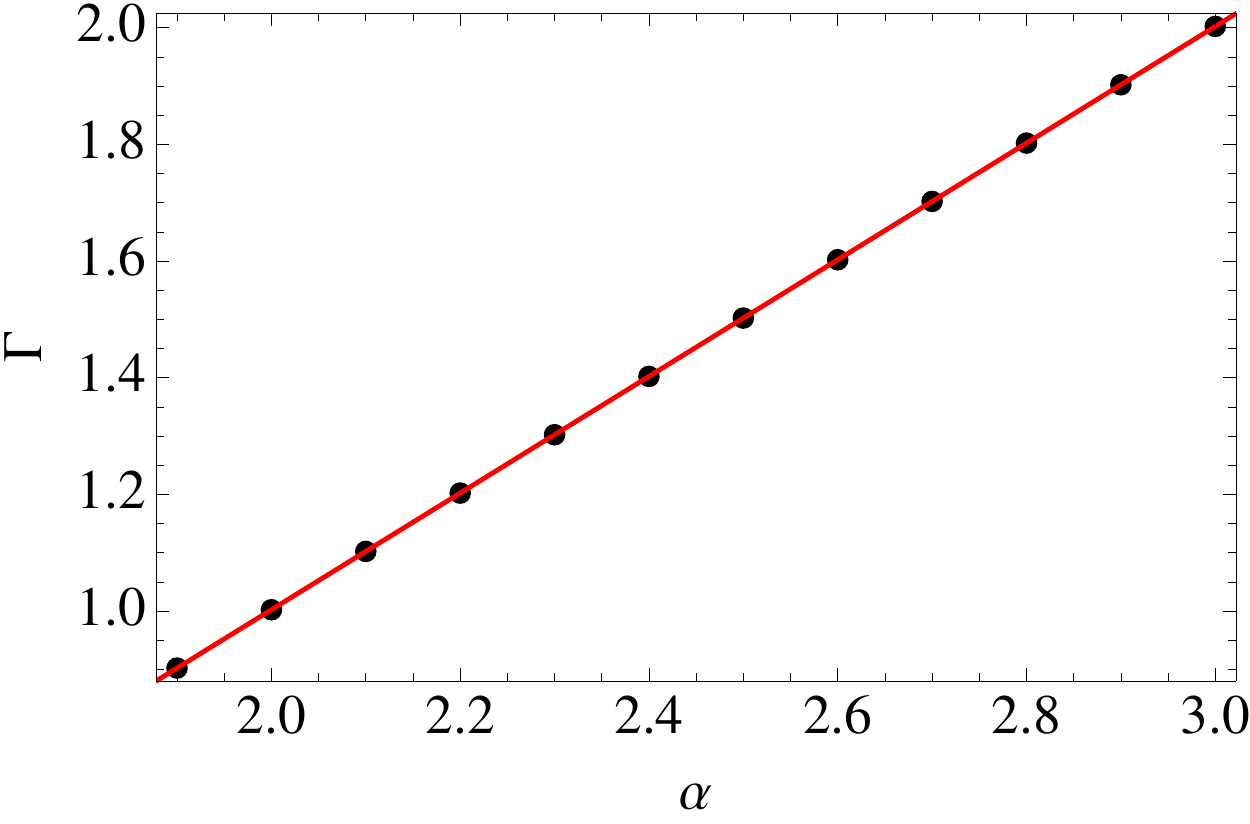}
\includegraphics[width=0.47\textwidth]{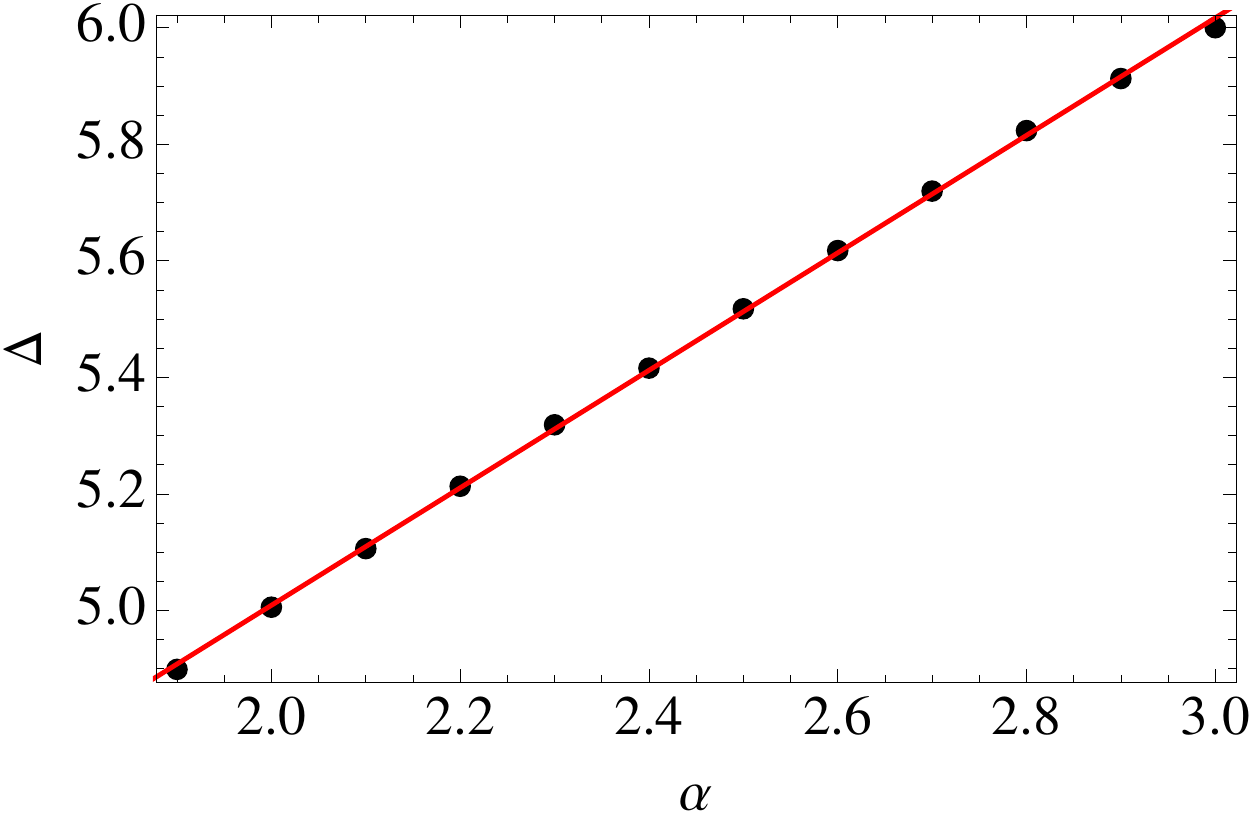}
\caption{\label{fig:spec} Power spectrum of the VEVs. Black dots 
correspond to the numerical result averaging over 35 realizations on a lattice of size $N_z=40$, $N_x=100$,
for a system with $L_x=3\pi/2$, and $k_*=62.7$ (corresponding to 47 modes in the sum \eqref{eq:noisecorr}).
On the left, for the charge density we plot $\Gamma$ versus $\alpha$, and the red line stands for the 
fit $\Gamma = 1.00\alpha-1.00$. On the right, for the quark condensate, we plot $\Delta$ versus $\alpha$, 
the red line shows the fit $\Delta = 1.01\alpha+2.99$.}
\end{center}
\end{figure}
We observe that they are very well
approximated by
\be
\Gamma\approx \alpha -1\,,\qquad \Delta\approx \alpha +3\,.
\label{eq:spcrenorm}
\ee
Notice that the power spectrum renormalization of the charge density agrees with that found in 
\cite{Arean:2013mta,Arean:2014oaa}, and implies that higher harmonics are enhanced with respect to the input
power spectrum. 
The opposite effect is observed for the quark condensate; the power spectrum renormalization we observe implies
that the higher harmonics are suppressed, and $c(x)$ is then smoothed out with respect to the input $\mu(x)$.
To illustrate these facts, in Fig.~\ref{fig:corremb} of App.~\ref{app:corremb} we plot the result of a simulation for an embedding with correlated noise.


\section{Conclusions and outlook} 
\label{sec:conclusions}

In this note we have presented a holographic model of (2+1)-dimensional charged matter with a disordered chemical potential.
The model is built upon a D3/D5 intersection at finite temperature and charge density, and the chemical potential consists of
random fluctuations about a nonzero baseline value.  We have worked in the probe limit where the setup reduces to inhomogeneous
embeddings of a D5-brane in a neutral black hole background. The construction of these embeddings has allowed us to 
study, both numerically and analytically, how the disorder affects the quark condensate and the charge density. In particular, we
have found that in the regime of moderate to large charge density, this quantity is enhanced by disorder, while the quark condensate is largely unaffected.

Arguably the main outcome of this work is the study of the DC conductivity for a top-down holographic model of charged matter
in the presence of disorder.
In the spirit of the membrane paradigm, the DC conductivity of our model can be expressed purely in terms of horizon data 
of the
background functions (those characterizing the embedding), and hence we could compute the conductivity of our disordered 
setup
without having to solve for the fluctuations. In the limit of weak disorder we obtained analytic expressions showing 
that disorder
enhances the conductivity at very small charge density, while decreasing it in the opposite regime, {\emph i.e.} 
large charge
density. 
Numerical simulations in agreement with the analytic predictions confirmed this behavior.
We have also studied the behavior of the DC conductivity as a function of the charge density, both through numerical
simulations and analytic approximations.
At weak noise the DC conductivity scales linearly with the charge density, 
with a slope that decreases as the strength of noise is increased, and which is always lower than that of the clean
system.
At strong noise, a  sublinear trend is observed in our numerical simulations.
Moreover, a careful analysis of our semi-analytical approximations confirmed that $\sdc$ becomes a
sublinear function of the spatial average of the charge density as this becomes large. 
This onset of sublinearities resembles that found in some regimes of graphene,
a context in which it was attributed to the effects of disorder~\cite{PhysRevLett.98.186806,PhysRevLett.107.156601}.

In a slightly tangential direction, we have also studied the effects of a noisy chemical 
potential characterized by a Fourier power spectrum.
This noise is correlated along distances of the order of the system size, and thus we refer to it as `correlated noise'.
In this scenario we have analyzed the spectral properties of the VEVs of the system: charge density and quark condensate,
observing what amounts to a spectral renormalization of sorts. 
For an input Fourier power spectrum 
of the chemical potential of the form $1/k^{2\alpha}$, the output power spectra of the VEVs are 
given by $1/k^{2\alpha +d}$, with $d\approx -2$ for the charge density, and $d\approx 6$ for the quark condensate. Therefore, the
higher wave numbers are enhanced in the spectrum of the charge density, which is then more irregular than the chemical
potential; while the opposite occurs for the quark condensate: the higher wave numbers are suppressed in the spectrum and the
quark condensate is a smoother function of the space direction $x$. This phenomenon was already observed in 
\cite{Arean:2013mta,Arean:2014oaa} for holographic models of superconductivity, and it deserves further study, specially in relation
with the universal response of the AdS geometry to time dependent sources found in~\cite{Buchel:2013gba}.

The analysis we have presented has revealed the D3/D5 intersection as a reasonably tractable top-down model of disordered
matter with the potential to reproduce behaviors observed in strongly coupled Condensed Matter systems.
Our results motivate a further exploration of disordered brane models, and there are several directions in which we plan
to make progress in the future. First, the most obvious step would be to extend our D3/D5 model to the case where the chemical
potential is disordered along both spatial directions.
Impurities in two-dimensional systems like graphene or thin superconducting films are randomly distributed along the two spatial
dimensions (although some correlation might still be present, see \cite{PhysRevLett.107.156601}), and hence considering two-dimensional disorder is the natural procedure. 
Moreover, according to the Harris criterion~\cite{Harris:1974zz}, that generalizes the standard power-counting criterion
to random couplings,
a chemical potential with random spatial inhomogeneities in 2 dimensions amounts to the introduction of marginally relevant
quenched disorder. 
Instead, as explained in~\cite{Garcia-Garcia:2015crx}, the present case where the chemical potential is disordered only along one direction corresponds  to a relevant noise. This fact would make an eventual extension of the present model beyond the probe limit both very interesting, and potentially very challenging.
In addition to changing the dimensionality of disorder, for applications to 
graphene near the charge neutrality point, it would be interesting to investigate further the case of strong noise
where the charge density changes sign along the system.
While this would be problematic in a scenario of massive embeddings, for which the topology could change when the charge 
density vanish, we have showed that it works for massless embeddings.
In this context it would be interesting to extend our results for the DC conductivity to larger values of the charge
density and the noise strength. 
 
Another interesting possibility would be to consider more sophisticated models where one could tune the total charge density,
and that of the impurities separately.\footnote{In the present model one could still think that by increasing the noise 
strength, the density of impurities grows. By defining an impurity as the region where the chemical potential 
exceeds its average by some percentage, it is clear that the number of impurities grows as the 
strength of noise is increased.}
A suitable framework could be that of Sen's tachyonic action for an overlapping brane-antibrane pair \cite{Sen:2003tm}, which was 
used in \cite{Arean:2015wea} to model superfluid phase transitions. One could deform the finite density configuration of 
\cite{Arean:2015wea} by switching on a disordered source for the scalar operator dual to the tachyon which would be akin
to a doping parameter.  
Finally, the definitive step forward would be to account for the backreaction of the flavor branes on the geometry, working in
the Veneziano limit in which both the number of color branes, $N_c$, and flavor branes, $N_f$, are large, but $x=N_f/N_c$ is kept
fixed. This limit, which is more tractable when the flavor branes are smeared along their orthogonal 
directions~\cite{Bigazzi:2005md,Nunez:2010sf},
would allow us to study
the RG flow of the dual disordered theory, and hopefully to characterize its ground state. Additionally, the whole set of transport 
coefficients would be accessible (recall that heat transport is dual to metric fluctuations, and hence inaccessible in the probe limit),
and the computed results reliable down to very low temperature. 
Solutions dual to brane intersections in the Veneziano limit and with a nonzero density of (2+1)-dimensional 
flavor degrees of freedom have been constructed in~\cite{Faedo:2015urf}. Interestingly, their IR geometry is a hyperscaling violating
Lifshitz-like geometry. A compelling and challenging, problem is then to introduce disorder in these solutions, again through the
chemical potential, and study the fate of the IR fixed point.

\section*{Acknowledgments}
We thank M. Baggioli, J. Erdmenger, L. Pando Zayas, M. P\'erez-Victoria, and I. Salazar for fruitful discussions. We also thank L. Pando Zayas
for comments on the manuscript and useful insights on the literature. Finally, we thank the Referee for a careful 
report that resulted in an improvement of the results presented in this paper.
D.A. would like to thank the Banff International Research Station, and the University of Michigan for hospitality
during the completion of this work,
and the FRont Of pro-Galician Scientists for unconditional support.
J.M.L. would like to thank the University of Pennsylvania for hospitality during early stages of this work.
The work of D.A. is supported by the German-Israeli Foundation (GIF), grant 1156. The work of J.M.L. is supported in part by the Ministry of Economy and Competitiveness (MINECO), under grant number FPA2013-47836-C3-1/2-P (fondos FEDER), and by the Junta de Andaluc{\'\i}a grant FQM 101 and FQM 6552.

\appendix
\section{The homogeneous case}
\label{app:homcase}
In this appendix we shall briefly review some interesting results of the homogeneous D3/D5 system that are
useful for the analysis in this paper.\\

For an embedding where $\chi$ and $\phi$ depend only on $z$ the DBI action \eqref{eq:action} takes
the form
\begin{equation}
{{\cal S}\over N_f\,T_{D5}\,L^6}\equiv\bar{\cal S}=-\int dt\,d^2x\,dz\,d\Omega_2\, {f\over z^4}\,
\sqrt{h\, (1-\chi^2)\,\left[  1-\chi^2+z^2 \chi'^2  -\frac{z^4(1-\chi^2)}{f^2} h\,\phi'^2\right]}\,.
\label{eq:homact}
\end{equation} 
In this action, $\phi$ is a cyclic coordinate. Thus, its conjugate generalized momentum is a conserved quantity,
\begin{equation}
d=\frac{\delta \bar {\cal S}}{\delta \phi'}=\frac{h^{\frac{3}{2}}}{f}
\frac{(1-\chi^2)^{\frac{3}{2}}\phi'}{\sqrt{ 1-\chi^2+z^2 \chi'^2  -\frac{z^4(1-\chi^2)}{f^2} h\,\phi'^2}}\,.
\label{eq:genmom}
\end{equation} 
Taking into account the asymptotic behavior of $\phi$ and $\chi$ given in Eqs. \eqref{eq:embUV}, 
one finds that $d$ is exactly the charge density $\rho$.

In order to obtain the equation of motion of $\chi$ it is useful
to compute the Legendre transformed action with respect to $\phi'$
\begin{equation}
\hat {\cal S} = \bar {\cal S}- \int dt\,d^2x\,dz\,d\Omega_2\,\, \rho\, \phi',
\end{equation} 
which expressed in terms of the variables $\chi$, $\chi'$ and $\rho$ reads
\begin{equation}
\hat{\cal S}=-\int dt\,d^2x\,dz\,d\Omega_2\,\frac{f\sqrt{h(1-\chi^2)(1-\chi^2+z^2\chi'^2)}}{z^4}\,
\sqrt{1+\frac{\rho^2 z^4}{h^2(1-\chi^2)^2}}\,,
\end{equation} 
and leads to the equation of motion
\begin{align}
&-\frac{d}{dz}\left[\frac{f\chi'\sqrt{h(1-\chi^2)}}{z^2\sqrt{1-\chi^2+z^2\chi'^2}}\sqrt{1
+\frac{\rho^2 z^4}{h^2(1-\chi^2)^2}}  \right]\nonumber\\
&=\frac{f \sqrt{h}\chi}{z^4 \sqrt{(1-\chi^2)(1-\chi^2+z^2\chi'^2)}}\sqrt{1
+\frac{\rho^2 z^4}{h^2(1-\chi^2)^2}}\left[ 2-2\chi^2+z^2\chi'^2
-\frac{2 \rho^2 z^4(1-\chi^2+z^2\chi'^2)}{\rho^2 z^4+h^2(1-\chi^2)^2}  \right]\,.
\label{eq:EoMChi}
\end{align}
Recall that when integrating this equation one must impose regularity at the horizon, which in view of the
IR expansions \eqref{eq:bhexpansion} implies $\chi'(1)=0$.
This condition fixes one of the two constants of integration
and therefore the quark mass $m$ and the quark condensate $c$ are not independent anymore.
Only numerical solutions of Eq. \eqref{eq:EoMChi} are known, and thus one cannot express $c$ as an analytic
function of $\mu$. However, as we will see in the next two sections, semi-analytical expressions
can be obtained in the small and large charge limits.

\subsection{Small charge limit}

For $\rho\to0$, Eq. \eqref{eq:EoMChi} can be expanded in powers of $\rho^2$ as
\begin{align}
&-\frac{d}{dz}\left\{\frac{f\chi'\sqrt{h(1-\chi^2)}}{z^2\sqrt{1-\chi^2+z^2\chi'^2}}\left[1+\frac{\rho^2 z^4}{2 h^2(1-\chi^2)^2} \right] \right\}\nonumber\\
&=\frac{f \sqrt{h}\chi}{z^4 \sqrt{(1-\chi^2)(1-\chi^2+z^2 \chi'^2)}}\left[ 2-2\chi^2+z^2\chi'^2
-\rho^2 z^4\frac{2-2 \chi^2+3z^2 \chi'^2 }{2 h^2 (1-\chi^2)^2}  \right]+O(\rho^4)\,.
\label{eq:EoMChiRho}
\end{align}
This equation allows us to solve for $\chi$ as a power series in $\rho^2$. Let us define
\begin{equation}
\chi(z)=\chi_0(z)+\rho^2 \chi_1(z)+O(\rho^4)\,,
\label{eq:ExpChi}
\end{equation}
and use the  asymptotic UV expansion \eqref{eq:chiUV} to express the quark condensate $c$ as
\begin{equation}
c=\frac{1}{2}\chi''_0(0)+\frac{\rho^2}{2}\chi''_1(0)+O(\rho^4)\,.
\label{eq:CAs}
\end{equation}
Next, to solve for the functions $\chi_n$ we substitute Eq.~\eqref{eq:ExpChi} in the equation of motion \eqref{eq:EoMChiRho}. 
The resulting equation can be solved order by order in $\rho^2$. At order zero we obtain the following differential equation
for $\chi_0(z)$
\begin{align}
-\frac{d}{dz}\left[\frac{f\chi_0'\sqrt{h(1-\chi_0^2)}}{z^2\sqrt{1-\chi_0^2+z^2\chi_0'^2}}\right]=\frac{f \sqrt{h}\chi_0}{z^4 \sqrt{(1-\chi_0^2)(1-\chi_0^2+z^2\chi_0'^2)}}\left[ 2-2\chi_0^2+z^2\chi_0'^2 \right]\,,
\label{eq:EoMChi0}
\end{align} 
which is nothing else than the equation of motion for $\chi(z)$ at zero charge density.
The boundary conditions for a black hole embedding with fixed mass are $\chi_0'(0)=m$ in the UV, where $m$ is the mass of the
quarks, and $\chi_0'(1)=0$ at the horizon.
Solutions with these boundary conditions exist only when $m\lesssim1.41$, while for $m>1.41$ 
only Minkowski embeddings, for which the brane never reaches the black hole horizon, 
exist~\cite{Kobayashi:2006sb,Evans:2008nf}.
We will restrict the analysis to the case $m<1.41$, which is also the range we consider in 
Sec.~\ref{ssc:bkresults}.
Once the solution for $\chi_0$ has been found, 
one can iteratively find the equations for $\chi_n$ with $n\geq1$, and solve them with the boundary conditions
$ \chi_n'(0)=\chi_n'(1)=0$. The equation for $\chi_1(z)$, which determines the coefficient of the $\rho^2$ scaling of $c$
in Eq.~\eqref{eq:CAs}, takes the form
\be
g_2 \chi_1''+ \frac{d g_2}{d z} \chi_1'+\left[ \frac{d g_1}{dz}+g_0 \right]\chi_1+\left[ \frac{d g_{-1} }{dz} +g_{-2} \right]=0\,,
\label{eq:EoMChi1}
\ee
where the functions $g_i$ are defined as
\begin{align}
&g_1=-\frac{f\sqrt{h} \chi_0 \chi_0'^3}{\sqrt{1-\chi_0^2}\left( 1-\chi_0^2+z^2 \chi_0'^2 \right)^{\frac{3}{2}}}\,,\quad
g_2=\frac{f\sqrt{h}(1-\chi_0^2)^{\frac{3}{2}}}{z^2(1-\chi_0^2+z^2 \chi_0^2)^{\frac{3}{2}}}\,,\nonumber\\
&g_0=\frac{f\sqrt{h}\left[ z^4 \chi_0'^4+3 z^2 \chi_0'^2(1-\chi_0^2)^2+2(1-\chi_0^2)^3 \right]}{z^4\left(1-\chi_0^2\right)^{\frac{3}{2}}\left(1-\chi_0^2+z^2 \chi_0'^2 \right)^{\frac{3}{2}}}\,,\nonumber\\
&g_{-1}=\frac{fz^2\chi_0'}{2h^{\frac{3}{2}}(1-\chi_0^2)^{\frac{3}{2}}\sqrt{1-\chi_0^2+z^2\chi_0'^2}}\,,\quad
g_{-2}=-\frac{f \chi_0 \left( 2-2\chi_0^2+3 z^2 \chi_0'^2 \right)}{2 h^{\frac{3}{2}}(1-\chi_0^2)^{\frac{5}{2}}\sqrt{1-\chi_0^2+z^2\chi_0'^2}}\,.
\end{align}

In order to express $c$ as a function of the chemical potential we need to determine the dependence of
$\rho$ on $\mu$.
The chemical potential can be expressed as 
\begin{equation}
\mu=-\int_0^1 dz \,\phi'\,,
\end{equation}
where we have taken into account that $\phi(1)=0$, and $\phi(0)=\mu$.
Since $\phi'(z)$ as function of $\rho$ and $\chi(z)$ is
\begin{equation}
\phi'=-\frac{\delta \hat {\cal S}}{\delta \rho}\,,
\end{equation}
one finds
\begin{equation}
\mu=\rho\int_{0}^1 dz\,\frac{ f\sqrt{1-\chi^2+z^2 \chi'^2}}{\sqrt{h(1-\chi^2)[h^2 (1-\chi^2)^2+\rho^2 z^4]}}\,.
\label{eq:muvsrhomass}
\end{equation}
The limit $\rho \to 0$ is easily found using Eq.~\eqref{eq:ExpChi},
\begin{equation}
\mu=\rho\int_{0}^1 dz\,\frac{ f\sqrt{1-\chi_0^2+z^2 \chi_0'^2}}
{h^{\frac{3}{2}}(1-\chi_0^2)^{\frac{3}{2}}}\left[1+O(\rho^2)\right]\,.
\label{eq:muvsrhosmrhomass}
\end{equation} 
Taking into account Eq.~\eqref{eq:CAs}, we can express $c$ as function of $\mu$ for a fixed mass and $\mu \to 0$,
\begin{equation}
c=\frac{1}{2}\chi''_0(0)+\frac{\chi''_1(0)}{2\left[\int_{0}^1 dz\,
\frac{ f\sqrt{1-\chi_0^2+z^2 \chi_0'^2}}{h^{\frac{3}{2}}(1-\chi_0^2)^{\frac{3}{2}}}\right]^2}\,\mu^2+O(\mu^4)\,.
\label{eq:cvsmusmmu}
\end{equation} 
Therefore, we see that at small enough $\mu$, the quark condensate scales quadratically with the chemical
potential. This is confirmed by the numerical analysis shown in Fig.~\ref{fig:massembrhoc}
where we considered an embedding with $m=0.5$. 
In Eq.~\eqref{eq:cvsmum05smrho} we show the explicit form of Eq.~\eqref{eq:cvsmusmmu}, after numerically
solving for $\chi_0$ and $\chi_1$ from Eqs.~(\ref{eq:EoMChi0}, \ref{eq:EoMChi1})
 for an embedding
with $m=0.5$.  The fit to the numerical data at low $\rho$ presented in Fig.~\ref{fig:massembrhoc} agrees very well with the 
prediction of Eq.~\eqref{eq:cvsmum05smrho}.

\subsection{Large charge limit}

The limit $\rho \to \infty$ can be analyzed in a similar way. However, there are some differences as
Eq.~\eqref{eq:EoMChi} depends on $\rho^2\,z^4$, and thus there is always a neighborhood of $z=0$, 
where $\rho^2 z^4$ stays finite in the limit $\rho\to\infty$. 
This makes $z=0$ a special point. If one tries to expand
$\chi(z)=\chi_0(z)+\chi_1(z) \rho^{-1}+O\left(\rho^{-2}\right)$, nontrivial solutions to the resulting equations cannot
satisfy $\chi'(1)=0$ and  $\chi(0)=0$ simultaneously. To address this problem we perform the change 
of variables $s=\sqrt{\rho}\,z$, and $\bar\chi=\sqrt{\rho}\,\chi$. The resulting equation can be solved 
perturbatively with the desired boundary conditions as we now see. 
The expansion of the equation after the change of variables in powers of $\rho^{-1}$ is
\begin{align}
-\frac{d}{ds}\left\{ \frac{\sqrt{1+s^4} \bar\chi'}{s^2} -\frac{\left[(1+s^4)\bar \chi'^2 -2s^2 \bar \chi^2 \right] \bar \chi'}{2\rho \sqrt{1+s^4}}\right\}=\frac{2\bar \chi}{s^4 \sqrt{1+s^4}}&\nonumber\\
-\frac{\left[ 2\bar\chi^2+s^2(1+s^4)\bar \chi'^2 \right]\bar \chi}{\rho(1+s^4)^{\frac{3}{2}}}+O(\rho^{-2}),&
\label{eq:LargeRho}
\end{align} 
with the boundary conditions
\begin{align}
&\bar \chi'(0)=m,\nonumber\\
&\bar \chi'\left(\sqrt{\rho}\right)=0.
\end{align}
Eq.~\eqref{eq:LargeRho} can be solved in powers of $\rho^{-1}$, hence we expand
\begin{equation}
\bar\chi(s)=\bar\chi_0(s)+\bar\chi_1(s)\rho^{-1}+O(\rho^{-2})\,.
\label{eq:LargeRhoAs}
\end{equation} 
Inserting this expansion in Eq.~\eqref{eq:LargeRho}, we find an equation for every $\bar\chi_n$ that can be 
solved iteratively. 
In terms of the original fields and variables we can write
\begin{equation}
\chi(z)=\frac{1}{\sqrt{\rho}}\left[\bar\chi_0\left(\sqrt{\rho}z\right)
+\bar\chi_1\left(\sqrt{\rho}z\right)\rho^{-1}+O\left(\rho^{-2}\right)\right]\,,
\label{eq:LargeRhoAs2}
\end{equation} 
which leads to the following expression for $c$ at large $\rho$,
\begin{equation}
c=\frac{\sqrt{\rho}}{2}\, \bar \chi''_0(0)+\frac{1}{2\sqrt{\rho}} \bar \chi''_1(0)
+ O\left(\rho^{-\frac{3}{2}}\right).
\label{eq:Cchi}
\end{equation} 
The equation determining $\bar \chi_0$ reads
\begin{align}
s^2(1+s^4)\,\bar \chi_0''-2s\,\bar\chi_0'+2\bar \chi_0=0\,.
\end{align} 
We are interested in solutions satisfying
$\bar\chi'_0(0)=m$, and $\displaystyle \lim_{s\to\infty} \bar\chi_0(s) <\infty$. 
They are given by
\begin{equation}
\bar \chi'_0(s)=m\,s+m\,\frac{1+i}{2}\frac{s}{K(-1)}\,
F\left( i\,\mathrm{arcsinh}\left( \frac{1+i}{\sqrt{2}}\, s\right),-1 \right)\,,
\end{equation} 
where $F(\phi,n)$ is the elliptic integral of the first kind, and $K(n)$ is the complete elliptic integral of 
the first kind. Plugging this solution into Eq.~\eqref{eq:Cchi} we obtain
\begin{equation}
c=- \frac{m}{\sqrt{2}K(-1)}\sqrt{\rho}+O\left( \rho^{-\frac{1}{2}} \right).
\label{eq:LargeRhoC}
\end{equation} 
Finally to write $c$ as a function of $\mu$, we use \eqref{eq:LargeRhoAs2} in Eq.~\eqref{eq:muvsrhomass}
and arrive to 
%
\begin{equation}
\mu=\rho\int_{0}^1 dz\,\frac{ f}{\sqrt{h(h^2+\rho^2 z^4)}}\left[1+O(\rho^{-1})\right].
\end{equation} 
This integral is solved in Eq.~\eqref{eq:murhoev} below (for massless embeddings). 
Here, we anticipate the asymptotic behavior \eqref{eq:murholgrho} when $\rho\to\infty$, and write
\begin{equation}
\mu = {\Gamma({1\over4})\,\Gamma({5\over4})\over\sqrt{\pi}}\,\sqrt{\rho}+O(\rho^0)\,.
\label{eq:muvsrholgrhomass}
\end{equation} 
Combining this with Eq.~\eqref{eq:LargeRhoC} we finally arrive to the following expression for $c$ as a function
of $\mu$ in the $\mu\to\infty$ limit
\begin{equation}
c=\frac{\sqrt{\pi}}{\sqrt{2}\,\,\Gamma\left(\frac{1}{4}  \right)\Gamma\left(\frac{5}{4}  \right)K(-1)}\,m\, \mu +O(\mu^0)\,,
\label{eq:cvsmulgmu}
\end{equation} 
which for the case $m=0.5$ results in the expression \eqref{eq:cvsmum05lgrho}, which agrees reasonably well with the
fit to the numerical data in Fig.~\ref{fig:massembrhoc}.

\subsection{Massless embeddings}
In the following we will restrict the analysis
to the massless homogeneous embedding where $\chi$ is identically zero, and $\phi$ is a function of $z$ only.
In this case, the DBI action \eqref{eq:homact} simplifies further to
\be
\bar{\cal S}=-\int dt\,d^2x\,dz\,d\Omega_2\, f\,z^{-4}\,\sqrt{h\left(1-{z^4\over f^2}\,h\,
\phi'^2\right)}\,,
\label{eq:homactm0}
\ee
and Eq.~\eqref{eq:genmom} results in the following expression for the charge density
\be
\rho={h^{3\over2}\over f}\,{\phi'\over\sqrt{1-{z^4\over f^2}\,h\,\phi'^2}}\,.
\label{eq:eomphih}
\ee
It is then straightforward to solve for $\phi'$ in terms of $\rho$,
which allows us to express the chemical potential as the following integral
\begin{align}
\mu&=-\int_0^1 dz \,\phi'\nonumber \\
&=\rho\,\int_0^1 dz\, {f\over \sqrt{h\left(z^4\,\rho^2+h^2\right)}}\,,
\label{eq:murhoint}
\end{align}
and after applying the following change of variables
\be
z^2={2\over u^2+\sqrt{u^4-4}}\,,
\label{eq:ztou}
\ee
the integral can be evaluated analytically as
\begin{eqnarray}
\mu&=&\int_{\sqrt{2}}^\infty du\,{\rho\over\sqrt{\rho^2+u^4}}\nonumber \\
&=&{\rho\over\sqrt{2}}\,_{1}F_2\left({1\over4},{1\over2},{5\over4};-{\rho^2\over4}\right)\,,
\label{eq:murhoev}
\end{eqnarray}
where $_{1}F_2$ is the hypergeometric function. This hypergeometric function can be easily
expanded in powers of $\rho$ both in the large and small $\rho$ limits, resulting in
\begin{subequations}
\begin{eqnarray}
&&\mu={\rho\over\sqrt{2}}+ O(\rho^3)\,;\qquad (\rho\ll1)\,,\label{eq:murhosmrho}\\
&&\mu = {\Gamma({1\over4})\,\Gamma({5\over4})\over\sqrt{\pi}}\,\sqrt{\rho}
-{4\sqrt{2}\,\Gamma({5\over4})\over\Gamma({1\over4})}+O(\rho^{-2})\,;\qquad (\rho\gg1)\,,
\label{eq:murholgrho}
\end{eqnarray}
\label{eq:murholim}
\end{subequations}
It will be useful to solve for $\rho$ instead,
\begin{subequations}
\label{eq:rhomulim}
\begin{eqnarray}
&&\rho\approx\sqrt{2}\mu\,;\qquad (\rho\ll1)\,,\label{eq:rhomusrho} \\
&&\rho\approx 0.291\mu^2+0.823\mu+0.58\,;\qquad (\rho\gg1)\,.
\label{eq:rhomulrho}
\end{eqnarray}
\end{subequations}

\subsubsection{DC Conductivity}

We shall now turn our attention to the DC conductivity, which in Eq. \eqref{eq:sigdcm0} was expressed
in terms of the behavior of $\phi$ at the horizon. For an homogeneous embedding, that equation reduces to
\be
\sigma_{\rm DC}={1\over \sqrt{1-(a^{(2)})^2/2}}\,,
\label{eq:sigdchom}
\ee
where $a^{(2)}$, which is defined through the IR expansion \eqref{eq:phibh}, is now independent of $x$,
and using \eqref{eq:eomphih}
can be easily expressed in terms of $\rho$ as
\be
a^{(2)}=\pm {\sqrt{2}\rho\over \sqrt{4+\rho^2}}\,.
\label{eq:a2rho}
\ee
Finally one can write the DC conductivity in terms of the charge density as
\be
\sdc={1\over2}\,\sqrt{4+\rho^2}\,.
\label{eq:sdchom}
\ee
It is illustrative to recall both the large and small $\rho$ limits of this result
\begin{eqnarray}
&&\sdc = 1+{\rho^2\over8}+O(\rho^4)\;;\quad (\rho\sim0)\,,\\
\label{eq:sdchlwrho}
&&\sdc={\rho\over2}+O(1/\rho)\;;\quad (\rho\to\infty)\,.
\label{eq:sdchhgrho}
\end{eqnarray}


\section{Numerical coda}
\label{app:numstability}

In this appendix we present additional numerical results for the study of the dependence of $\sdc$ on
the charge density performed in Sec.~\ref{ssec:sdcvsrho0}.

\begin{figure}[htb]
\begin{center}
\includegraphics[width=0.49\textwidth]{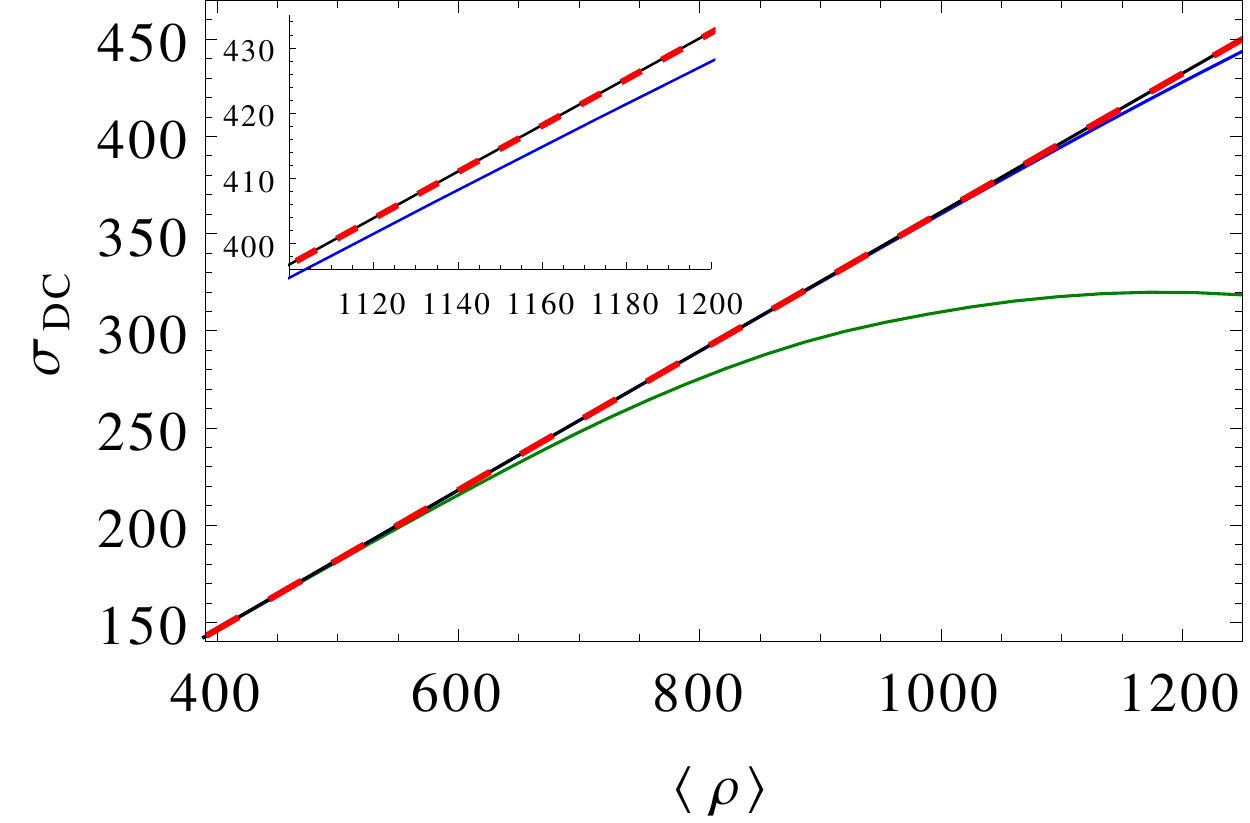}
\includegraphics[width=0.49\textwidth]{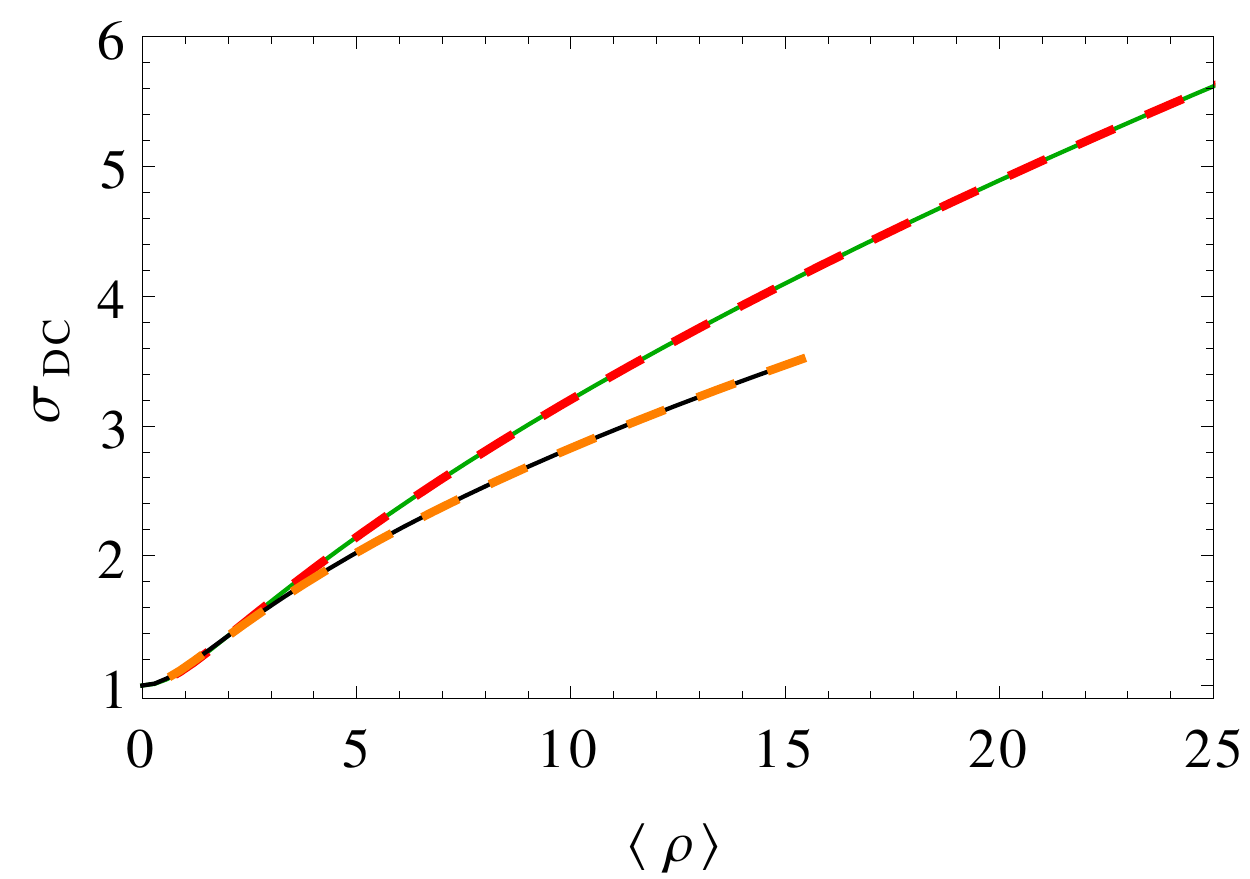}
\caption{\label{fig:numstability}DC conductivity versus charge density. 
On the left we present data 
at $w=3$ for lattices of size
$60\times60$ (green solid line), $80\times80$ (blue solid line), $100\times100$ (red dashed line)
and $120\times120$ (black solid line). 
We have set $L_x=20\pi$, $k_*=1$, and averaged over 25 realizations.
On the right we plot the results for lattices of size $100\times 100$ (dashed lines) and $120\times 120$ 
(solid lines) at strong noise: $w=6$ (green and red lines) and $w=8$ (black and orange lines).
We have set $L_x=20\pi$, $k_*=1$, and averaged over 43 and 57 realizations for $w=6$ and $w=8$ respectively.}
\end{center}
\end{figure}

First, in order to assert the reliability of the numerical simulations generating 
Figs.~\ref{fig:sdcvsrho} and \ref{fig:sdcvsrhoStw}
we study the stability of the value of $\sdc$ against the lattice size. 
In Fig~\ref{fig:numstability} left,
we plot $\sdc$ versus $\langle\rho\rangle$ at $w=3$ for lattices of size
$60\times60$, $80\times80$, $100\times100$ and $120\times120$. One can observe that for lattices of size
$100\times100$ and larger the
results have converged and become stable against the increase of lattice size.
On the right panel of Fig.~\ref{fig:numstability} we perform a similar exercise for $w=6$ (green and red lines),
and $w=8$ (black and orange lines). In this case we just compare lattices of size $100\times100$,
and $120\times120$. It becomes clear that for the ranges of $\langle\rho\rangle$ under study the results have converged
too.

\begin{figure}[htb]
\begin{center}
\includegraphics[width=0.7\textwidth]{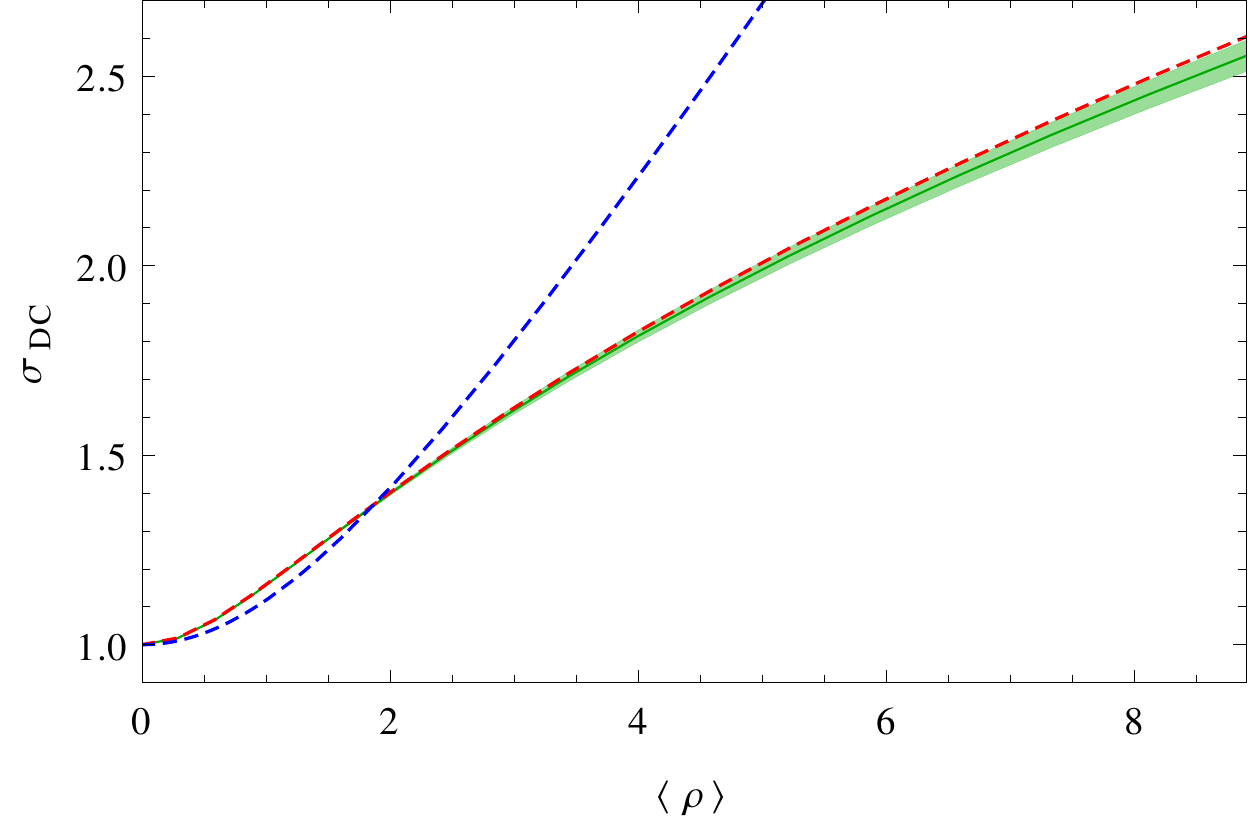}
\caption{\label{fig:sigvsrhow10}DC conductivity versus charge density at $w=10$ (green solid line).
The red dashed line shows the prediction of Eq.~\eqref{eq:sdcall} after averaging over 1000 realizations, and the blue dashed line the
results for the clean system.
We have used a grid of size $120\times 120$, set $k_*=1$, $L_x=20\pi$, and averaged over $30$ realizations.
The shaded bands depict the error of the average.} 
\end{center}
\end{figure}

We close this appendix by extending the analysis in Fig.~\ref{fig:sdcvsrhoStw} to
the case of a stronger noise with $w=10$.
In Fig.~\ref{fig:sigvsrhow10} we plot our numerical result for
$\sdc$ vs $\langle\rho\rangle$ (green solid line), and compare it to the
semi-analytical prediction \eqref{eq:sdcall} (red dashed line).
Although for this noise strength our numerics do not reach large values of the charge density,
this result seems to confirm the observation made below
Fig.~\ref{fig:sdcvsrhoStw} that for $w=8$ and higher, the numerical results fall slightly 
below the semi-analytical prediction for large enough $\langle\rho\rangle$. 

\section{Embeddings with correlated noise}
\label{app:corremb}
\begin{figure}[htb]
\begin{center}
\includegraphics[width=0.32\textwidth]{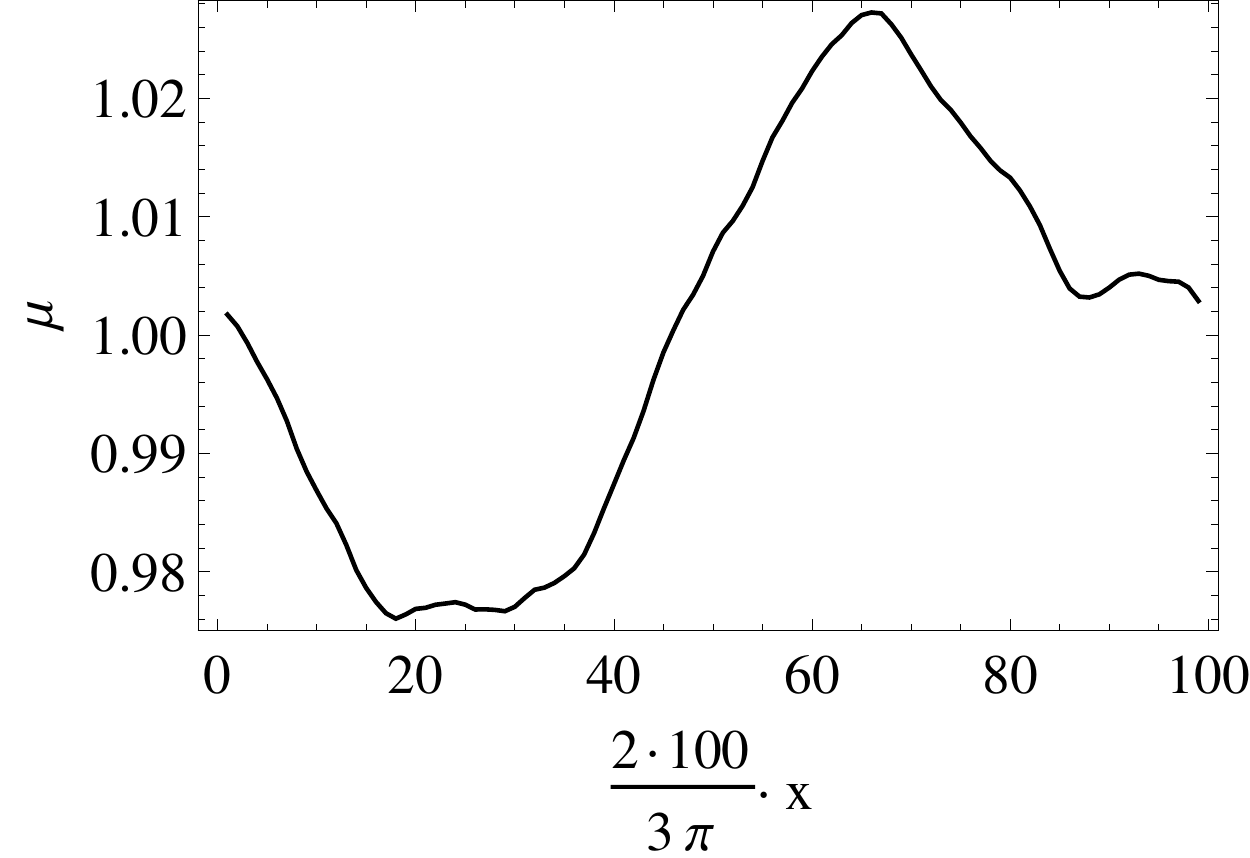}
\includegraphics[width=0.32\textwidth]{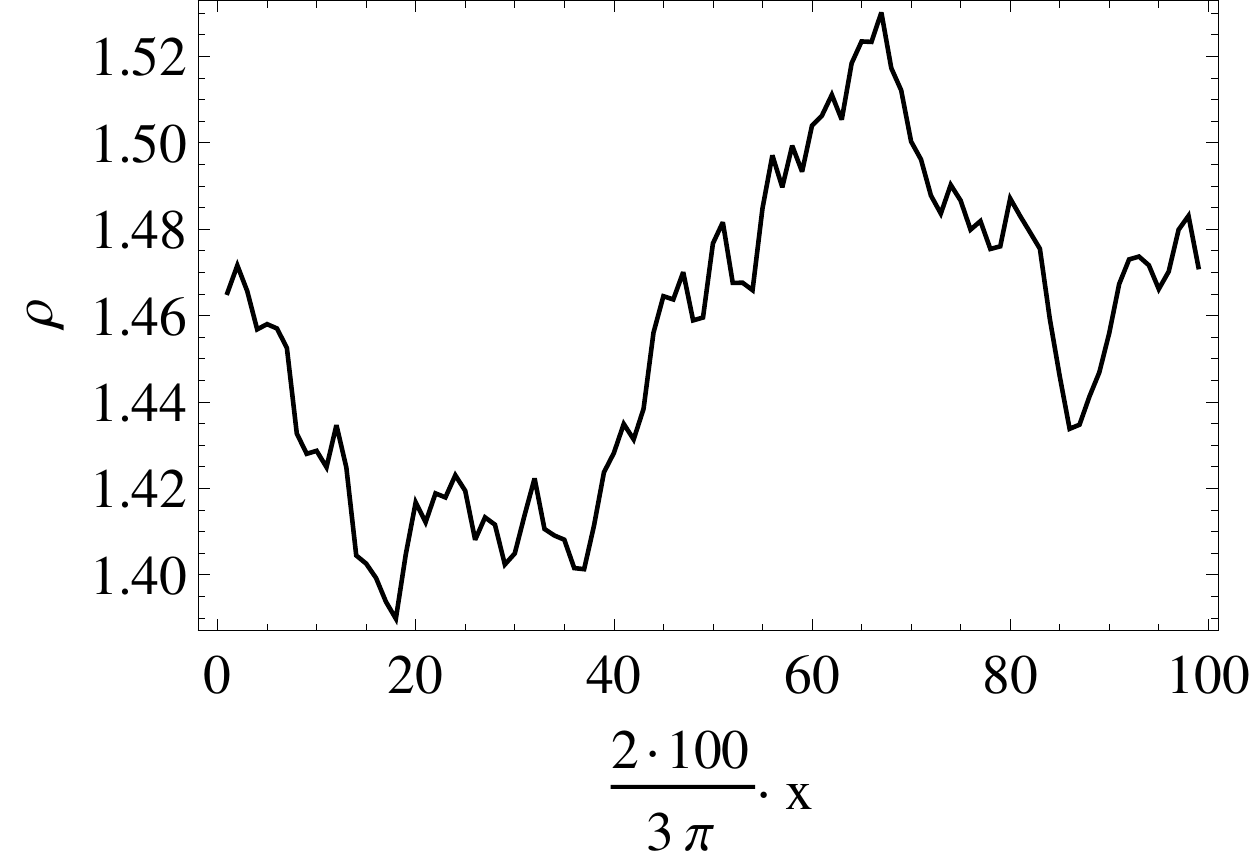}
\includegraphics[width=0.32\textwidth]{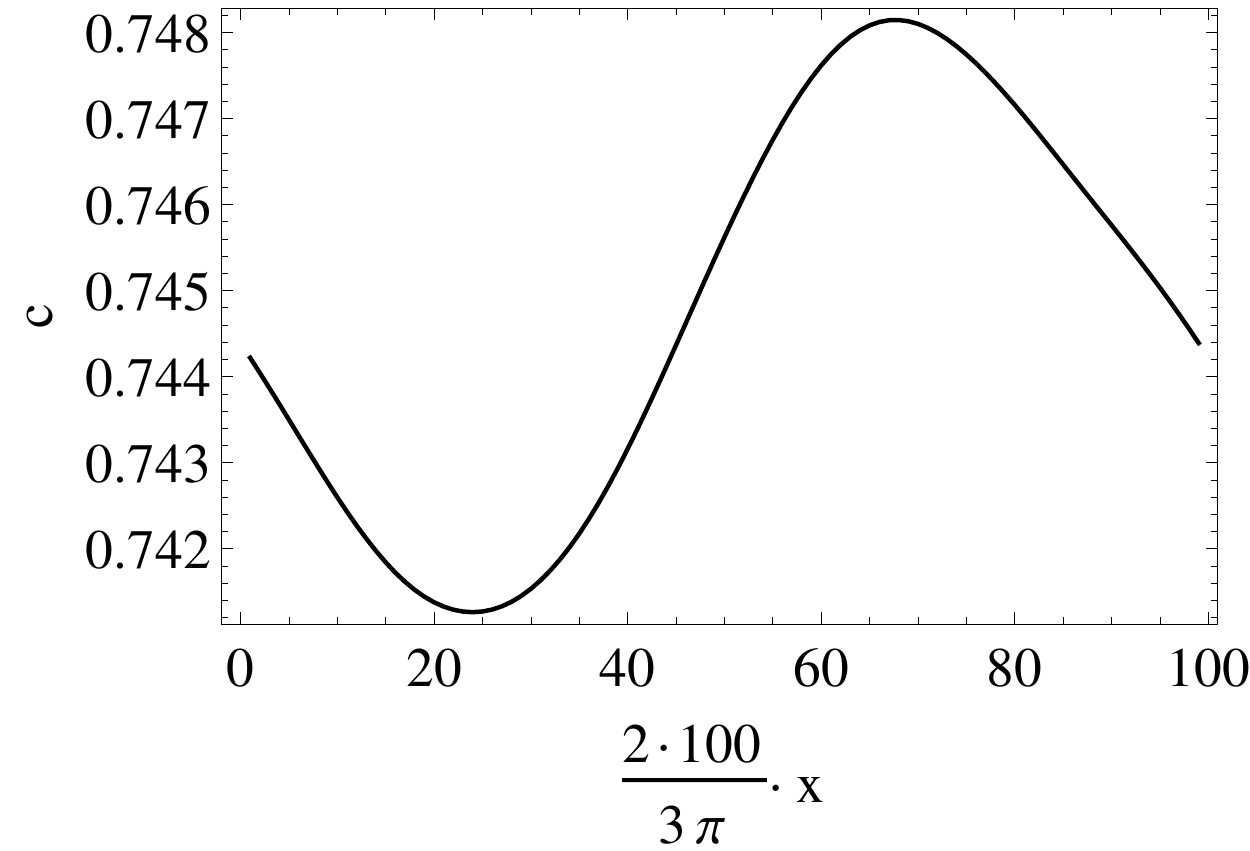}
\caption{\label{fig:corremb}
Massive embedding. The three panels display, from left to right, the chemical potential 
$\mu(x)$, the charge density $\rho(x)$, and the quark condensate $c(x)$ for a simulation  
with $\mu_0=1$, $m=0.5$, $w=1$, $\alpha=2$, $L_x=3\pi/2$, and $k_*=62.7$ (corresponding to 47 modes)
on a lattice of size $N_z=40$, $N_x=100$.} 
\end{center}
\end{figure}

In this appendix we show the result of one simulation for the massive embeddings with correlated noise
\eqref{eq:noisecorr} considered in Sec.~\ref{sec:spectral}. In Fig.~\ref{fig:corremb} we plot the
input chemical potential $\mu(x)$, together with the output charge density $\rho(x)$, and quark condensate $c(x)$.
These plots illustrate neatly the fact that while the charge density is roughened as a consequence
of the enhancement of the higher wave numbers in the sum \eqref{eq:noisecorr}, the  quark condensate is 
smoothed out due to the suppression of those same higher modes.

\bibliography{noisy}

\end{document}